\documentclass[manuscript]{acmart}
\usepackage{array} 

\usepackage{latexsym}
\usepackage{longtable}
\usepackage{multirow}
\usepackage{graphicx}
\usepackage{subcaption}
\usepackage{fdsymbol}

\usepackage{xcolor}
\usepackage{mdframed} 
\usepackage{color}

\RequirePackage{microtype}
\RequirePackage{etoolbox}
\RequirePackage{booktabs}
\RequirePackage{refcount}
\RequirePackage{totpages}
\RequirePackage{environ}
\usepackage{graphicx}
\newmdenv[
  linecolor=black,
  backgroundcolor=gray!20, 
  frametitlebackgroundcolor=gray!20, 
  frametitlerule=false,
  leftmargin=10pt,
  rightmargin=10pt,
  innerleftmargin=10pt,
  innerrightmargin=10pt,
  innertopmargin=10pt,
  innerbottommargin=10pt
]{myframe}
 
\AtBeginDocument{%
  }

\renewcommand\footnotetextcopyrightpermission[1]{}  
\begin{document}

\title{Evaluating Conversational Recommender Systems via Large Language Models}
\subtitle{A User-Centric Framework}

\author{Nuo Chen}
\orcid{0000-0001-8600-8203}
\affiliation{%
  \institution{The Hong Kong Polytechnic University}
  \country{HK, China}
}
\email{pleviumtan@outlook.com}

\author{Quanyu Dai}
\orcid{0000-0001-7578-2738}
\authornote{Corresponding author}
\affiliation{%
  \institution{Huawei Noah's Ark Lab}
  \country{China}}
\email{daiquanyu@huawei.com}

\author{Xiaoyu Dong}
\orcid{0000-0002-4788-9279}
\affiliation{%
  \institution{The Hong Kong Polytechnic University}
  \country{HK, China}
}

\author{Piaohong Wang}
\orcid{0009-0008-8393-7101}
\affiliation{%
  \institution{City University of Hong Kong}
  \country{HK, China}}

\author{Qinglin Jia}
\orcid{0000-0002-3583-6719}
\affiliation{%
  \institution{Huawei Noah's Ark Lab}
  \country{China}}

\author{Zhaocheng Du}
\orcid{0000-0002-1811-129X}
\affiliation{%
  \institution{Huawei Noah's Ark Lab}
  \country{China}}

\author{Zhenhua Dong}
\orcid{0000-0002-2231-4663}
\affiliation{%
  \institution{Huawei Noah's Ark Lab}
  \country{China}}

\author{Xiao-Ming Wu}
\orcid{0000-0002-3130-0554}
\affiliation{%
  \institution{The Hong Kong Polytechnic University}
    \country{HK, China}
  }
\email{xiao-ming.wu@polyu.edu.hk}

\renewcommand{\shortauthors}{Chen et al.}

\begin{abstract}
Conversational recommender systems (CRSs) integrate both recommendation and dialogue tasks, making their evaluation uniquely challenging. Existing approaches primarily assess CRS performance by separately evaluating item recommendation and dialogue management using rule-based metrics. However, these methods fail to capture the real human experience, and they cannot draw direct conclusions about the system's overall performance. As conversational recommender systems become increasingly vital in e commerce, social media, and customer support, the ability to evaluate both recommendation accuracy and dialogue management quality using a single metric, thereby authentically reflecting user experience, has become the principal challenge impeding progress in this field.

In this work, we propose a user-centric evaluation framework based on LLMs for CRS, namely Conversational Recommendation Evaluator (\textbf{CoRE})~\footnote{This article is a follow-up study of~\citet{chen2025evaluating}, which has been accepted by HCRS@TheWebConf 2025 but has not been publicly published. Sections \ref{ch:factors} and \ref{ch:benchmark} of this paper are based on \citet{chen2025evaluating}, and the remaining sections comprise newly introduced research content.}. 
CoRE consists of two main components: (1) \textbf{LLM-As-Evaluator.} Firstly, we comprehensively summarize 12 key factors influencing user experience in CRSs and directly leverage LLM as evaluator to assign a score to each factor. (2) \textbf{Multi-Agent Debater.} Secondly, we design a multi-agent debate framework with four distinct roles (common user, domain expert, linguist, and HCI expert)  to discuss and synthesize the 12 evaluation factors into a unified overall performance score.

Furthermore, we apply the proposed framework to evaluate four CRSs on two benchmark datasets. The experimental results show that CoRE aligns well with human evaluation in most of the 12 factors and the overall assessment. Especially, CoRE's overall evaluation scores demonstrate significantly better alignment with human feedback compared to existing rule-based metrics. 

Through the integration of large language models driven multi-factorial quantitative assessment and a multi agent debate mechanism, CoRE not only delivers scoring results that closely match human annotations but also establishes an end to end reproducible user centric evaluation benchmark for conversational recommender systems, thereby promoting the coordinated advancement of evaluation standardization and system optimization in the domain of CRS evaluation.
\end{abstract}



\received{20 February 2007}
\received[revised]{12 March 2009}
\received[accepted]{5 June 2009}


\maketitle

\section{Introduction}
\begin{figure*}[t]
  \centering
  \includegraphics[width=.99\linewidth]{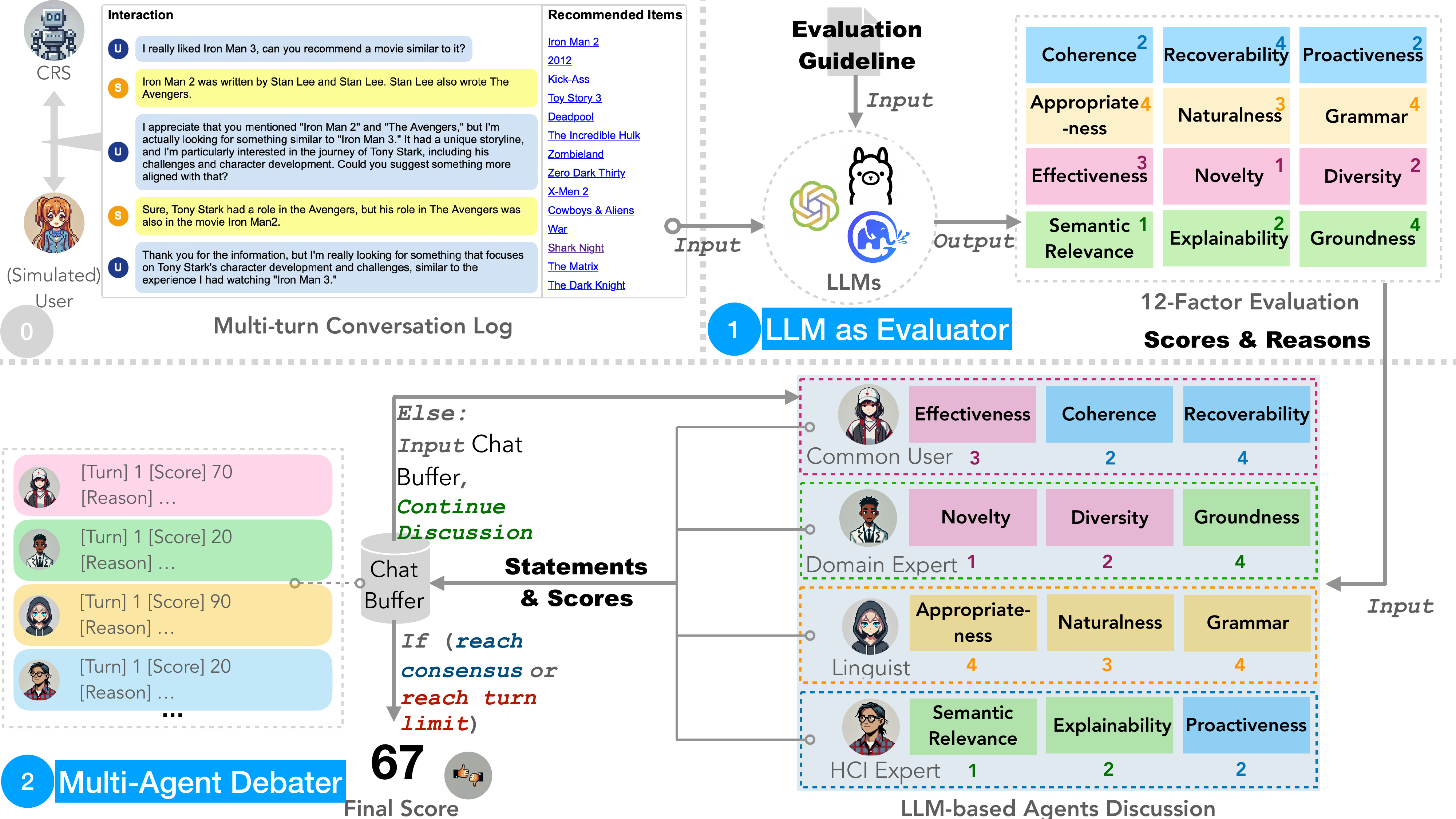}
  \caption{The overview of our method. We evaluate user interactions with conversational systems through a large language model across 12 factors. The scores and justifications for the twelve factors are used as input to the Multi-Agent Debater, where each agent receives inputs corresponding to the specific factors associated with its designated role. After several rounds of debate, the agents produce a final overall score.}
  \vspace{-1em}
  \label{fig:flow}
\end{figure*}

Conversational recommendation (CRS), which refers to systems that provide recommendations to users through multiple rounds of conversations~\cite{li2018towards}, has attracted considerable attention because of its increasingly important role in people's daily lives~\cite{fu2020tutorial}. Among them, those that do not rely on predefined templates but instead generate dialogue responses in an \textit{ad hoc} manner are referred to as \textit{chit-chat CRSs} (e.g.,~\cite{chen2019kbrd,wang2022barcor,wang2022unicrs,mele2021,pramod2022}). In this paper, unless stated otherwise, the term CRS refers specifically to chit-chat CRS.

\begin{figure*}[t]
  \includegraphics[width=0.58\linewidth]{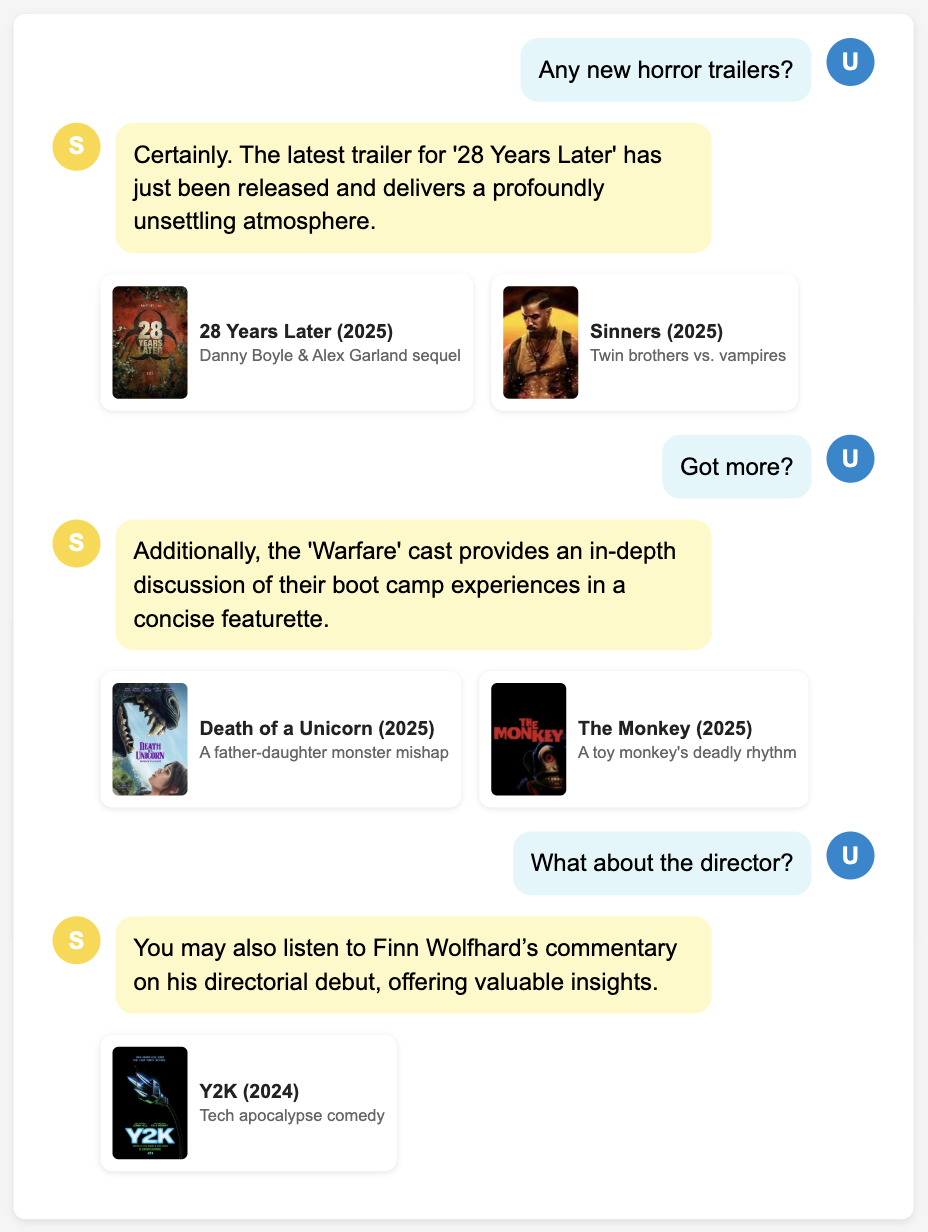}
  \caption{An example of a user (U) interacting with a chit‑chat–style conversational recommender system (S). The CRS uses natural language to suggest movies to the user, while also presenting a separate list of recommended items at each turn. The items in that list don’t necessarily match exactly those mentioned in the conversation.}
  \label{fig:crs_demo}
\end{figure*}

Unlike traditional Recommender Systems (RSs), which model user interests based on implicit feedback such as click-through rates or browsing history and generate item recommendations accordingly~\cite{schafer2007collaborative, ekstrand2011collaborative, koren2008}, Conversational Recommender Systems (CRSs) employ language models to collect and interpret explicit user feedback through interactive dialogue. Notably, CRSs not only generate natural language responses but can also present a separate list of recommended items~\citep{jannach2021, GAO2021100}. 
Figure~\ref{fig:crs_demo} illustrates a multi-turn interaction where the user requests horror movie recommendations and information about a director, and the system responds accordingly. Based on the dialogue, a separated list of inferred relevant items is displayed on the right.

From a task-centered perspective, in contrast to traditional RSs that focus exclusively on generating item suggestions through user preference modeling, CRSs are additionally required to perform \textit{dialogue management} to deliver recommendations within the flow of conversation~\cite{lei2020conversational,sun2018}. \textit{Dialogue Management},  which refers to maintaining coherent and goal-oriented multi-turn interactions by tracking the dialogue state (e.g., conversation history) and selecting the system’s next action (e.g., asking, clarifying, or providing information) based on user input and context, constitutes a core component of the conversational recommendation task alongside item recommendation~\citep{xu-seneff-2010-dialogue,ge-xu-2015-dialogue}. Moreover, in contrast to general dialogue or question-answering systems, the dialogue management in CRSs also involves delivering recommendations as part of the conversation.

The intrinsic complexity of the conversational recommendation task, which stems from the integration of dialogue management and recommendation generation, not only distinguishes CRSs from both traditional recommender systems and conventional dialogue systems, but also presents substantial challenges for the evaluation of CRSs~\citep{jannach2022}. On the one hand, CRSs must purposefully drive users to express their potential needs in dialogue, ensuring the natural and comfortable user dialogue experience; and on the other hand, CRSs also need to infer the user's item preferences through dialogue and provide accurate recommendations in conversation and in the separated list. Nevertheless, current evaluation practices of CRSs fail to address the integrated nature of dialogue flow management and user preferences modeling, instead treating item recommendation and dialogue management as isolated tasks. Specifically, most studies adopt task-specific rule-based metrics, such as Recall for recommendation accuracy and BLEU for dialogue quality~ \cite{chen2019kbrd,wang2022barcor,wang2022unicrs,zhang2024empirical,feng2023large}. This ignores the nature of conversational recommendation in which dialogue management is integrated with user preference modelling, resulting in a biased evaluation.   

The first issue with most current evaluation paradigms is fragmented evaluation. 
The fragmented evaluation approach overlooks the interdependencies between recommendation and conversational dynamics, limiting its ability to holistically and fairly evaluate CRSs' performance. This shortcoming directly hinders researchers from fairly evaluating the system's overall performance. Specifically, the fragmented evaluation methods struggle to address the practical overall CRSs evaluation problem:

    \emph{If System A excels in recommendation but underperforms in dialogue management, while System B is the opposite, which system performs better overall?}

Another issue with most current evaluation paradigms is their misalignment with user experience. Empirical studies argued that rule-based evaluation metrics, such as Recall and BLEU, often fail to align with actual user experiences~\citep{reiter-2018-structured,chen2017meta,cremonesi2012, Manzoor2024ChatGPT}. These metrics are insufficient for accurately reflecting the quality of conversational recommendations and may instead introduce biased or unreasonable estimates \cite{rosario2023grading}.

Some prior research in dialogue systems and CRS evaluation has identified several key factors, such as accuracy, novelty, proactiveness and explainability, that influence user experience through empirical studies~\citep{cai2020,cai2022,siro2023, jin2021, jin2024}. Meanwhile, some studies (e.g., \citet{sakai2023swan, langevin21}) have proposed abstract frameworks aiming to integrate multiple dimensions of system performance into a unified assessment that can be used for CRS evaluation. But these efforts have yet to yield standardized metrics for the direct assessment of CRS performance.

On the other hand, recent studies argue that Large Language Models (LLMs), demonstrating considerable natural language understanding capabilities, have significant potential to align with human preferences in text quality evaluation~\citep{liu-etal-2023-g,fu-etal-2024-gptscore,chen-etal-2023-exploring-use,chiang2023large,wang2024pandalmautomaticevaluationbenchmark,gao2023human}. Thus, this positions LLMs as a promising tool for intelligent evaluation of CRSs. However, despite this potential, only a few studies~\cite{wang2023-rethink,huang2024concept} have explored LLM-based evaluation for conversational recommendation tasks. Moreover, none of these studies have addressed the holistic assessment of CRSs integrating recommendation with dialogue management, leaving a gap in methodologies for unified performance measurement.

To address this gap, in this work, we propose a user-centric evaluation framework based on LLMs for CRSs, namely \textbf{Co}nversational \textbf{R}ecommendation \textbf{E}valuator (\textbf{CoRE}). As shown in Figure~\ref{fig:flow}, the framework consists of two sequential components, with the upper right part representing the first part, and the lower section representing the second part.

The first part is \textbf{LLM-As-Evaluator}: LLMs are employed as evaluators to assess conversation logs between users and CRSs. The evaluation focuses on twelve key factors influencing user experience in conversational recommendation, such as \textit{coherence} and \textit{explainability}. Each factor is assigned a score ranging from 0 to 4, accompanied by detailed reasoning. This component is depicted in the top right part of Figure~\ref{fig:flow}.

The second part is \textbf{Multi-Agent Debater}:
We design a multi-agent debate framework to derive a unified overall performance score based on the evaluation scores and explanations of the twelve factors. Specifically, four agents with distinct roles (i.e., a common user, a domain expert, a linguist, and an HCI expert) representing evaluators with different backgrounds and specialized knowledge, engage in discussions and collaboratively determine a final overall score ranging from 0 to 100. This design overcomes the limitations of relying solely on a single annotator, which may introduce bias and instability, and aligns with best practices that emphasize collaboration among multiple annotators~\citep{van-der-lee-etal-2019-best,karpinska2021perils}. This component is depicted in the bottom part of Figure~\ref{fig:flow}.

Based on our proposed framework, we formulate four \textbf{Research Questions} (RQs):

\begin{itemize}
    \item \textbf{RQ1}: What are the score correlations among the twelve evaluation factors in the CoRE framework as produced by human annotators versus LLM-based evaluators?
    \item \textbf{RQ2}: To what extent are the evaluation results of the twelve key factors by LLMs consistent with those provided by human evaluators?
    \item \textbf{RQ3}: Compared to existing metrics (e.g., recall, persuasiveness), how consistent are the evaluation results of our proposed method with those provided by human evaluators?
    \item \textbf{RQ4}: How do the major conversational recommender systems perform under our evaluation framework?
\end{itemize}

To answer the above research questions, we conducted comprehensive experiments on two widely-used benchmark datasets, ReDial~\citep{li2018towards} and OpenDialKG~\citep{moon-etal-2019-opendialkg}, involving four representative CRSs, including both neural-based models (BARCOR~\citep{wang2022barcor}, KBRD~\citep{chen2019kbrd}, UniCRS~\citep{wang2022unicrs}) and a GPT-based model (CHATCRS~\citep{wang2023-rethink}). 
We adopted a modified user simulator based on~\citet{wang2023-rethink} to generate multi-turn dialogue logs that realistically mimicked user interactions by providing vague requests and ensuring a dynamic conversation. The generated logs were then evaluated using various large language models (e.g., GPT-4o-mini, GLM-4-Air, the Llama-3 family, the Mistral family, and the Qwen-3 family) across the twelve human-centered factors. Then the Multi-Agent Debater based on GPT-4o-mini was introduced to aggregate factor-wise scores provided by GPT-4o-mini into a unified overall assessment score. Additionally, to validate the proposed framework, we conducted a human evaluation where ten independent annotators rated a total of 80 randomly selected dialogue logs from the test sets. Each evaluator scored the logs on the same twelve factors and provided an overall score, allowing for direct comparison with the LLM-based assessments. 

From experimental results, we have the following key findings: (1) The key factors in the proposed framework show a degree of independence under human evaluation, but LLM-based evaluators tend to blur those boundaries to different degrees: some models introduce only moderate coupling, while others create much stronger inter-factor interdependencies. (2) The proposed framework~(CoRE), with appropriate LLM serving as the evaluator, aligns well with human evaluation in most of the key factors influencing user experience in conversational recommendations. (3) CoRE, with appropriate LLM serving as the evaluator, aligns well with human evaluation in terms of CRS overall performance evaluation. (4) LLMs have strong potential of being used directly as CRS. Neural network-based CRS could exhibit deficiencies in some key user experience factors such as Semantic Relevance, Explainability, and Proactiveness, which negatively impact the overall user interaction experience.

The main contributions of our work are as follows:
\begin{itemize}
    \item To the best of our knowledge, we are the first to propose a user-centered, LLM-based framework for the overall evaluation of CRSs. 
    Our work fills a key research gap in the evaluation of CRSs: While numerous user studies have explored user-centered evaluation approaches, there remains a lack of practical and computable evaluation measures for assessing CRS performance. Our proposed method contributes to advancing CRS evaluation practices toward a more human-centered direction.
    \item We conducted human evaluations of user-CRS interaction logs and showed that, compared to existing evaluation metrics, our framework aligns with human evaluation better. The collected human feedback also serves as a valuable resource for future research on CRS evaluation. 
    \item We systematically assessed four CRSs based on the proposed framework, benchmarking their performance across various factors related to user experience. This evaluation enhances the understanding of different aspects of CRS capabilities and promotes the development of more human-centered conversational recommendation systems.
\end{itemize}

The structure of this paper is as follows. Section \ref{ch:rw} reviews prior work on conversational recommender systems and evaluation methods for conversational recommendation as well as studies on text assessment using large language models. Section \ref{ch:preliminaries} formally defines our research context by characterizing the key evaluation factors for CRS and comparing them to offline evaluation metrics. Section \ref{ch:method} presents our framework, comprising an LLM as evaluator module and a multi agent debater component. Section \ref{ch:rqs} further details our research questions and highlights their interrelationships. Section \ref{ch:experiment} describes the experimental setup, including dataset selection, the choice of CRS models for evaluation, baseline metrics, the procedure for obtaining simulated user–CRS dialogue logs, and the process for gathering human annotations. Section \ref{ch:result} reports the results of these experiments. Section ~\ref{ch:discuss} discusses some limitations of our work and potential directions for future research. Finally, Section \ref{ch:conclusion} summarizes the principal findings, discusses limitations of the current work, and outlines directions for future research.

\section{Related Work}
\label{ch:rw}

\subsection{Conversational Recommender Systems}
Dialogue systems are designed to provide social support or functional services to human users through natural language interactions. Key areas in this field have primarily focused on dialogue context comprehension~\citep{chen-etal-2022-unidu} and response generation techniques~\citep{roller-etal-2021-recipes}. Recommender Systems aim to help users filter and select items that meet their information needs. To achieve this goal, traditional recommender systems generate recommendations based on implicit feedback from past user-item interactions~\citep{schafer2007collaborative,ekstrand2011collaborative, koren2008, bell2007,koren2009,covington2016}. As the pace of modern life accelerates, traditional recommender systems face inherent limitations due to their static interaction paradigms and insufficient adaptability to users’ evolving preferences. In response, Conversational Recommender Systems (CRSs) have emerged as a promising research frontier, integrating intelligent dialogue agents with recommendation engines to enable dynamic, context-aware, and user-centric interactions~\citep{Ren2020CRSAL}. Taking a different approach to traditional RSs, CRSs aim to model the relationship between user preferences and candidate item representations to deliver accurate recommendations, while simultaneously generating coherent responses that are contextually aligned with the recommended items~\citep{jannach2021,GAO2021100,zamani2023,li2018towards,liang2021learning}. As CRSs interact with users through multi-turn dialogues, CRSs are also regarded as a specialized form of dialogue systems~\citep{deng2025proactive, li2021seamlessly,Ren2020CRSAL}.


In recent years, many implementations of CRS have emerged~\citep{chen2019kbrd,zhou2020,wang2022barcor, wang2022unicrs,wang2023-rethink, deng2023,ni2023, lei2020,tu2022,shang2023}. In these implementations, the interaction patterns between users and CRSs can be divided into two main categories: One line of work relies on handcrafted response templates and predefined interactive actions (e.g., asking users' preferences for item attributes or making recommendations) to interact with users~\citep{lei2020,tu2022,zhou2020,shang2023}; Another line of work focuses on interacting with users through more flexible natural language conversations (chit-chat)~\citep{chen2019kbrd,wang2022unicrs,wang2022barcor}. 

In our study, we focus on the evaluation of chit-chat based CRS, and we selected four representative CRSs for evaluation, in which three (BARCOR~\citep{wang2022barcor}, KBRD~\citep{chen2019kbrd} and UniCRS~\cite{wang2022unicrs}) are based on neural networks, while one (CHATCRS~\citep{wang2023-rethink}) directly utilizes GPT as a CRS.

\begin{itemize}
    \item \textbf{BARCOR}~\citep{wang2022barcor}: It proposes a unified conversational recommender system framework built upon BART, which jointly models item recommendation and response generation within a single end-to-end architecture. This unified design facilitates parameter sharing and enables coherent alignment between recommended items and generated responses.
    \item \textbf{KBRD}~\citep{chen2019kbrd}: It incorporates DBpedia~\footnote{https://www.dbpedia.org/} to enhance the semantic representation of entities mentioned in dialogues by grounding them in structured knowledge using Relational Graph Convolutional Networks (R-GCN). The resulting entity embeddings are combined with dialogue representations to enhance both recommendation accuracy and response generation.
    \item \textbf{UniCRS}~\citep{wang2022unicrs}: It leverages knowledge-enriched prompts within a pre-trained language model (e.g., DialoGPT) to encode task-specific instructions informed by knowledge graph semantics. This approach unifies recommendation and response generation, allowing for joint training in a shared parameter space.
    \item \textbf{CHATCRS}~\citep{wang2023-rethink}: It employs ChatGPT~\footnote{https://chatgpt.com/} as the conversational agent by leveraging predefined instructions, enabling the system to collect user preferences, generate recommendations, and provide explanations, without any further model training.
\end{itemize}

We selected the above systems because many studies~\citep{wang2023-rethink, wang2023-improve, zhu2024,zhu2024reliable, Bernard_2025,Bernard_SIGIRAP} have evaluated them or used them as baselines, which helps to connect our work with existing evaluations of conversational recommender systems.

\subsection{Evaluating Conversational Recommender Systems}
Evaluation is a fundamental aspect of developing search and recommender systems, encompassing both \textit{user-centric} assessments, which capture the effectiveness of the system from the user's perspective, such as user satisfaction~\citep{Bernard_2025}; as well as \textit{system-centric metrics} such as Precision, Recall, and inference time~\citep{siro2023,AlMaskari2010Review}. The evaluation of CRS also encompasses both user-centric and system-centric approaches, and can be further categorized into three main scenarios:
\begin{itemize}
    \item Lab-based \textbf{User studies} examine perceived system quality, evaluating aspects like recommendation relevance, perceived transparency, and ease of use~\citep{pu2011}.
    \item Online metrics, such as \textbf{A/B testing}, are commonly used in real-world settings to assess whether the system achieves its intended objectives, typically through metrics such as user engagement or conversion rates~\citep{jannach2019}.
    \item \textbf{Offline evaluations}, conducted without user involvement, rely on objective metrics such as recommendation accuracy (Precision, Recall, etc.) or recommendation diversity. 
\end{itemize}

Due to the high cost of A/B testing and user studies, offline evaluation is commonly adopted in practice to assess CRSs. The most commonly used metrics for evaluating CRS include Recall, Precision, \textit{etc.}, for assessing recommendation task performance, and BLEU, ROUGE, Perplexity, Distinct-N, \textit{etc.}, for evaluating dialogue management task performance. Nevertheless, empirical studies indicate that these rule-based metrics do not align with human preferences~\citep{chen2017meta,reiter-2018-structured,cremonesi2012}. One key contribution of our study is the use of large language models (LLMs) as evaluators to enable user-centric offline evaluation.


In the context of user-centered evaluation for CRSs, previous studies have argued that several factors may influence users’ assessment of their interactions with the system, from both the recommendation and dialogue perspectives
. These factors include: \textbf{\textit{(perceived) accuracy of recommendation}}~\citep{contreras2021shopping,Manzoor2024ChatGPT,jin2021,zhang2024navigating}, \textbf{\textit{diversity of recommended items}}~\citep{jannach2020escaping,shani2011evaluating,jin2021,zhang2024navigating}, \textbf{\textit{novelty of recommended items}}~\citep{jannach2020escaping,shani2011evaluating,jin2021,zhang2024navigating}, \textbf{\textit{explainability of recommendation}}~\citep{sakai2023swan,jin2021,zhang2024navigating,zhang2024llm}, \textbf{\textit{user intent understanding and response capability}}  (e.g., maintaining a coherent dialogue flow, recovering from errors)~\citep{sakai2023swan,langevin21,jin2021,zhang2024navigating}, \textbf{\textit{the quality of system-generated text}} (e.g., fluency, linguistic appropriateness and naturalness)~\citep{jin2021, jannach2022, langevin21}, \textbf{\textit{system informativeness for decision-making}} (e.g., information sufficiency, information accuracy)~\citep{sakai2023swan, jannach2022}, \textbf{\textit{proactiveness of the system}}~\citep{kraus-etal-2020-comparison,deng2025proactive,christakopoulou2016}.
Nevertheless, these studies did not propose any computable metric for evaluating CRS performance. 

 Some prior work has explored both human-annotated and LLM-based approaches for evaluating CRSs from a user-centered perspective. ~\citet{manzoor2022infact} uses crowd-sourced pairwise comparisons to assess CRS response quality, and~\citet{yang2024} proposed a behavior alignment method that relies on human-annotated data. However, these methods rely on collecting human data, which is costly. ~\citet{wang2023-rethink} utilized LLMs to assess whether the items recommended by CRS meet user needs, while ~\citet{huang2024concept} leveraged LLMs to evaluate the social attributes of CRS. ~\citet{zhang2024llm} leveraged LLMs to assess the quality of explanation texts generated by CRSs. However, these studies did not propose a method for assessing the overall performance of CRS. To the best of our knowledge, our work is the first to propose a unified, LLM-based framework for holistic, user-centered evaluation of CRSs. Our work bridges the gap between user studies on CRSs and current evaluation practices in the field. We extracted twelve factors influencing users’ interaction experiences with conversational recommender systems from prior studies, used LLMs to assign evaluation scores to each factor, and then aggregated these into an overall score. By proposing a feasible evaluation framework, our study contributes to advancing CRS evaluation practices toward a more human-centered paradigm.


\subsection{Large Language Models as Evaluators for Text Quality}

Text quality evaluation plays a critical role in advancing natural language generation (NLG) tasks. Human evaluation remains a fundamental approach for assessing the quality of text generated by machine learning models~\citep{guzman2015how,gillick2010non}. However, its high cost, subjectivity and limited reproducibility pose significant challenges to ensuring fair and consistent comparisons across different natural language processing (NLP) models and algorithms~\citep{gillick2010non, clark2021all, karpinska2021perils}. In contrast, automated evaluation metrics such as BLEU and ROUGE offer advantages in terms of low cost and high reproducibility. However, empirical studies have shown that these metrics often exhibit poor alignment with human preferences~\citep{reiter-2018-structured}.

In recent years, researchers have leveraged Large Language Models for evaluating text quality in various tasks and argued that LLMs, demonstrating significant natural language understanding capabilities, have the potential to align with human preferences. For example, ~\citet{gao2023human} argued that ChatGPT demonstrates strong capability in annotation tasks involving Likert-scale scoring, and it could outperform commonly used automatic evaluation metrics, showing high alignment with human judgments.  

Some work leveraged existing LLMs as evaluators, experimenting with zero-shot prompting and in-context learning (ICL) methods, and have explored evaluation paradigms such as Likert-scale scoring and pair-wise comparison, as well as different output formats: having LLMs directly emit numeric scores or preferred candidates, or leveraging implicit output probabilities. ~\citet{chiang2023large} introduced a method wherein they present LLMs with the exact human evaluation instructions, sample texts, and rating questions, and parse the models’ generated responses into numeric quality scores as a reproducible surrogate for expert human assessment. ~\citet{lin-chen-2023-llm-eval} proposed a framework that leverages LLMs to automatically fill evaluation forms and score open-domain conversational system responses across multiple dimensions (e.g., grammar, relevance, and appropriateness). ~\citet{chen-etal-2023-exploring-use} investigated the effectiveness of ChatGPT as the evaluator for reference-free text quality evaluation by comparing explicit numeric score generation, implicit probability-based scoring, and pairwise comparison paradigms, finding that explicit score generation delivers the most reliable performance across diverse NLG tasks. ~\citet{fu-etal-2024-gptscore} introduced a LLM-based method leveraging zero-shot prompting and ICL to perform  multi-aspect evaluation of texts by interpreting conditional generation probabilities as quality scores. ~\citet{liu-etal-2023-g} leveraged GPT-4’s chain-of-thoughts to auto-generate detailed evaluation steps and employs a form-filling paradigm with probability-weighted scoring to assess NLG output quality. ~\citet{chan2023chateval} proposed a multi-agent debate framework that orchestrates a team of LLMs to collaboratively discuss and score generated responses, thereby mimicking human evaluation processes for NLG tasks. 

Some work directly trained an LLM to serve as an evaluator.~\citet{wang2024pandalmautomaticevaluationbenchmark} proposed an LLM-based evaluator called PandaLM for assessing LLM-generated text. The evaluation is conducted from perspectives such as relative conciseness, clarity, instruction adherence, comprehensiveness, and formality. ~\citet{ye2025learning} proposed a method prompting a pre-trained LLM to self-generate contrastive preference judgments for text with natural-language rationales. 

Nevertheless, to our knowledge, few studies have explored the use of LLMs for evaluating conversational recommendation tasks. In this study, we employ LLMs to evaluate CRS–user interaction logs. Our evaluation scope covers not only the text quality typically assessed in NLG tasks, but also the quality of the items recommended by the system—namely, whether they meet user needs and exhibit diversity and novelty—and additional dimensions of interest in traditional or conversational recommendation tasks, such as whether the system provides explanations for its recommendations. The first part of our approach (LLM-As-Evaluator) prompts the LLM to follow our custom evaluation instructions and employs a chain-of-thought process to score the interaction logs step by step, producing a 0–4 Likert-scale rating. The second part of our approach (Multi-Agent Debater) has LLMs assume roles with different domain expertise and, based on the scores and rationales produced in the first part, engage in multi-round discussions to derive a final overall score on a 0–100 scale.

\section{Preliminaries}
\label{ch:preliminaries}





Based on previous practices in CRS evaluation~\citep{wang2023-rethink, wang2023-improve,zhu2024,zhu2024reliable}, in chit-chat-based conversational recommendation, we formalize a single interaction session as a conversation log
\[
L = \{D, R\},
\]
where: $D$ denotes the full user–system dialogue transcript, capturing the sequential exchange of messages; $R$ denotes the corresponding sequence of recommended item lists provided by the system at each dialogue turn.

The dialogue component $D$ is written as
\[
D = \bigl(d^1_u,\; d^1_s,\; d^2_u,\; d^2_s,\;\dots,\; d^T_s,\; d^{T+1}_u\bigr),
\]
where $d^i_u$ denotes the user’s utterance at turn $i$, $d^i_s$ denotes the system’s utterance at turn $i$, and $T$ denotes the total number of complete system response turns.

By convention, we assume each dialogue begins with a user utterance $d^1_u$ and ends with a final user utterance $d^{T+1}_u$, ensuring that recommendation actions can be consistently aligned with preceding user inputs.

The recommendation component $R$ is given by
\[
R = \bigl(r^1_s,\; r^2_s,\;\dots,\; r^T_s\bigr),
\]
where each
\[
r^i_s = \bigl(p^i_1,\; p^i_2,\;\dots,\; p^i_K\bigr)
\]
is the ordered list of $K$ items (e.g., products, movies, songs) suggested by the system immediately after its $i$-th utterance $d^i_s$. The index $i$ ranges from $1$ to $T$, so that there is exactly one recommendation list per system turn.

To evaluate the overall performance of a conversational recommender system, we define a metric
\[
M: L \;\longmapsto\; m,\quad m \in \mathbb{R},
\]
which maps the entire conversation log $L$ to a real-valued score $m$. 

Taking Figure~\ref{fig:crs_demo} as an example, the interaction between user and system can be formalized as a conversation log $L = \{D, R\}$, and the dialogue component
\[
D = \bigl(d^1_u, d^1_s, d^2_u, d^2_s, d^3_u, d^3_s, d^4_u\bigr)
\]
captures the sequence of utterances, namely the user queries ``Any new horror trailers?'', ``Got more?'', and ``What about the director?'' (represented by \(d^1_u, d^2_u, d^3_u\)) together with the corresponding system responses (represented by \(d^1_s, d^2_s, d^3_s\)). Here we define $d^4_u$ as empty ($d^4_u=\varnothing$). The recommendation component
\[
R = \bigl(r^1_s, r^2_s, r^3_s\bigr)
\]
captures, for each system turn \(i \in \{1,2,3\}\), an ordered list of items \(r^i_s\): 
\[
\begin{aligned}
r^1_s &= \{\text{``28 Years Later (2025)'', ``Sinners (2025)''}\},\\
r^2_s &= \{\text{``Death of a Unicorn (2025)'', ``The Monkey (2025)''}\},\\
r^3_s &= \{\text{``Y2K (2024)''}\}.
\end{aligned}
\]

In previous CRS evaluation practices, researchers typically decomposed the conversation log \(L\) and obtained dialogue-related metric scores \(m_{\textrm{Dial}}\) by applying mappings of the form \(D \;\longmapsto\; m_{\textrm{Dial}} \quad( m_{\textrm{Dial}} \in \mathbb{R})\), where metrics such as BLEU or ROUGE were used. Similarly, recommendation-related metric scores \(m_{\textrm{Rec}}\) were obtained through mappings \(L \;\longmapsto\; m_{\textrm{Rec}} \quad (m_{\textrm{Rec}} \in \mathbb{R})\), using metrics such as Recall or Precision. However, such evaluation approaches fail to provide a holistic assessment of \(L\) as a whole, which constitutes one of the research gaps addressed by this study.

\section{CoRE: Our Proposed Method}
\label{ch:method}
In this section, we introduce our proposed framework, \textbf{Co}nversational \textbf{R}ecommendation \textbf{E}valuator (\textbf{CoRE}). 
It consists of two sequential components: (1) \textbf{LLM-As-Evaluator.} Employing LLMs to evaluate CRS across twelve distinct factors that influence user experience one by one. This evaluation is based on the carefully designed assessment criteria, allowing the LLM to generate both scores and detailed reasons for each factor.
(2) \textbf{Multi-Agent Debater.} Obtaining an overall evaluation score for the CRS through a multi-agent debate framework, which discusses and synthesizes the evaluation scores of the above twelve factors into a unified score.

Figure~\ref{fig:flow} shows the overall evaluation process of our method. In this study, the user-system dialogue log displayed in the top-left corner of Figure~\ref{fig:flow} is collected via interactions between the user simulator and the CRS. However, our method is also applicable for evaluating user-system dialogue logs collected in real-world scenarios. 
\subsection{LLM-As-Evaluator for Twelve Factors}
\label{ch:factors}
\label{12factors}
\begin{figure}[t]
  \includegraphics[width=1.0\linewidth]{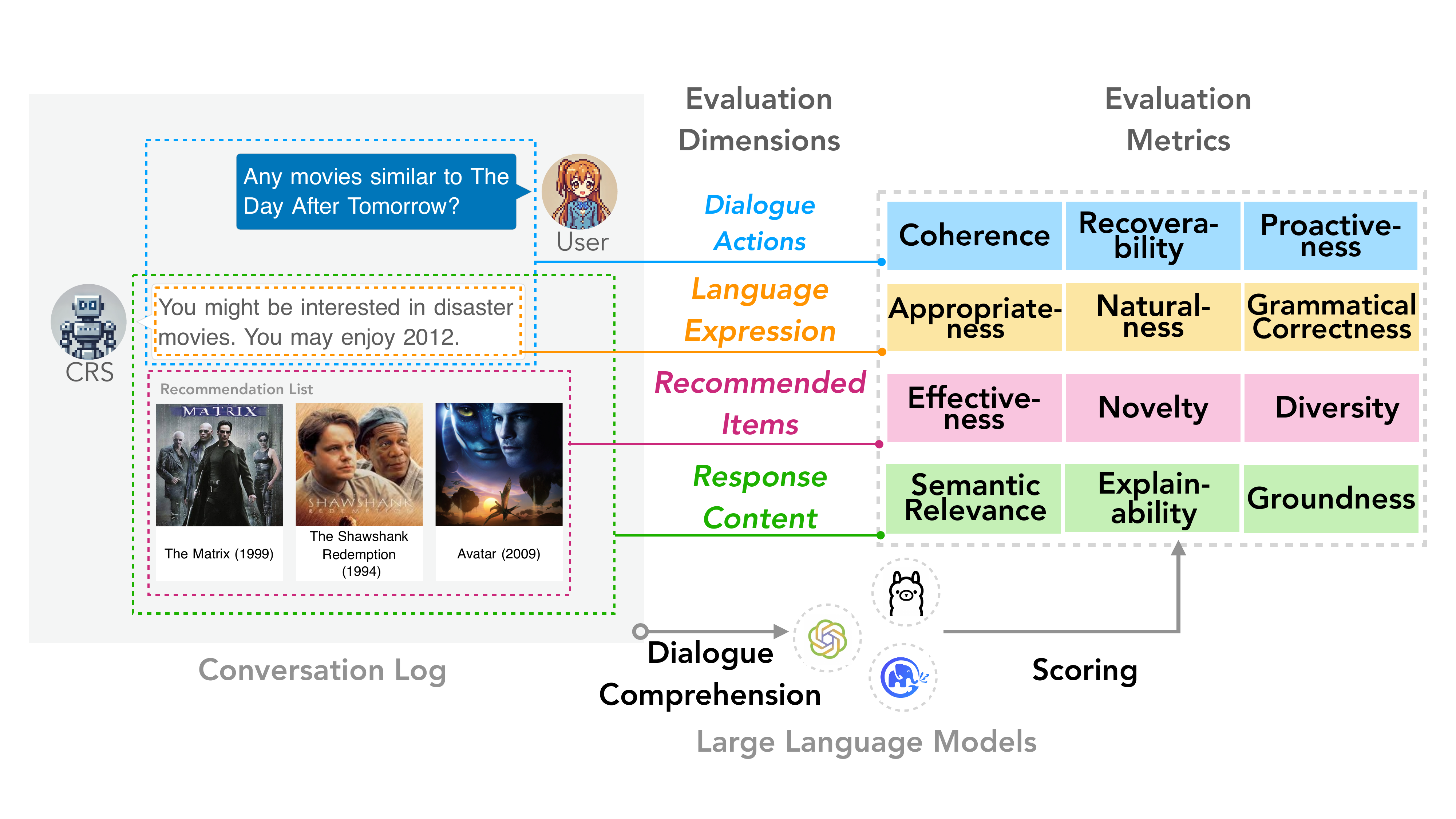}
  \caption{The first part of our proposed evaluation framework encompasses four dimensions, covering a total of twelve factors. 
  }
  \vspace{-1em}
  \label{fig:dimensions}
\end{figure}
\begin{table*}[h]
\caption{The concepts of the twelve factors from four evaluation dimensions.}
\label{tab:factor_concept}

\resizebox{\textwidth}{!}{%
    \begin{tabular}{|m{2.2cm}|m{2.2cm}|m{11.1cm}|}
\hline
Dimension &  Factor  & Concept\\ 
\hline
\multirow{3}{=}[-3.8em]{Dialogue Actions} & Coherence & The system's ability to understand the user's intention and take action correspondingly, ensuring a logical and relevant flow of conversation. For example, when the user requests a recommendation, the system provides one; when the user asks the system to introduce a specific item, the system is able to do so.
\\ 
\cline{2-3} & Recoverability & The system's ability to recognize and correct mistakes based on user feedback, adjusting its responses accordingly. This involves the system effectively understanding user feedback or corrections and adjusting its responses accordingly in the subsequent conversation.\\ 
\cline{2-3} & Proactiveness &  The system's ability to actively shape and guide the conversation. This involves the system not only responding to the user's queries but also initiating topics, asking about user preferences, and suggesting relevant information or follow-up questions.
\\ 
\hline
\multirow{3}{=}[-1.4em]{Language Expression} & Grammatical Correctness & The degree to which the text adheres to standard grammatical rules, including proper sentence structure, subject-verb agreement, tense consistency, and correct usage of words and phrases.
\\ 
\cline{2-3} & Naturalness & How closely the system-generated text resembles native speaker expressions in vocabulary and grammar, ensuring it is easy to understand for most people.\\ 
\cline{2-3} & Appropriateness & The system's sensitivity to cultural, ethical, and social norms. The language used should be polite, respectful, and free from offensive or insulting content. 
\\ 
\hline
\multirow{3}{=}[-1.8em]{Recommended Items} & Effectiveness & How well the system’s recommended items align with the user’s expressed interests in a conversation. 
\\
\cline{2-3} & Novelty & The degree of freshness of an item. The higher the novelty, the less familiar the recommended item is to the user. 
\\ 
\cline{2-3} & Diversity & The variety of items presented by the system, ensuring that the recommended list includes a range of different, yet relevant options. It is measured by the presence of different features of items in the recommendations.
\\ 
\hline
\multirow{3}{=}[-1.7em]{Response Content} & Semantic Relevance & The degree of connection between the content generated by the system and the items mentioned in the recommendation list provided by the system, or to what extent is the content discussed by the system closely related to the items in the recommendation list.
\\
\cline{2-3} & Explainability & The system's ability to provide clear and detailed explanations for its recommendations. The explanation should be persuasive, helping the user understand why certain items are suggested and how well these items match their preferences.
\\ 
\cline{2-3} & Groundness & The factual accuracy of system-generated text, ensuring that the descriptions of items are correct, based on verifiable facts, and free from false or misleading information. 
\\ 
\hline
\end{tabular}
}
\vspace{-1em}

\end{table*}
\begin{table*}[h]
\caption{The instruction template provided to the LLMs for 12-factor evaluation.}
\label{tab:inst}
\centering
\begin{myframe}
You are a professional evaluator of conversational recommendation systems. I'll provide you with the conversation history between the user and the system, which is divided into two parts: \texttt{<history>} and \texttt{<interaction>}.

The history and interaction parts occurred within the same conversation and are in sequential order, with the history part ending and the interaction part beginning right after. The history gives you context for the conversation, which you only need to understand, not rate.
You need to rate the system's content \texttt{<system>} in the \texttt{<interaction>} part according to the guidelines below.

*IMPORTANT* Only rate the system's responses \texttt{<system>} in the \texttt{<interaction>} part. Do not rate any other parts (like the user's messages or the history part)!

You will assess the \{\textcolor{blue}{FACTOR}\} of the recommender system's response following the guidance below. Please note that you only need to evaluate the system's response in \texttt{<interaction>}. Do not evaluate the user's dialogue, do not evaluate the system's responses in \texttt{<history>}.

\hspace*{2em}\texttt{<evaluation\_guideline>}

\hspace*{4em}\texttt{<definition>}

\hspace*{6em}\{\textcolor{blue}{FACTOR\_DEFINITION}\}

\hspace*{4em}\texttt{</definition>}

\hspace*{4em}\texttt{<evaluation\_standard>}

\hspace*{6em}\{\textcolor{blue}{EVALUATION\_STANDARD}\}

\hspace*{4em}\texttt{</evaluation\_standard>}

\hspace*{4em}\texttt{<evaluation\_step>}

\hspace*{6em}\{\textcolor{blue}{EVALUATION\_STEP}\}

\hspace*{4em}\texttt{</evaluation\_step>}

\hspace*{2em}\texttt{</evaluation\_guideline>}

\{\textcolor{red}{CONVERSATION\_LOG}\}

Please rationale your rating according to the guideline first, then put the score you assigned in the format of \texttt{<rating>}\{\emph{score}\}\texttt{</rating>}.
\end{myframe}
\end{table*}
\begin{table*}[h]
\caption{The definition, evaluation standard and evaluation steps of Coherence.}
\label{tab:coh_guide}
\centering
\begin{tabular}{|m{1.5cm}|m{13.5cm}|}
\hline
Definition & 
Coherence refers to the extent to which the system correctly understands the user's intention in a conversation and responds correspondingly. 
A coherent response directly addresses the user's query or statement, maintaining a logical and relevant flow of conversation. 
This includes correctly interpreting the user's intentions, such as asking for clarification when needed, making suitable recommendations, 
or providing clear explanations.
\\ \hline
Evaluation Standard & 
1. If you believe that each response from the system aligns with the user's intention, give 4 points.

2. If there is 1 response that does not align with the user's intention, give 3 points. 

3. If there are 2 responses that do not align with the user's intention, give 2 points.

4. If there are 3 responses that do not align with the user's intention, give 1 point.

5. If more than 4 responses do not align with the user's intention, give 0 points.
\\ \hline
Evaluation Steps & 
1. Read the User Input and System Response: Carefully review the user's query or statement and the system's response.

2. Assess Understanding: Determine whether the system accurately understood the user's intention or question.

3. Consider Logic and Clarity: Ensure the response is logical, clear, and directly addresses the user's needs or questions.

4. Assign a Score: Based on the coherence of the response, assign a score from 0 to 4, reflecting how well the system understood and responded to the user's input.

5. Provide Justification: Include a brief explanation for the score, noting any specific issues in the response related to coherence (if any).
\\ \hline
\end{tabular}
\end{table*}

As we mentioned earlier, CRSs are responsible for both item recommendation and dialogue management, which means they share characteristics with both dialogue systems and recommender systems. However, the fact that CRSs provide recommendations through conversation also gives them unique characteristics of their own. The integration of multi-turn conversations and item recommendations endows CRSs with unique characteristics that distinguish them from both general conversational systems and recommender systems.


In this section, we aim to explore the following question: \textit{\textbf{what factors should we consider when evaluating CRSs and how should we score them?}}

Prior research has indicated that a range of factors may influence how users assess their interactions with the system, encompassing both recommendation quality and dialogue effectiveness. These factors include: \textit{(perceived) accuracy of recommendation}~\citep{contreras2021shopping,Manzoor2024ChatGPT,jin2021,zhang2024navigating}, \textit{diversity~\citep{jannach2020escaping,shani2011evaluating,jin2021,zhang2024navigating} of recommended items}, \textit{novelty of recommended items}~\citep{jannach2020escaping,shani2011evaluating,jin2021,zhang2024navigating}, \textit{explainability of recommendation}~\citep{sakai2023swan,jin2021,zhang2024navigating,zhang2024llm}, {\textit{user intent understanding and response capability}}  (e.g., maintaining a coherent dialogue flow, recovering from errors)~\citep{sakai2023swan,langevin21,jin2021,zhang2024navigating}, \textit{the quality of system-generated text} (e.g., fluency, linguistic appropriateness and naturalness)~\citep{jin2021, jannach2022, langevin21}, \textit{system informativeness for decision-making} (e.g., information sufficiency, information accuracy)~\citep{sakai2023swan, jannach2022}, {\textit{proactiveness of the system}}~\citep{kraus-etal-2020-comparison,deng2025proactive,christakopoulou2016}.

Inspired by the previous studies, we heuristically summarize and list twelve factors that affect user satisfaction with conversational recommender systems, and categorize them into four dimensions based on the evaluation subjects involved in user-system interactions. Table~\ref{tab:factor_concept} presents the concept of each factor.

\textbf{Dialogue Actions.} This dimension evaluates whether the system can take actions that align with the user's intentions during the conversation. The key factors in this dimension include \textit{Coherence}, \textit{Recoverability}, and \textit{Proactiveness}. These factors are also commonly used to assess general conversational agents. The primary evaluation subject is the system's actions within the dialogue.

\textbf{Language Expression.} This dimension examines whether the system's response text closely mirrors human-like expression while maintaining linguistic appropriateness. The factors under this dimension include \textit{Grammatical correctness}, \textit{Naturalness}, and \textit{Appropriateness}. These factors are also employed in evaluating natural language generation tasks. The main evaluation subject is the text generated by the system.

\textbf{Recommended Items.} This dimension assesses whether the items in the system's recommendation list satisfy user needs while maintaining novelty and diversity. The relevant factors here are \textit{Effectiveness}, \textit{Novelty}, and \textit{Diversity}. These factors are widely used in evaluating traditional recommender systems. The primary evaluation subject is the recommendation lists provided by the system.

\textbf{Response Content.} This dimension evaluates the CRS's ability to seamlessly integrate text generation with recommendation tasks. The main evaluation subjects involve both the system-generated text and the recommended items. The factors in this dimension include \textit{Semantic Relevance}, \textit{Explainability}, and \textit{Groundness}. These factors are unique to conversational recommendation tasks, distinguishing them from traditional text generation, dialogue agent, or recommendation tasks.

The example in Figure~\ref{fig:dimensions} shows how the twelve factors correspond to evaluation subjects in the user-CRS interaction. The blue dashed line surrounds the dialogue between the user and the CRS, indicating that the \textbf{dialogue actions} dimension focuses on evaluating the conversational actions and strategies between the user and the CRS. In the example, the user's query ``Any movies similar to \textbf{The Day After Tomorrow}'' expresses an intent to receive a movie recommendation, and the system correctly adopts the strategy of providing a recommendation in response. The orange dashed line surrounds the system-generated text, which refers to the \textit{language expression} used by the system when responding to the user. In the illustration, the system responds using appropriate language in a relatively natural and grammatically correct manner. The purple dashed line surrounds the \textbf{recommended items}, showcasing the three movies recommended by the system: \textit{The Matrix}, \textit{The Shawshank Redemption}, and \textit{Avatar}. These items are diverse in genre, but they are not novel enough and do not align with the user's request for movies similar to \textit{The Day After Tomorrow}. The green dashed line surrounds both the system's text and the recommended items, emphasizing the \textbf{response content} dimension, which evaluates the relevance between the system's text and the recommended items. For instance, the system's response text has little correlation with the recommended movie list, resulting in a low semantic relevance.


To obtain evaluation scores for the twelve factors, we leverage LLMs as evaluators by providing them with dialogue logs and detailed evaluation guidelines for each factor, instructing them to generate scores ranging from 0 to 4 along with the corresponding reasoning for each factor. Table~\ref{tab:inst} illustrates the structure of the instruction provided to the LLM evaluators. Table~\ref{tab:coh_guide} presents the evaluation guideline when \textcolor{blue}{FACTOR} is set to \textit{\textcolor{blue}{Coherence}}, covering \textcolor{blue}{FACTOR\_DEFINITION}, \textcolor{blue}{EVALUATION\_STANDARD}, and \textcolor{blue}{EVALUATION\_STEP}. For details on the evaluation guideline for each factor, refer to Appendix~\ref{ch:eval_instr}. An example of \textcolor{red}{CONVERSATION\_LOG} can be found in Figure~\ref{fig:log_example}.

\subsection{Multi-Agent Debater for Overall Evaluation}
\label{ch:multiagent}

Even when we have collected metric scores across the twelve factors to evaluate a CRS, synthesizing these into a comprehensive assessment of the system’s overall effectiveness remains a challenging task. This section seeks to answer the following question: \textbf{\textit{how can we derive a unified measure of overall performance for a CRS based on the evaluations from the twelve factors?}}



One intuitive and heuristic approach is to compute the average score across the different factors. However, this method has inherent limitations, as each factor may exert a varying degree of influence on the overall user experience. Additionally, the factors may interact with one another in complex ways, further complicating the evaluation process.



Moreover, previous research has shown that relying on a single annotator can introduce bias and instability into the evaluation process~\cite{van-der-lee-etal-2019-best,karpinska2021perils}, and best practices recommend collaboration among multiple annotators. 

Therefore, we design a multi-agent debate framework in which multiple LLM agents engage in discussions and debates to derive a unified evaluation of the system's overall performance as inspired by~\citet{chan2023chateval}.


Specifically, we heuristically define four representative agent roles, each emphasizing a distinct set of factors relevant to their expertise: Common User, Domain Expert, Linguist, and HCI Expert. The Common User prioritizes the system's performance in terms of Effectiveness, Coherence, and Recoverability. The Domain Expert focuses on Novelty, Diversity, and Groundness. The Linguist evaluates aspects such as Appropriateness, Naturalness, and Grammatical Correctness. Lastly, the HCI Expert examines Semantic Relevance, Explainability, and Proactiveness. For this classification, we provide the following rationale.

\textbf{Ordinary users} value the system's efficiency in meeting their needs, expecting relevant recommendations and coherent dialogues that align with their intentions, helping to support a smooth experience and enhance satisfaction~\citep{nielsen2000designing,jin2021,siro2023}. 

\textbf{Domain experts}, particularly movie experts, expect the system to offer valuable, diverse insights and are more sensitive to the accuracy of the information due to their specialized knowledge.  The justification for introducing the Domain Expert is as follows. A specific challenge in evaluating CRSs is verifying the Groundness (factual veracity) of the generated item descriptions and attribute details. Average users often lack the encyclopedic knowledge required to detect subtle factual errors or fabricated details about niche items. A domain expert (e.g., a film critic or bibliophile) acts as a necessary gatekeeper for content accuracy. Furthermore, beyond accuracy, domain experts are uniquely positioned to evaluate the \textit{Novelty} and \textit{Diversity} of recommendations. They can distinguish between generic, popularity-biased suggestions and high-quality, serendipitous discoveries from the \textit{long tail}, ensuring the system provides value beyond what users already know.

The \textbf{Linguist} is crucial for assessing communicative competence. Since user trust in CRS is heavily dependent on the naturalness and pragmatic appropriateness of the generated dialogue, this role ensures the system adheres to linguistic norms that lay users perceive but cannot rigorously quantify. Unlike traditional recommender systems that rely on static graphical interfaces, CRSs interact via Natural Language Generation (NLG). The quality of linguistic delivery is not merely a stylistic preference but a functional prerequisite for user trust. Research in human–computer conversation indicates that systems should understand and adopt users’ language, presenting information in a natural and logically ordered manner to facilitate smooth interactions~\cite{langevin21}. Moreover, pragmatic competence, that is, the ability to use language appropriately in social contexts, is as critical as grammatical correctness~\citep{DAVIES20072308}. A Linguist is necessary to evaluate nuances that lay users often overlook or cannot articulate, such as tonal consistency, and sociolinguistic appropriateness. If a system generates accurate recommendations but employs unnatural phrasing or incoherent discourse markers, it risks diminishing the user's perception of system intelligence and trustworthiness, thereby undermining the acceptance of the recommendations.

\textbf{HCI experts}, on the other hand, focus on semantic relevance, explainability, and proactivity, as these factors are crucial for creating user-centered systems and fostering a positive user experience. From a Human-Computer Interaction perspective, a CRS functions as an interactive decision-support system rather than a passive information retrieval tool. The evaluation of such systems requires a specialized focus on interaction dynamics that mediate user decision-making processes. An HCI expert is essential to assess critical usability dimensions such as Explainability (how well the system justifies its choices to foster trust) and Proactiveness (the system's ability to employ mixed-initiative strategies to guide users). While a common user evaluates the final utility of an item, an HCI expert critiques the interaction flow and cognitive load. They ensure that the system's dialogue strategy effectively reduces the user's decision effort and provides sufficient transparency, which are determinants of long-term user satisfaction and system adoption.

The factors of interest for each role differ from those outlined under each dimension in Section~\ref{12factors}. This distinction arises because the classification in Section~\ref{12factors} is organized based on the evaluation subject, while the current framework is structured around the potential concerns of users with varying backgrounds. By emphasizing the factors most relevant to each role, we ensure that the evaluation of the CRS addresses a wide range of user needs, from the user-centric concerns of ordinary users to the expert-level reviews of domain experts, linguists, and HCI professionals.

The debate process is as follows:


\textbf{Round 1}: We provide each role with the factor-wise scores and corresponding reasons relevant to their interests, which can be obtained through the method described in Section~\ref{12factors}. Each role is then asked to assess, from their own perspective, how likely they are to use the system. They are instructed to assign an overall score ranging from 0 (indicating they would never use the system under any circumstances) to 100 (indicating they would always use the system), along with a detailed justification for their decisions. Once all roles have provided their scores and explanations, their overall scores and justifications are added to the \textit{chat buffer}. If all roles provide the same score, the discussion ends; otherwise, it proceeds to the next round.


\textbf{Round 2 and Beyond}: Each role is provided with the content in the chat buffer and tasked with the following:
(1) Reassess their overall score and provide justification, considering their own interests while also taking into account the opinions of other roles.
(2) In cases where significant discrepancies exist among the scores assigned by different roles, attempt to persuade others by presenting well-reasoned arguments.
Once all roles have contributed, their updated overall scores and justifications are appended to the \textit{chat buffer}. If all roles converge on the same score, the discussion ends; Otherwise, the process continues until the predefined round limit is reached.


Finally, we compute the average scores across the four roles to determine the final score. 

The lower part of Figure~\ref{fig:flow} illustrates the process of the multi-agent discussion. Table~\ref{tab:debate_inst} presents the instruction template for LLM debates. The content of FACTOR\_RESULT can be found in Table~\ref{tab:factor_result_sample}.  The content of TASK\_DESC can be found in Table~\ref{tab:task_inst}.  The content of CONVERSATION\_LOG can be found in Figure~\ref{fig:log_example}.  

Appendix~\ref{ch:agent_inst} provides additional details about the multi-agent debate framework, including the role-playing instructions for each character. The content of CHAT\_BUFFER can be found in Table~\ref{tab:agent_result_sample}.

\begin{table*}[h]
\caption{The instruction template provided to the LLMs for multi-agent debate.}
\label{tab:debate_inst}
\centering
\begin{myframe}
\textcolor{blue}{\{ROLE\_DESC\}}

\textcolor{blue}{\{TASK\_DESC\}}

\textcolor{red}{\{CONVERSATION\_LOG\}}

\textcolor{magenta}{\{FACTOR\_RESULT\}}

\textcolor{magenta}{\{CHAT\_BUFFER\}}

\end{myframe}
\end{table*}
\begin{table*}[h]
\caption{An example of \{FACTOR\_RESULT\}}
\label{tab:factor_result_sample}
\centering
\begin{myframe}

\{

    \quad"dimension\_name": "Groundness",
    
   \quad "evaluation\_result": "The system's response misunderstood...
    
    \quad Given that there are three obvious factual errors in the recommendations, 
    
    \quad I would rate the system's response as follows:
    
    \quad<rating>1</rating>",
    
    \quad"score": "1"
    
\}

\end{myframe}
\end{table*}

\section{Research Questions}
\label{ch:rqs}
In this study, we build upon the Conversational Recommendation Evaluator (CoRE) framework by integrating two complementary components, an LLM‐as‐evaluator module and a multi‐agent debating mechanism, to achieve a comprehensive, user‐centric assessment of conversational recommender systems. Building on this methodological foundation, we then empirically investigate four interrelated research questions through systematic experiments and analyses:
\begin{itemize}
     \item \textbf{RQ1}: What are the score correlations among the twelve evaluation factors in the CoRE framework as produced by human annotators versus LLM-based evaluators?
    \item \textbf{RQ2}: To what extent are the evaluation results of the twelve key factors by LLMs consistent with those provided by human evaluators?
    \item \textbf{RQ3}: Compared to existing metrics (e.g., recall, persuasiveness), how consistent are the evaluation results of our proposed method with those provided by human evaluators?
    \item \textbf{RQ4}: How do the major conversational recommender systems perform under our evaluation framework?
\end{itemize}

The four questions presented are interconnected, forming the multi-layered research framework of this study. One purpose of formulating \textbf{RQ1} is to examine the pairwise correlations among the twelve CoRE dimensions as rated by human annotators and thereby assess the extent to which each factor contributes uniquely to the characterization of conversational quality. Low correlation scores indicate that each factor’s score captures a distinct construct rather than redundant information shared with other factors. Another purpose of \textbf{RQ1} is to evaluate the extent to which LLM-based evaluators, guided solely by the provided instructions and dialogue logs, can independently score each CoRE factor in strict accordance with the prescribed criteria, rather than producing scores that are inadvertently coupled across multiple factors due to various biases. The objective of \textbf{RQ2} is to examine the extent to which the scores assigned by LLMs across the twelve key factors correspond to those provided by human evaluators; a higher degree of correspondence would indicate that the LLMs accurately capture human preferences in their assessments. \textbf{RQ3} seeks to evaluate the degree to which the overall scores produced by the multi‐agent debate component of our proposed framework correspond with the holistic user experience judgments provided by human evaluators. This alignment constitutes one of the most critical criteria for validating the performance of any CRS evaluation metric. \textbf{RQ4} shifts the focus from the evaluation framework to the systems themselves, enabling our proposed framework to stand in for human evaluators in assessing representative conversational recommender systems (e.g., BARCOR, KBRD, UniCRS, CHATCRS) across all twelve key factors and in terms of their overall performance.
\section{Experimental Setup}
\label{ch:experiment}
\begin{figure*}[t]
  \includegraphics[width=0.75\linewidth]{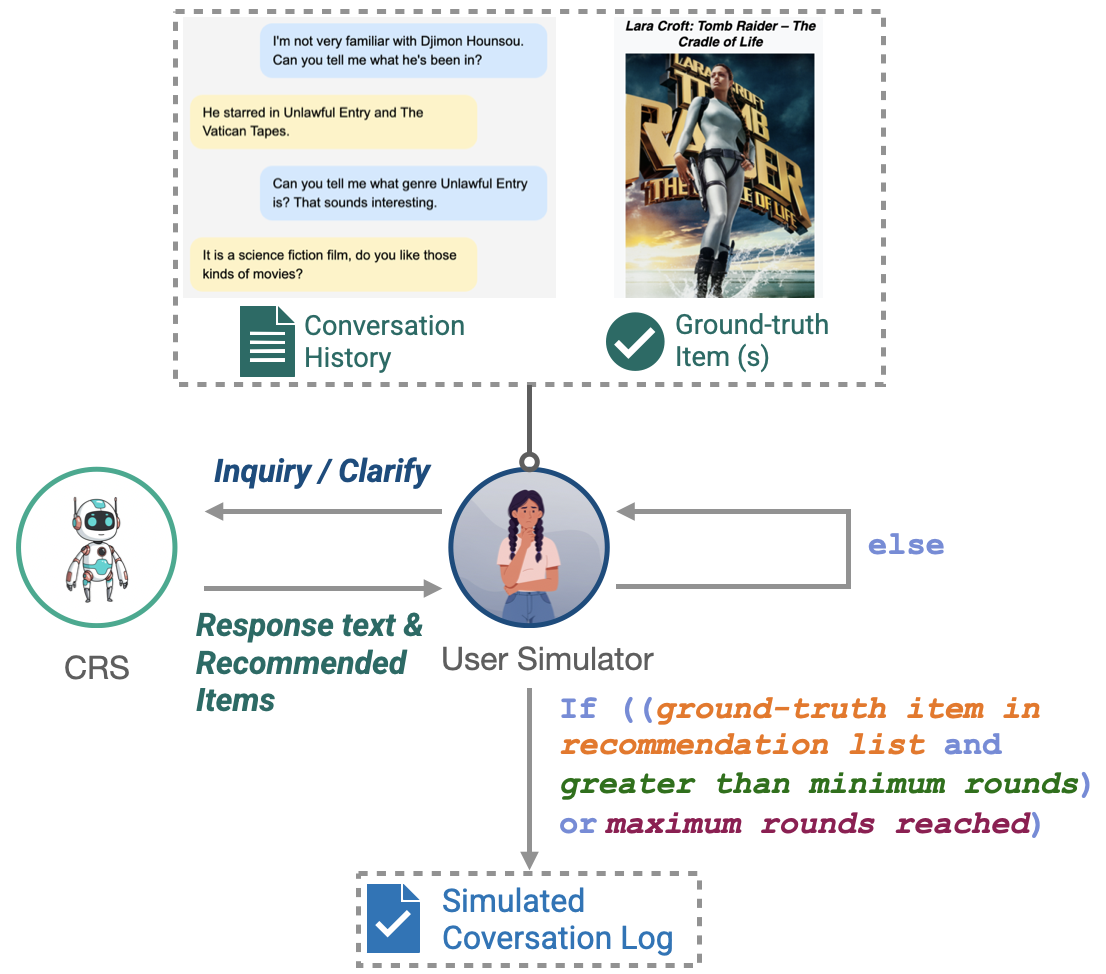}
  \caption{The interaction process between the user simulator and the CRS.}
  \label{fig:user_simulator}
\end{figure*}

To comprehensively evaluate the effectiveness and reliability of our proposed evaluation framework, we conduct a series of controlled experiments combining LLM-based assessments with human evaluation. This section details the datasets used for simulation, the CRS models under comparison, the baseline performance metrics, and the implementation of the user simulator. We further describe the procedure for generating and scoring dialogue logs using LLMs and a multi-agent debate mechanism, as well as the protocol for collecting human feedback. This setup allows us to systematically investigate the extent to which LLM-based evaluators align with human judgments across a range of recommendation scenarios.

\begin{table}[h]
\caption{Summary of CRS Datasets}
\label{tab:crs_datasets}
\centering
\begin{tabular}{ccc}
\hline
\textbf{Dataset} & \textbf{\#Dialogues}  & \textbf{Domains} \\ \hline
ReDial           & 10,006                          & Movie            \\ 
OpenDialKG       & 13,802                          & Movie, Book, Sports, Music \\ 
\hline
\end{tabular}
\end{table}
\subsection{Dataset}
Following previous work~\citep{wang2023-rethink,wang2023-improve,zhu2024,zhu2024reliable}, we select ReDial~\citep{li2018towards} and OpenDialKG~\citep{moon-etal-2019-opendialkg} as the benchmark datasets~\footnote{We used these datasets following the terms of their licenses.}. ReDial primarily focuses on movie recommendations, and OpenDialKG encompasses a broader range of domains, including movies, books, sports, and music. Table~\ref{tab:crs_datasets} shows a summary of the two datasets. 
In the experiment, we evaluate CRSs using the test sets of two datasets: 3,341 records from the ReDial evaluation set and 1,187 records from the OpenDialKG evaluation set. 

\subsection{CRS Model}
Following previous work~\citep{wang2023-rethink,wang2023-improve,zhu2024,zhu2024reliable}, we evaluate the following CRSs: BARCOR~\citep{wang2022barcor}, KBRD~\citep{chen2019kbrd}, UniCRS~\cite{wang2022unicrs} and CHATCRS~\citep{wang2023-rethink}. For BARCOR, KBRD, and UniCRS, which are implemented via neural network, we use the pre-trained models provided by~\citet{wang2023-rethink}; For CHATCRS, which is implemented using LLM, we utilize the GPT-4o-mini-0718 (also denoted as GPT-4o-mini in following text) as the LLM model.

\subsection{Baseline Metrics}
We select Recall@\{1, 5, 10\} and Persuasiveness as our baseline metrics. Recall is used to evaluate recommendation tasks, while Persuasiveness, proposed by~\citet{wang2023-rethink}, serves as a metric for assessing conversational recommendation tasks. Text-quality metrics, like BLEU are not chosen, as our logs involve multi-turn user-system interactions, whereas the dataset only provides single-turn references.

\subsection{User Simulator}
We build a user simulator based on~\citet{wang2023-rethink} with modification to collect user-system dialogues. The history context of the dialogue and the ground truth item are provided as input to the simulator, and the simulator is required to describe relevant information of the ground truth item in the conversation, without directly revealing the ground truth item. When generating simulated dialogues, each session in the original data contains one or more ground-truth items. A hit is counted as long as the CRS's recommended list includes at least one of these ground-truth items. By doing so, we simulate a real user's behavior in a CRS when describing vague or ambiguous demands. To ensure sufficient system dialogue text, we ensure that each dialogue session between the user simulator and CRS includes 3 to 5 turns. In each round of dialogue, the system generates response text and a list of recommended items. If the system identifies the ground-truth item, and more than three rounds of dialogue have occurred, the session ends; If the system has already identified the ground-truth item but the conversation turns are fewer than 3, we require the simulator to ask the CRS to introduce the ground-truth item; If more than five rounds of dialogue have occurred and the system has still not identified the ground-truth item, the session also ends.  We use GPT-4o-mini to implement the simulator. After each conversation, in accordance with the method of \citet{wang2023-rethink}, we direct the simulator to provide a Persuasiveness score. For each record in the dataset, we have the user simulator interact with four different systems. As a result, we obtain $3,341 \times 4 = 13,364$ logs on ReDial and $1,187 \times 4 = 4,748$ logs on OpenDialKG. 
Figure~\ref{fig:log_example} shows an example of the conversation log after processing.

\subsection{Evaluation Process}
\begin{figure}[t]
    \centering

    \begin{subfigure}[t]{0.48\linewidth}
        \centering
        \includegraphics[width=\linewidth]{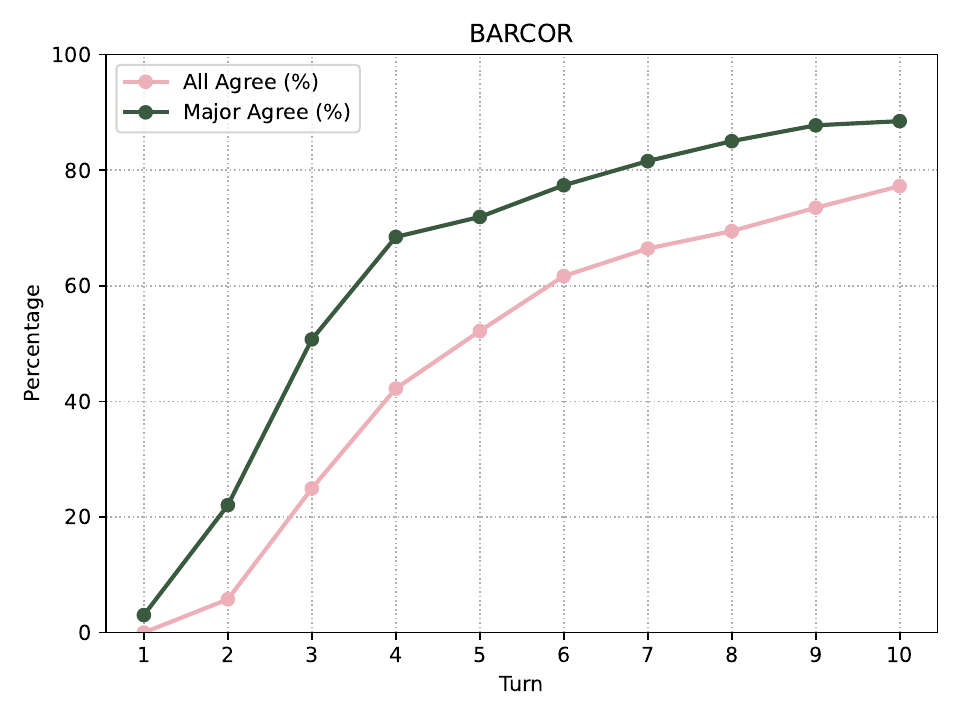}
        \caption{BARCOR}
        \label{fig:barcor_agreement}
    \end{subfigure}
    \hfill
    \begin{subfigure}[t]{0.48\linewidth}
        \centering
        \includegraphics[width=\linewidth]{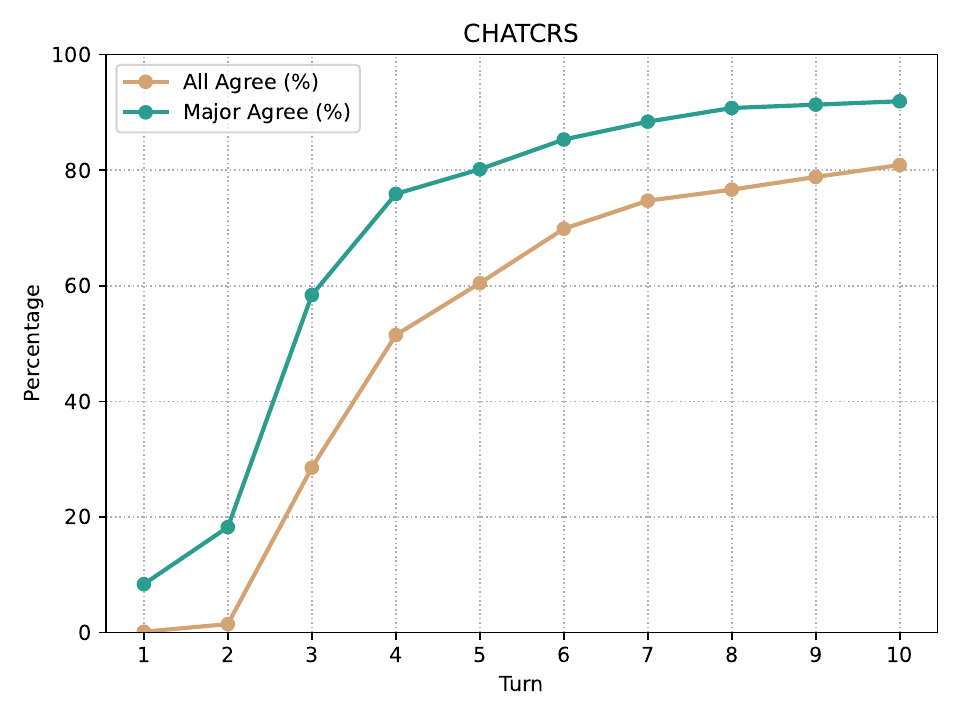}
        \caption{CHATCRS}
        \label{fig:chatcrs_agreement}
    \end{subfigure}

    \vspace{0.6em}

    \begin{subfigure}[t]{0.48\linewidth}
        \centering
        \includegraphics[width=\linewidth]{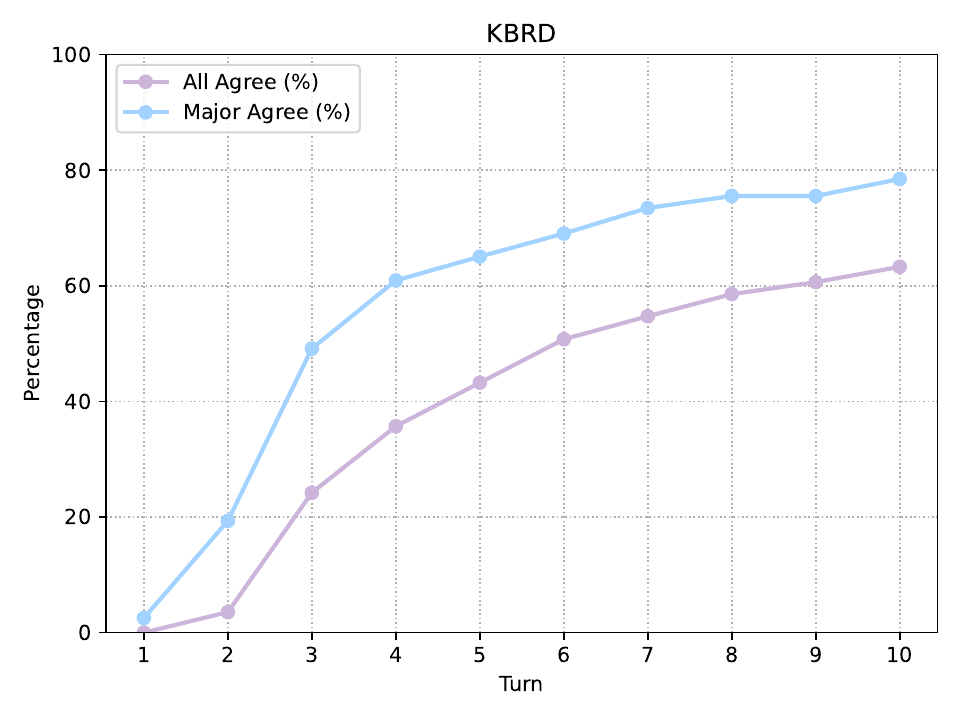}
        \caption{KBRD}
        \label{fig:kbrd_agreement}
    \end{subfigure}
    \hfill
    \begin{subfigure}[t]{0.48\linewidth}
        \centering
        \includegraphics[width=\linewidth]{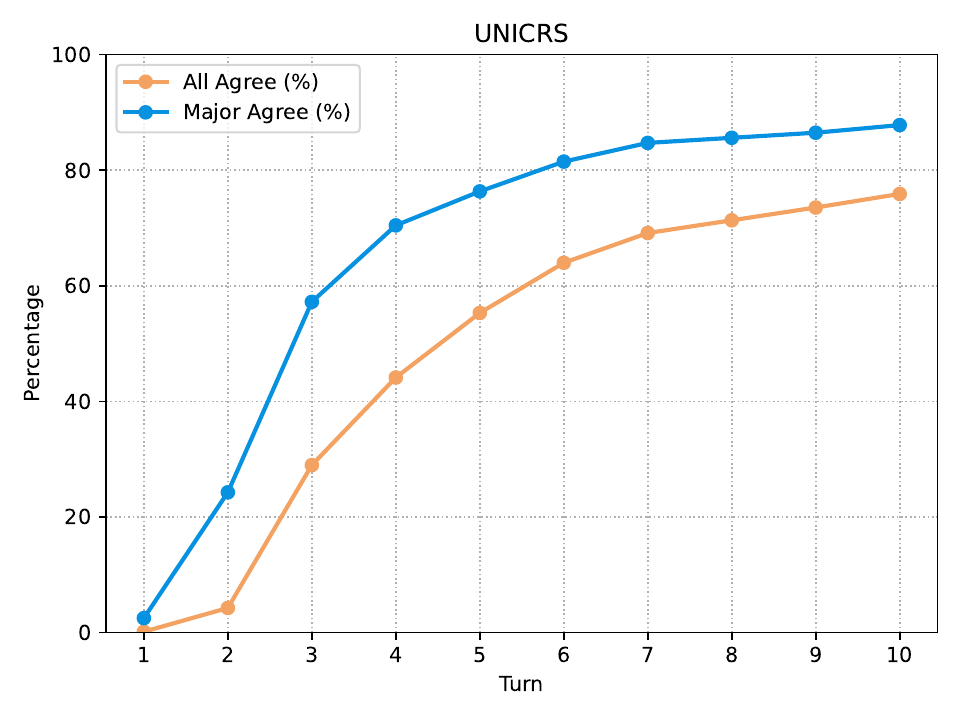}
        \caption{UNICRS}
        \label{fig:unicrs_agreement}
    \end{subfigure}

    \caption{
    Turn-wise evolution of agreement ratios for four conversational recommendation models.
    Each subfigure shows the proportion of \emph{All Agree} (all four evaluators assign the same score) and \emph{Major Agree} (at least three evaluators assign the same score) judgments as a function of the number of turns. 
    }
    \label{fig:agreement_by_turn}
\end{figure}

After obtaining dialogue logs through user simulation, we post-processed the logs by concatenating the recommendation list given by CRS in each round into a session-level list: $r_\text{session} = \text{concat}(R)$. $r_\text{session}$ is used to evaluate Effectiveness, Diversity, Semantic Relevance, Novelty, and is also used to calculate Recall. Figure~\ref{fig:log_example} shows an example of a processed log, where all items recommended by the system during the dialogue are combined into the recommendation list.

We then utilized LLMs to evaluate the conversation logs, based on the twelve factors listed in Section~\ref{ch:factors}. For each factor, we have described its definition, scoring criteria, and scoring steps in the instruction; A score between 0 and 4, as well as the scoring reason was obtained. For details on the scoring, refer to Appendix~\ref{ch:eval_instr}. 

We selected the following LLMs as evaluators, including two business models and eight open-source models: GPT-4o-mini-0718, GLM-4-Air, Llama-3-8B, Llama-3-70B, Ministral-3B, Ministral-8B, Ministral-14B, Qwen-3-8B, Qwen3-14B, Qwen-3-32B.

For the multi-agent debater, we use GPT-4o-mini to implement the four agents, taking the twelve-factor scores and reasonings provided by GPT-4o-mini as the initial input. 
The discussion proceeds for a maximum of four rounds.

The rationale is as follows. To identify an appropriate truncation point for the discussion rounds, we sampled approximately 15 percent of the dialogue logs from each system, resulting in a total of about 2,700 randomly selected records. For each record, four evaluators engaged in ten rounds of discussion, and the score assigned by each evaluator was recorded at every round. Figure~\ref{fig:agreement_by_turn} presents the turn-wise evolution of agreement ratios for four conversational recommendation models. From Figure~\ref{fig:agreement_by_turn}, one can observe that all figures reveal a logarithmic-like upward trend in evaluator consensus across the ten discussion turns 
. At the initial stage, unanimous agreement was negligible across all models, with metrics virtually at zero. During the first four discussion turns, the proportions of unanimity and majority consensus increase rapidly with each successive round of discussion. From the fourth discussion turn onward, the rate of increase in agreement exhibits a noticeable deceleration. Within the ten discussion turns, as the discourse progressed, all systems exhibited substantial growth in both unanimity and majority consensus. Furthermore, from a model-wise perspective, CHATCRS emerged as the most capable model for achieving convergence in the early stages, starting from a low baseline but eventually surpassing 50 percent in unanimous agreement and exceeding 75 percent in majority consensus by the fourth turn. BARCOR and UniCRS followed a similar positive trend, with both models securing unanimous agreement rates exceeding 40 percent and reaching approximately 70 percent for majority agreement. In contrast, KBRD displayed a more gradual rate of convergence, concluding this phase with unanimous consensus hovering around 35 percent and majority agreement nearing 60 percent. In summary, setting the maximum number of discussion rounds to four represents a cost–benefit trade-off and is also consistent with prior practice, as adopted in previous work such as \citet{chan2023chateval}.

\subsection{Human Feedback Collection} 
\begin{figure*}[t]
  \includegraphics[width=0.9\linewidth]{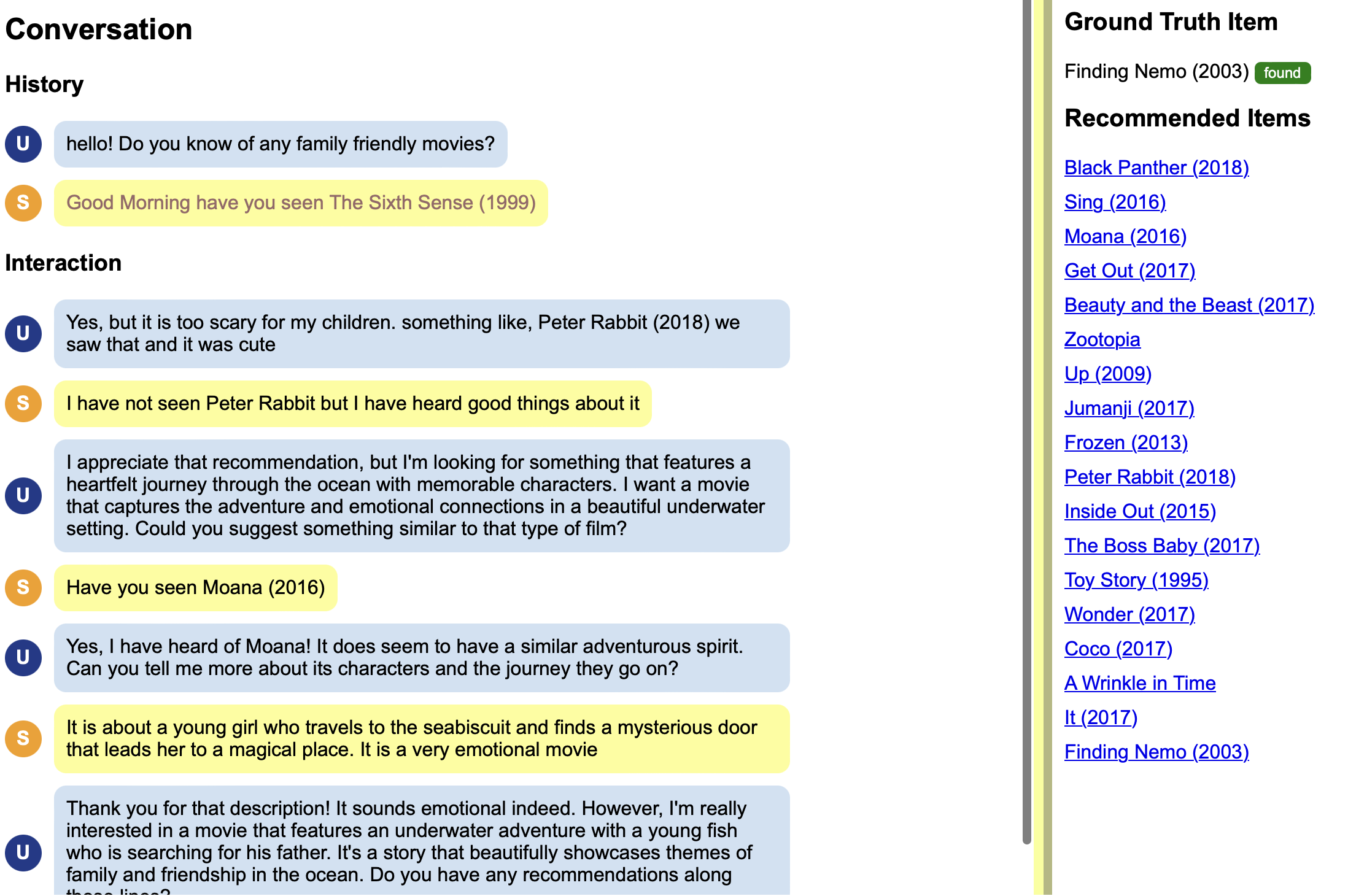}
  \caption{The screenshot of the interface used for human annotation.}
  \label{fig:anno_screenshot}
\end{figure*}
We hired ten human annotators for the annotation task. All annotators were university students or graduate students and had an English proficiency level equivalent to CEFR C1 or above. Each annotator received a payment of approximately \$40.

We randomly selected 10 records from the test sets of OpenDialKG and ReDial, respectively. For each record, every system has a corresponding user-system conversation log. As a result, we obtained a total of $2 \times 10 \times 4 = 80$ conversation logs. Each of the 10 evaluators was responsible for eight conversation logs. For each log, each evaluator received the same guidelines as the LLM evaluators on twelve factors 
. Additionally, they are required to give an overall rating on a scale from 0 (indicating they would never use the system under any circumstances) to 100 (indicating they would always use the system). For factor-wise scoring, we provided human evaluators and LLM evaluators with the same guidelines to examine the potential of LLMs as substitutes for human evaluation in cases where human evaluators are unavailable. For overall scoring, we did not provide human evaluators with detailed guidelines but instead asked them to make intuitive judgments. The goal was to assess CoRE's ability to align with human user experience.

Figure~\ref{fig:anno_screenshot} shows a screenshot of the interface given to human annotators.

\section{Result Analysis}
\label{ch:result}

In this section, we present a comprehensive analysis of the experimental results to assess the validity, reliability, and applicability of our proposed evaluation framework. The analysis is structured around four key research questions. We begin by examining the statistical independence and inter-factor correlations among the twelve evaluation factors, comparing patterns observed in human annotations and LLM-based assessments (\textbf{RQ1}). Next, we assess the alignment between LLM-generated factor-wise scores and human judgments, identifying model-specific strengths and limitations in approximating human evaluations (\textbf{RQ2}). We then evaluate the coherence of overall system-level judgments produced by our framework, demonstrating that multi-agent aggregation achieves significantly higher agreement with human ratings than conventional metrics such as Recall or Persuasiveness (\textbf{RQ3}). Finally, we benchmark the performance of four representative CRSs across both individual factors and multi-agent perspectives, revealing their relative strengths and weaknesses in dialogue quality and recommendation effectiveness (\textbf{RQ4}).

\subsection{Validity of the Proposed Framework}
\label{ch:validity}

\subsubsection{Factor-wise correlation analysis}
\begin{figure}[htbp]
  \includegraphics[width=0.99\linewidth]{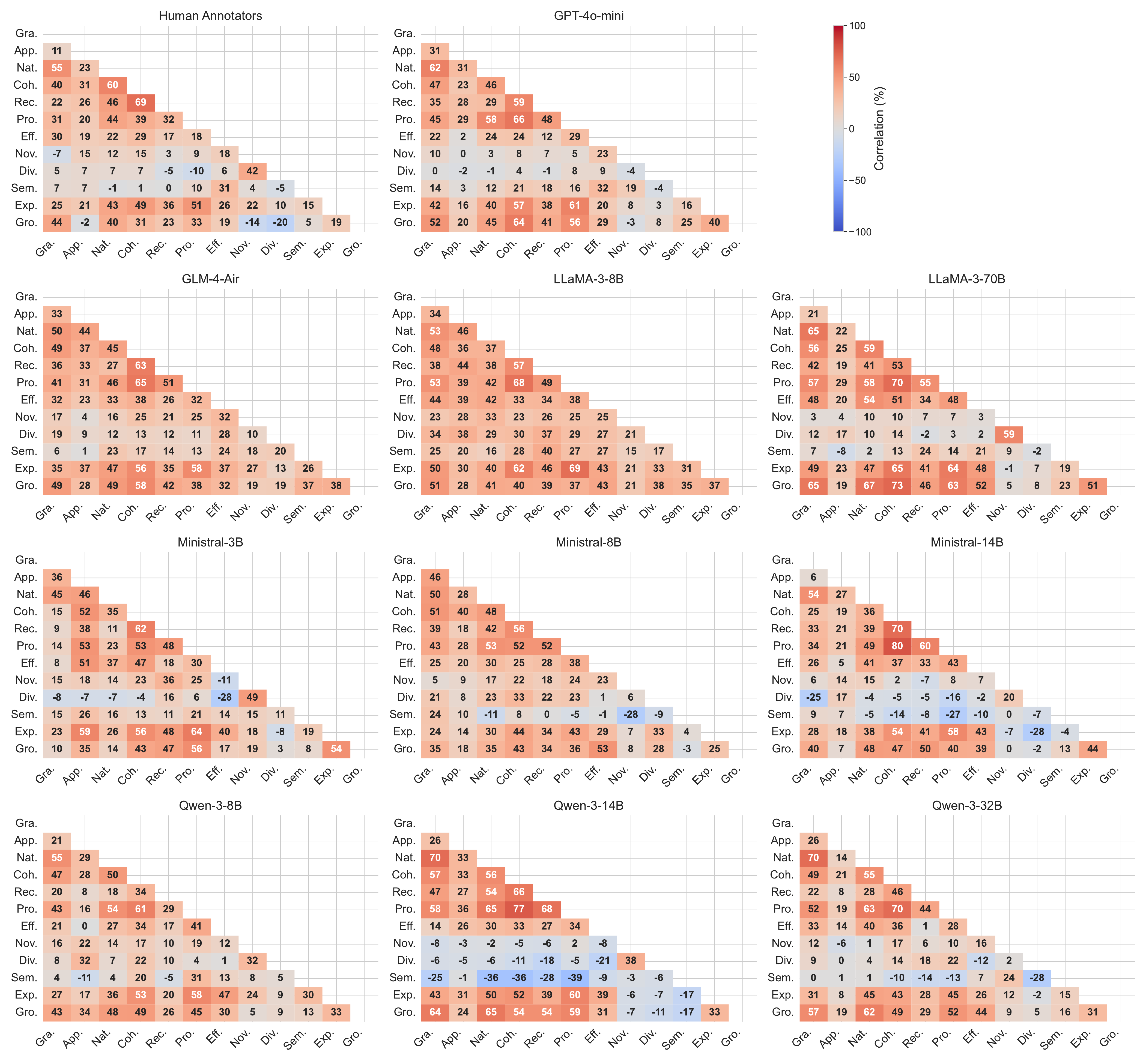}
  \caption{Kendall's $\tau_b$ correlation between factor scores (expressed in percentages) for different evaluators: (a) human annotators; (b) GPT-4o-mini; (c) GLM-4-Air; (d, e) LLaMA-3 family (8B and 70B); (f, g, h) Ministral family (3B, 8B, and 14B); and (i, j, k) Qwen family (8B, 14B, and 32B). Coh. represents Coherence; Rec. represents Recoverability; Pro. represents Proactiveness; Gra. represents Grammatical Correctness; Nat. represents Naturalness; App. represents Appropriateness; Eff. represents Effectiveness; Nov. represents Novelty; Div. represents Diversity; Sem. represents Semantic Relevance; Exp. represents Explainability: Gro represents Groundness.
  }
  \vspace{-1em}
  \label{fig:factorwise_corr}
\end{figure}

For the twelve factors in the proposed framework, we collected the scores given by human evaluators for each factor, the individual scores assigned by each of the ten LLM evaluators for each factor, based on a sample of 80 records. We computed the Kendall's $\tau_b$ correlation scores between the factor scores. Figure~\ref{fig:factorwise_corr} shows the Kendall's $\tau_b$ correlation matrices (coefficients in percentages).

As shown in subfigure (a) of Figure~\ref{fig:factorwise_corr}, in general, scores on each factor given by human annotators show a low pairwise correlation with those on other factors. The average absolute correlation value of human evaluators across the 66 factor‐pairs is approximately 0.22. Further analysis reveals that, out of the 66 factor‐pairs in (a), 62 pairs have an absolute correlation below 0.5, and only 4 pairs reach or exceed 0.5. Moreover, more than a half (33) of the factor-pairs exhibit an absolute correlation below 0.2. Specifically, the only pairs with correlations $\geq$ 0.5 are \textit{Coherence}–\textit{Recoverability} (0.69), \textit{Naturalness}–\textit{Coherence} (0.60), \textit{Grammar}–\textit{Naturalness} (0.55), and \textit{Proactiveness}–\textit{Explainability} (0.51). These factor pairs with relatively high positive correlations originate from within the language and dialogue factors. The overall weak correlations observed across factors underscore their conceptual distinctness. This suggests that the twelve factors in the proposed framework function largely as orthogonal constructs, where each factor captures a unique aspect of conversational recommendation performance with minimal redundancy.

The average absolute correlation value of LLM evaluators across the 66 factor-pairs is around 0.28. The overall mean correlation of the LLMs is higher than that of human evaluators, indicating that the models are more prone to mutual influence between different evaluation metrics (e.g., texts perceived as grammatically correct are often also considered more natural). The magnitude of inter-factor correlations varies by model architecture.  The \textbf{Llama}-based models and \textbf{GLM-4-Air} demonstrate a high degree of metric coupling, with average absolute correlations of  $0.362$ for Llama-3-8B and $0.307$ for Llama-3-70B, and $0.305$ for GLM-4-Air, respectively. Intermediate values are observed for \textbf{GPT-4o-mini} and \textbf{Ministral} series, with average absolute correlations of $0.250$ for GPT, $0.274$ for Ministral-3B, $0.264$ for Ministral-8B, and $0.250$ for Ministral-14B, respectively. The Qwen3 series exhibits a polarized pattern in terms of metric coupling: while Qwen3-8B and Qwen3-32B demonstrate high factor distinctness with some of the lowest average absolute correlations ($0.248$ and $0.244$, respectively), Qwen3-14B shows a higher correlation ($0.307$). This variance highlights that certain models are more prone to inter-factor coupling than others.

For the majority of LLMs, there are relatively strong positive correlations among dialogue related and language related factors, including \textit{Grammar (Grammatical Correctness)},\textit{Naturalness}, \textit{Coherence}, \textit{Proactiveness}, and \textit{Appropriateness}. For instance, \textbf{Qwen3-14B} exhibits a substantial correlation between \textit{Grammar} and \textit{Naturalness} ($0.78$), while \textbf{GPT} demonstrates an exceptionally high dependency between \textit{Coherence} and \textit{Proactiveness} ($0.84$). Similarly, \textbf{Llama-3-70B} shows a strong link between \textit{Coherence} and \textit{Proactiveness} ($0.80$).  This observation suggests that, despite providing the comprehensive definitions and rubrics, the assessments performed by LLMs can still be affected by factors beyond the specific factor being evaluated.


In summary, the finding addressed \textbf{RQ1} by demonstrating that, \textbf{\emph{the twelve proposed factors are fundamentally orthogonal, each capturing a unique construct. However, both human and LLM evaluators exhibit a certain degree of inter‐factor coupling in their ratings of dialogue- and language‐related factors, with the coupling being more pronounced for LLM evaluators.
}}  LLM evaluators also exhibit a higher degree of inter‐factor coupling in their ratings of recommendation items and response content, suggesting that their judgments may be influenced by factors beyond the specific factor being evaluated. This insight contributes to the practice of using LLMs for CRS evaluation by demonstrating that, while LLMs can be used as evaluators, care must be taken to design prompt structures and scoring guidelines that minimize unintended inter-factor dependencies.

\subsubsection{The Analysis of System Ranking Consistency}
\begin{figure*}[htbp]
  \centering
  \includegraphics[width=.99\linewidth]{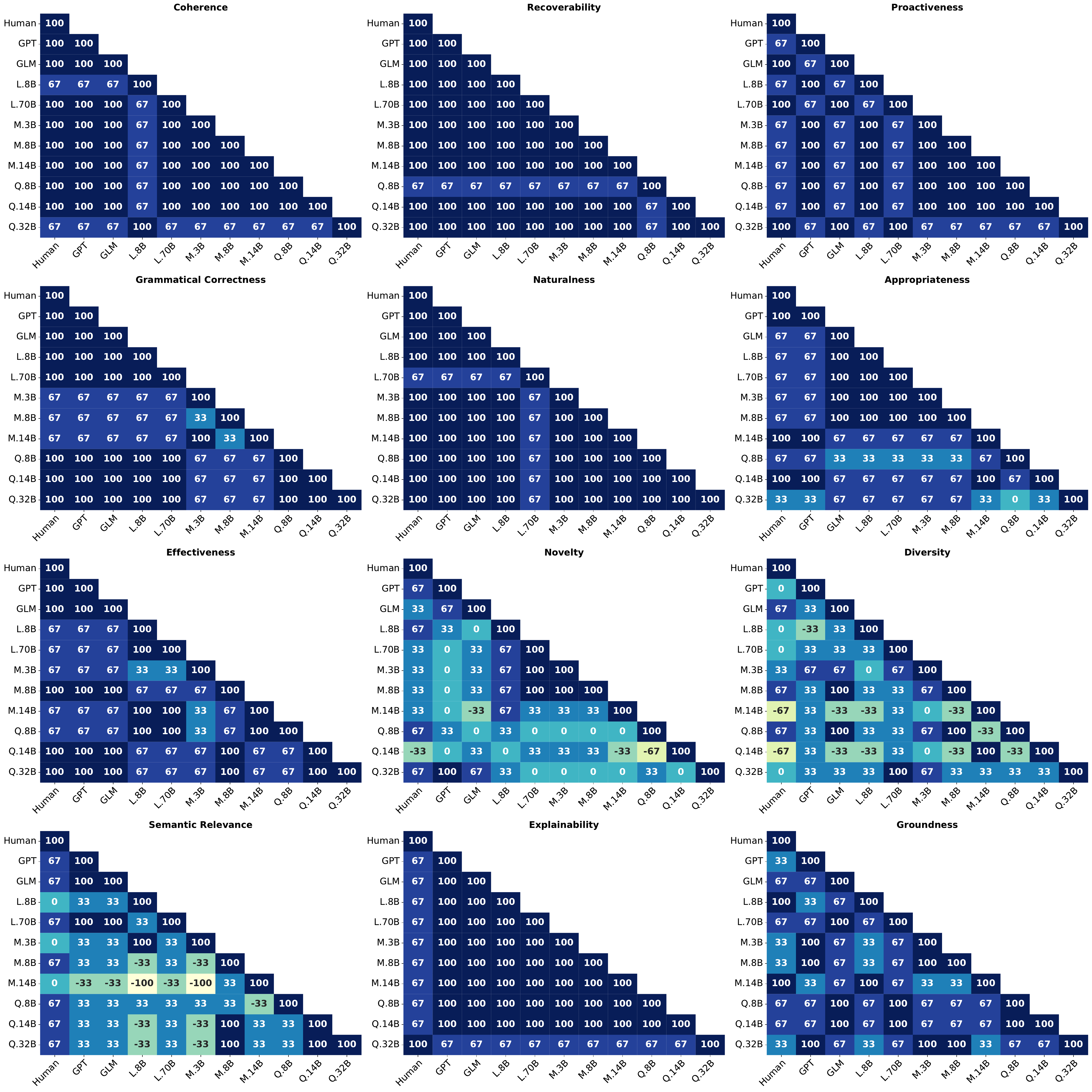}
\caption{Kendall’s $\tau_b$ correlation (expressed in percentages) between system rankings across eleven evaluators among twelve evaluation factors. Each cell represents the pairwise correlation of system rankings, where a value of 100 indicates perfect ranking alignment and negative values signify inverse ranking trends. L.{8B, 70B} represents Llama-3-{8B, 70B}; M.{3B, 8B, 14B} represents Ministral-{3B, 8B, 14B}; and Q.{8B, 14B, 32B} represents Qwen-3-{8B, 14B, 32B}, respectively.}
  \label{fig:corr_sys_rank}
\end{figure*}
\begin{figure*}[htbp]
  \centering
  \includegraphics[width=.99\linewidth]{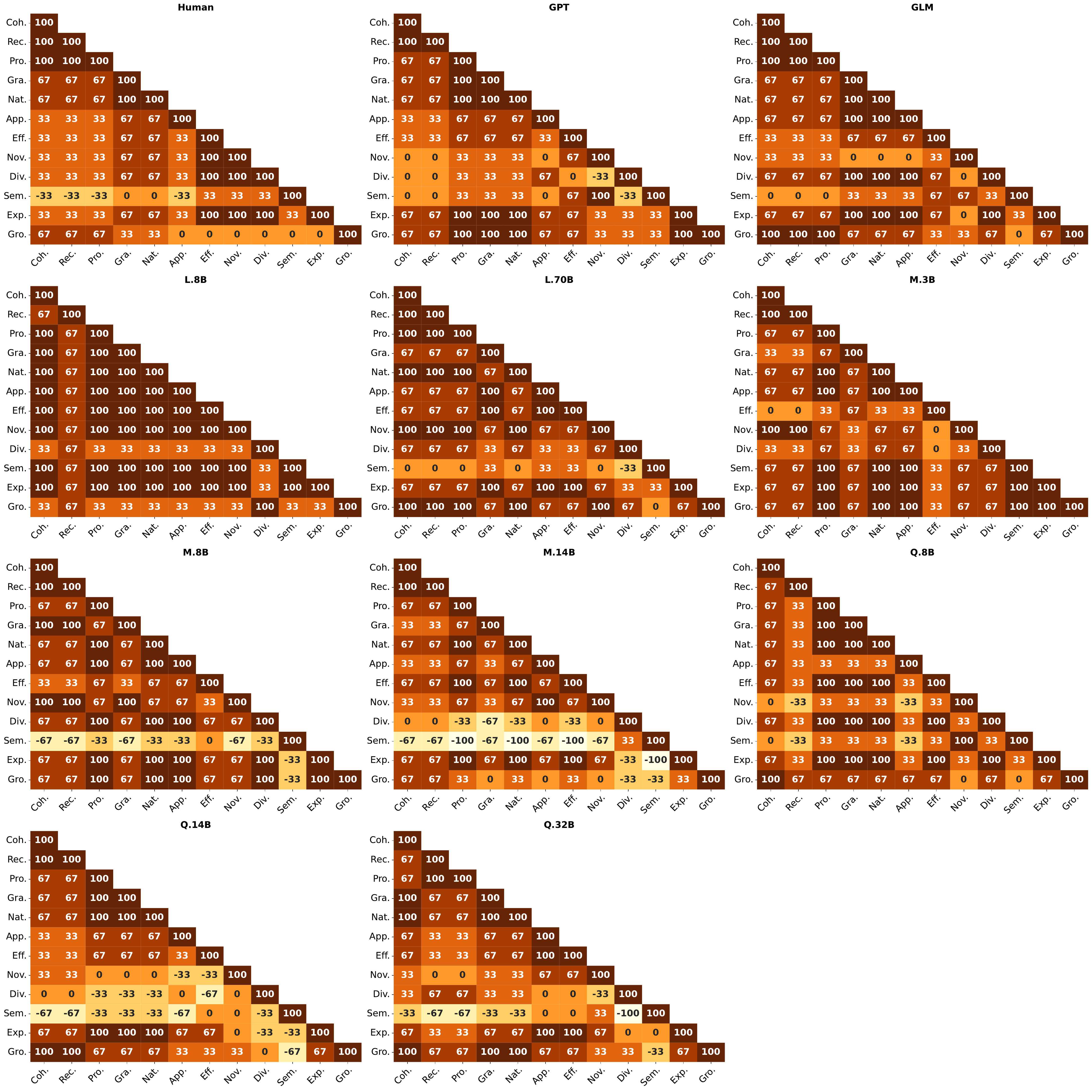}
\caption{Kendall’s $\tau_b$ correlation (expressed in percentages) between system rankings across twelve evaluation factors among eleven evaluators. Each cell represents the pairwise correlation of system rankings, where a value of 100 indicates perfect ranking alignment and negative values signify inverse ranking trends. L.{8B, 70B} represents Llama-3-{8B, 70B}; M.{3B, 8B, 14B} represents Ministral-{3B, 8B, 14B}; and Q.{8B, 14B, 32B} represents Qwen-3-{8B, 14B, 32B}, respectively. Coh. represents Coherence; Rec. represents Recoverability; Pro. represents Proactiveness; Gra. represents Grammatical Correctness; Nat. represents Naturalness; App. represents Appropriateness; Eff. represents Effectiveness; Nov. represents Novelty; Div. represents Diversity; Sem. represents Semantic Relevance; Exp. represents Explainability: Gro represents Groundness.}
  \label{fig:corr_sys_rank_per_factor}
\end{figure*}

Figure~\ref{fig:corr_sys_rank} and Figure~\ref{fig:corr_sys_rank_per_factor}  present an analysis of the consistency in ranking lists between humans and ten LLMs across twelve factors for four systems (BARCOR, CHATCRS, KBRD, UniCRS), measured using Kendall’s $\tau$. A score of 100 indicates perfect agreement in ranking order, while -100 indicates complete reversal of ranking order. 

Figure ~\ref{fig:corr_sys_rank} illustrates the consistency of ranking lists across different models for each factor. For approximately half of the evaluated factors, namely \textit{Coherence}, \textit{Naturalness}, \textit{Proactiveness}, \textit{Recoverability}, and \textit{Explainability}, robust positive correlations ($\tau > 0.66$) are observed in both inter-LLM and Human--LLM ranking comparisons. This pattern indicates a strong consensus: irrespective of whether the evaluator is a human or an LLM, including GPT, GLM, Qwen, Llama, and Ministral, highly similar system ranking lists are produced for these language- and dialogue-related factors. For one third of the factors, specifically \textit{Grammatical Correctness}, \textit{Appropriateness}, \textit{Effectiveness}, and \textit{Groundness}, similarly high positive correlations ($\tau > 0.66$) are also evident in both inter-LLM and Human-LLM rankings, with only a few isolated deviations. In contrast, for the factors of \textit{Novelty}, \textit{Diversity}, and \textit{Semantic Relevance}, substantial disagreement emerges in system rankings, both between human evaluators and LLMs and among the LLMs themselves. Notably, this discordance can escalate to strong negative correlations, as exemplified by the \textit{Novelty} factor, where the rankings produced by Qwen-3-8B and Qwen-3-14B exhibit a pronounced negative association.

Figure~\ref{fig:corr_sys_rank_per_factor}  illustrates the consistency of system rankings across different factors for the same model. In the upper-left triangular region of the correlation matrices, which encompasses fundamental quality factors such as \textit{Coherence}, \textit{Grammar}, \textit{Naturalness}, and \textit{Appropriateness}, both human evaluators and the vast majority of model-based evaluators (including the GPT-4o-mini, GLM-4-Air, Llama-3-8B, and Qwen series) exhibit a distinct deep brown coloration, ($\tau > 0.66$). This pattern indicates a high degree of concordance in the system rankings produced by these evaluators regarding basic dialogue and linguistic factors. For example, on \textit{Grammar} and \textit{Naturalness}, the ranking produced by GPT-4o-mini is the same: first \textit{CHATCRS}, second \textit{BARCOR}, third \textit{UniCRS}, and fourth \textit{KBRD}. However, substantial divergences appear when examining specific factors and model behaviors. Within the Human evaluation sub-plot, the \textit{Semantic Relevance} factor notably demonstrates negative ranking correlations with \textit{Coherence} and \textit{Grammatical Correctness} (e.g., $\tau = -0.33$ ). In contrast, the Llama-3-8B matrix is nearly saturated with values of 100, implying that the model produced identical system rankings across all 12 evaluation factors. Conversely, the matrices for Qwen-3-32B  and Ministral-14B reveal marked yellow regions in the lower-right quadrant indicative of negative values; most notably, Qwen-3-32B  exhibits a correlation of $\tau = -0.67$  between \textit{Semantic Relevance} and \textit{Grammar}, representing a significant deviation from the patterns observed in other large models.

In summary, from the perspective of inter-model consistency, for traditional and well defined language- and dialogue-related factors, LLMs produce highly consistent evaluation outcomes, although these are not necessarily fully aligned with human judgments. In contrast, for higher level tasks involving creativity and diversity, substantial divergence and uncertainty persist in the evaluation criteria, both among different LLMs and between LLMs and human evaluators.

\subsubsection{Validity of factor-wise evaluation using LLMs.} 
\begin{figure*}[htbp]
  \centering
  \includegraphics[width=.99\linewidth]{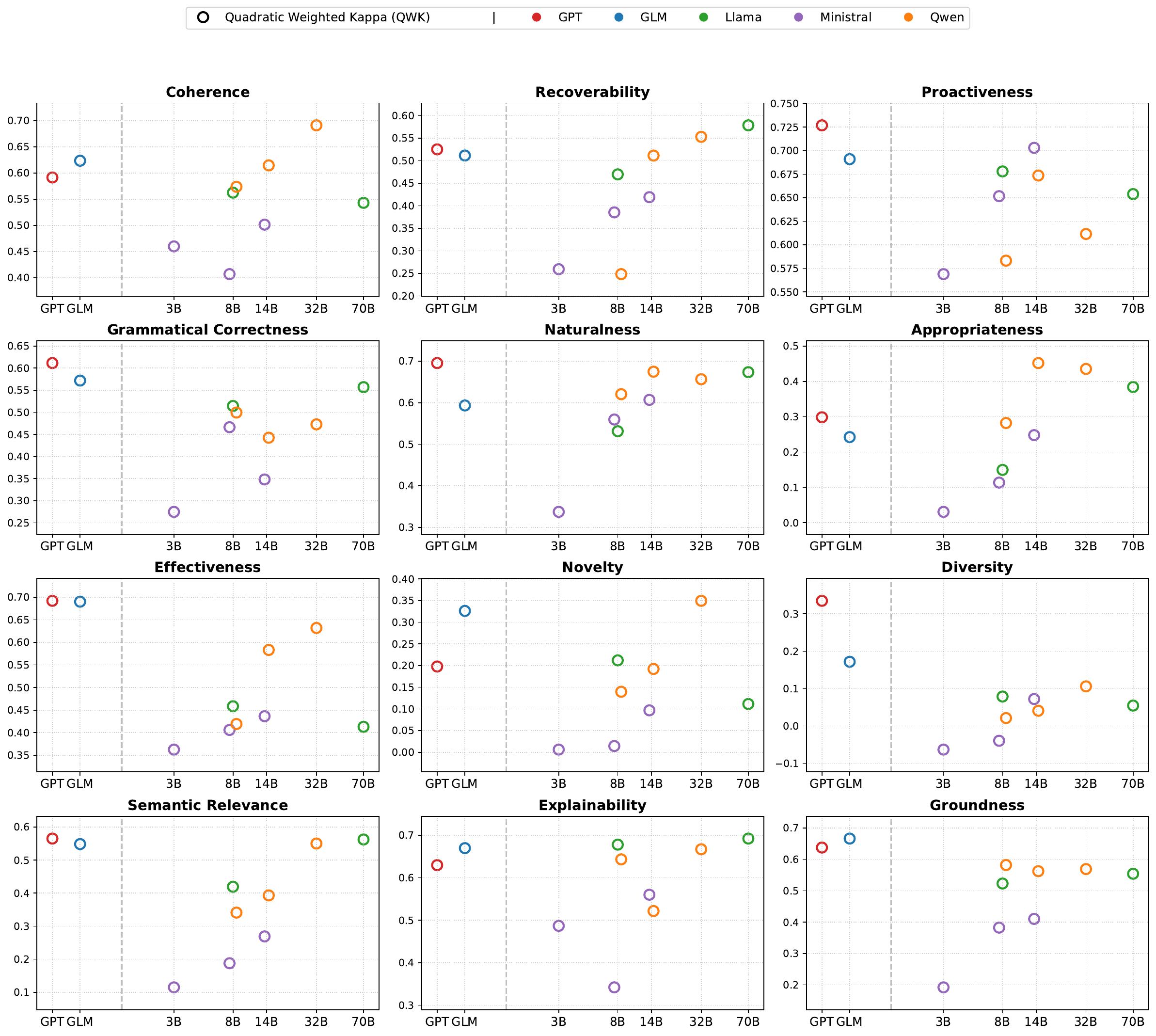}
  \caption{Agreement analysis between various large language models and human annotators across twelve evaluation factors. The x-axis illustrates the model family and parameter scale, while the y-axis denotes the Quadratic Weighted Kappa (QWK) score.}
  \label{fig:factor_corr_visualization}
\end{figure*}

Figure~\ref{fig:factor_corr_visualization} and Table~\ref{tab:factor_align} report the Quadratic Weighted Kappa (QWK) scores between LLM-based evaluators and human ratings across twelve conversational recommendation factors. The QWK score assesses the alignment between automated and human assessments, with higher values indicating stronger agreement. 

From an overall capability perspective, GPT-4o-mini demonstrates strong and consistent agreement with human evaluators across the evaluation factors.  GPT-4o-mini achieves QWK scores above 0.6 on \textit{Coherence}, \textit{Proactiveness},  \textit{Naturalness}, and \textit{Grammatical Correctness}. Particularly, GPT-4o-mini shows strong performance in  \textit{Proactiveness}  and \textit{Naturalness}, with a QWK score around 0.7, indicating a clear advantage in evaluating dialogue- and language-related factors. Qwen-3-32B and Llama-3-70B also exhibit robust overall capabilities and maintain competitive performance across the majority of factors. Notably, Qwen-3-32B demonstrates strong alignment in \textit{Coherence} ($0.69$) and \textit{Explainability} ($0.67$). On some factors, Qwen-3-32B’s performance is comparable to or even exceeds that of GPT-4o-mini, particularly in \textit{Coherence} ($0.69$ vs. $0.59$) and \textit{Novelty} ($0.35$ vs. $0.20$). GLM-4-Air, Qwen-3-8B, and Llama-3-8B also demonstrate strong performance on specific factors. In particular, GLM-4-Air performs notably well in \textit{Coherence} and \textit{Proactiveness}. Llama-3-8B shows relatively strong results in \textit{Proactiveness} and \textit{Explainability}, while Qwen-3-8B exhibits competitive performance in \textit{Explainability} and \textit{Naturalness}. In contrast, the Ministral models, particularly the 3B variant, show weaker overall performance. These models obtain lower QWK scores across multiple factors, with Ministral-3B performing the weakest, reflecting limitations in evaluating the language quality and dialogue behavior of CRSs, compared to larger and more capable models. However, the lower-tier classification of Ministral models is not absolute. Ministral-14B outperforms Qwen-3-14B in \textit{Proactiveness} ($0.70$ vs. $0.67$) and \textit{Explainability} ($0.56$ vs. $0.52$). Similarly, Ministral-8B demonstrates a superior evaluation alignment with human on \textit{Proactiveness} ($0.65$) compared to Qwen-3-8B ($0.58$). 

From the perspective of the relationship between parameter scale and performance, overall, model performance shows a clear positive association with parameter scale. For example, within the Qwen-3 series, performance on most evaluation metrics increases as model size grows from 8B to 14B and further to 32B, with particularly pronounced gains in \textit{Semantic Relevance} and \textit{Novelty}. A similar pattern is observed in the Ministral series, where the 14B model substantially outperforms the 3B variant across multiple factors. Notably, the 14B model achieves markedly higher scores in \textit{Proactiveness} at 0.7 and \textit{Naturalness} at 0.61, compared to 0.27 and 0.34, respectively, for the 3B model. However, this trend is not universal. On certain evaluation factors, smaller models within the same family can outperform their larger counterparts. In the Qwen-3 series, Qwen-3-8B achieves a higher QWK score than Qwen-3-14B in \textit{Grammatical Correctness}, with scores of 0.50 and 0.44, respectively. A similar pattern is observed for \textit{Explainability}, where Qwen-3-8B attains a substantially higher score of 0.64 compared to 0.52 for Qwen-3-14B. 

A comparison within identical parameter scales reveals that models within the same parameter class exhibit heterogeneity in their performance across different evaluation factors. For instance, among the 8B models, Llama-3-8B and Qwen3-8B show strong alignment with human judgments in \textit{Explainability}, with QWK scores of 0.68 and 0.64, respectively, whereas Ministral-8B lags behind at 0.34. A similar pattern is observed for \textit{Groundness}, where Qwen3-8B achieves a QWK of 0.58 and Llama-3-8B scores 0.52, compared to only 0.38 for Ministral-8B. A similar divergence is observed in the 14B class, where Qwen3-14B shows higher agreement with human evaluators on \textit{Coherence} and \textit{Semantic Relevance}, while Ministral-14B exhibits stronger alignment on \textit{Proactiveness} and \textit{Explainability}. These results indicate that, even at comparable parameter scales, differences in model architecture and training lead to distinct patterns of alignment with human judgments across evaluation factors, rather than reflecting the intrinsic performance of the CRSs themselves.

A cross-factoral comparison further reveals clear strengths and weaknesses in current model capabilities. \textit{Naturalness} and \textit{Proactiveness} consistently exhibit higher agreement between LLM-based evaluators and human judgments, with most models achieving scores above 0.6. This indicates that contemporary models align well with human evaluators when assessing these aspects of CRS performance. \textit{Coherence}, \textit{Explainability} and \textit{Groundness} also show stable performance for most major models except for Ministral family, suggesting that these aspects of evaluation are relatively well supported by current architectures. In contrast, \textit{Diversity} represents a substantial challenge for all evaluated models. Scores in this factor are uniformly low, with Ministral-3B even exhibiting a negative value of $-0.06$. Even the strongest model, GPT-4o-mini, achieves only $0.34$. \textit{Novelty} similarly remains a weak factor, with generally low scores across models. This pattern reflects a limitation for models to evaluate the diversity and novelty of item lists returned by CRSs.

Taken together, these findings provide strong empirical support for the modular design of the CoRE framework. By assigning the most suitable model to each factor based on its demonstrated strengths, the framework yields more faithful approximations of human evaluations than any monolithic approach. In practical deployment, this could be implemented via a simple weighted average or a more dynamic routing mechanism that selects the optimal model for each evaluation factor. The consistent gains observed from score aggregation reinforce the utility of combining LLM perspectives, particularly in complex multi-faceted evaluation tasks such as conversational recommendation.


In summary, the results addressed \textbf{RQ2}, suggesting that, \textbf{\emph{CoRE, when paired with appropriately selected LLM evaluators on a per-factor basis, achieves strong alignment with human evaluations across most key aspects of user experience in conversational recommendation, specifically in factors related to dialogue actions, language expression, and response content.}} This finding has significant implications: it supports a modular evaluation design where different models contribute to different parts of the evaluation process based on their strengths. This modularity enables more robust and reliable automated evaluation pipelines, and highlights the potential of LLM ensembling as a practical strategy for approximating human evaluations in multi-factoral settings. These insights advance the field’s understanding of how LLMs can be responsibly integrated into the evaluation process and pave the way for scalable, hybrid human–machine evaluation frameworks in conversational AI.

\subsubsection{Validity of overall evaluation using LLMs.}

Table~\ref{tab:agent_align} reports the Spearman correlation $r$ and Kendall correlation $\tau_b$ between human evaluation scores and a range of CRS evaluation metrics. CoRE (Avg.~\{GPT, LLaMa, GLM, Ministral, Qwen\}) denotes the average score across twelve evaluation factors obtained using a single large language model as the evaluator, while CoRE (Multi-Agent) corresponds to the final score produced by the multi-agent debate framework.

Across all single-model instantiations, CoRE consistently exhibits substantially stronger alignment with human judgments than traditional recommendation metrics. In particular, CoRE (Avg.~GPT-4o-mini) achieves $r = 0.663$ and $\tau_b = 0.505$, far exceeding Recall@1 ($r = 0.140$, $\tau_b = 0.127$), Recall@5 ($r = 0.107$, $\tau_b = 0.091$), and Recall@10 ($r = 0.079$, $\tau_b = 0.068$). Comparable levels of correlation are observed for other backbones, including LLaMa-3-70B ($r = 0.678$, $\tau_b = 0.511$), Qwen3-14B ($r = 0.665$, $\tau_b = 0.486$), and GLM-4-Air ($r = 0.632$, $\tau_b = 0.480$), indicating that although model capacity and architecture affect absolute alignment, the CoRE formulation itself remains robust across diverse evaluators.

In contrast, Persuasiveness shows only weak correlation with human scores ($r = 0.281$, $\tau_b = 0.269$), underscoring the limitations of single-dimensional heuristics that focus narrowly on rhetorical effectiveness rather than holistic conversational quality. Smaller evaluators such as Ministral-3B and Ministral-8B further exhibit reduced alignment, suggesting that evaluator capability remains a contributing factor when fine-grained judgment is required.

The CoRE multi-agent approach achieves the strongest overall alignment, reaching $r = 0.698$ and $\tau_b = 0.538$. This represents a relative gain of approximately 5.28 percent in Spearman \(r\) and 6.53 percent in Kendall \(\tau_b\) compared to CoRE (Avg. GPT). Relative to the strongest single-model baseline, this corresponds to absolute improvements of 0.020 in Spearman correlation and 0.027 in Kendall correlation. We attribute this gain to the structured debate mechanism, which reconciles divergent role-specific assessments and mitigates idiosyncratic evaluator biases through iterative argumentation and consensus building. By synthesizing twelve complementary factors spanning dialogue quality, user modeling, and recommendation effectiveness, CoRE produces a unified score that captures aspects of conversational recommender performance that neither recall-based metrics nor single-perspective judgments can adequately reflect. Overall, these results demonstrate that CoRE, particularly in its multi-agent formulation, provides a reliable, interpretable, and human-aligned metric for evaluating conversational recommender systems.

In summary, these results addressed \textbf{RQ3}, demonstrating that \textbf{\emph{CoRE, when paired with appropriately selected LLM evaluators and aggregated via a multi agent framework, provides a far more faithful measure of overall CRS performance than traditional recall‑based metrics or single‑factor scores.}} Our results contribute to the emerging practice on holistic CRS evaluation and suggest new directions for building human-aligned evaluation protocols through model coordination. In doing so, this study strengthens the foundation for principled, interpretable, and dynamic evaluation methods that go beyond static metrics and better reflect real-world user expectations.

\begin{table}[ht]
\centering
\caption{The alignment between our overall evaluation metrics and human evaluation is measured using Spearman’s $r$ and Kendall’s $\tau_b$.}
\begin{tabular}{lcc}
\toprule
 \textbf{Metric} & \textbf{$r$} & \textbf{$\tau_b$} \\
\hline
CoRE (Avg. GPT-4o-mini)      & 0.663 & 0.505 \\
CoRE (Avg. GLM-4-Air)        & 0.632 & 0.480 \\
CoRE (Avg. Llama-3-8B)       & 0.639 & 0.486 \\
CoRE (Avg. Llama-3-70B)      & 0.678 & 0.511 \\
CoRE (Avg. Ministral-3B)     & 0.562 & 0.420 \\
CoRE (Avg. Ministral-8B)     & 0.561 & 0.439 \\
CoRE (Avg. Ministral-14B)    & 0.600 & 0.451 \\
CoRE (Avg. Qwen3-8B)         & 0.635 & 0.478 \\
CoRE (Avg. Qwen3-14B)        & 0.665 & 0.486 \\
CoRE (Avg. Qwen3-32B)        & 0.618 & 0.450 \\
CoRE (Multi-Agent)        & \textbf{0.698 } & \textbf{0.538} \\
\midrule
Recall@10         & 0.079 & 0.068 \\
Recall@5         & 0.107 & 0.091 \\
Recall@1         & 0.140 & 0.127 \\
Persuasiveness     & 0.281 & 0.269 \\
\bottomrule
\end{tabular}
\vspace{-1em}
\label{tab:agent_align}
\end{table}
\begin{figure*}[htbp]
    \centering
    \begin{subfigure}[t]{0.32\textwidth}
        \centering
        \includegraphics[width=4.8cm]{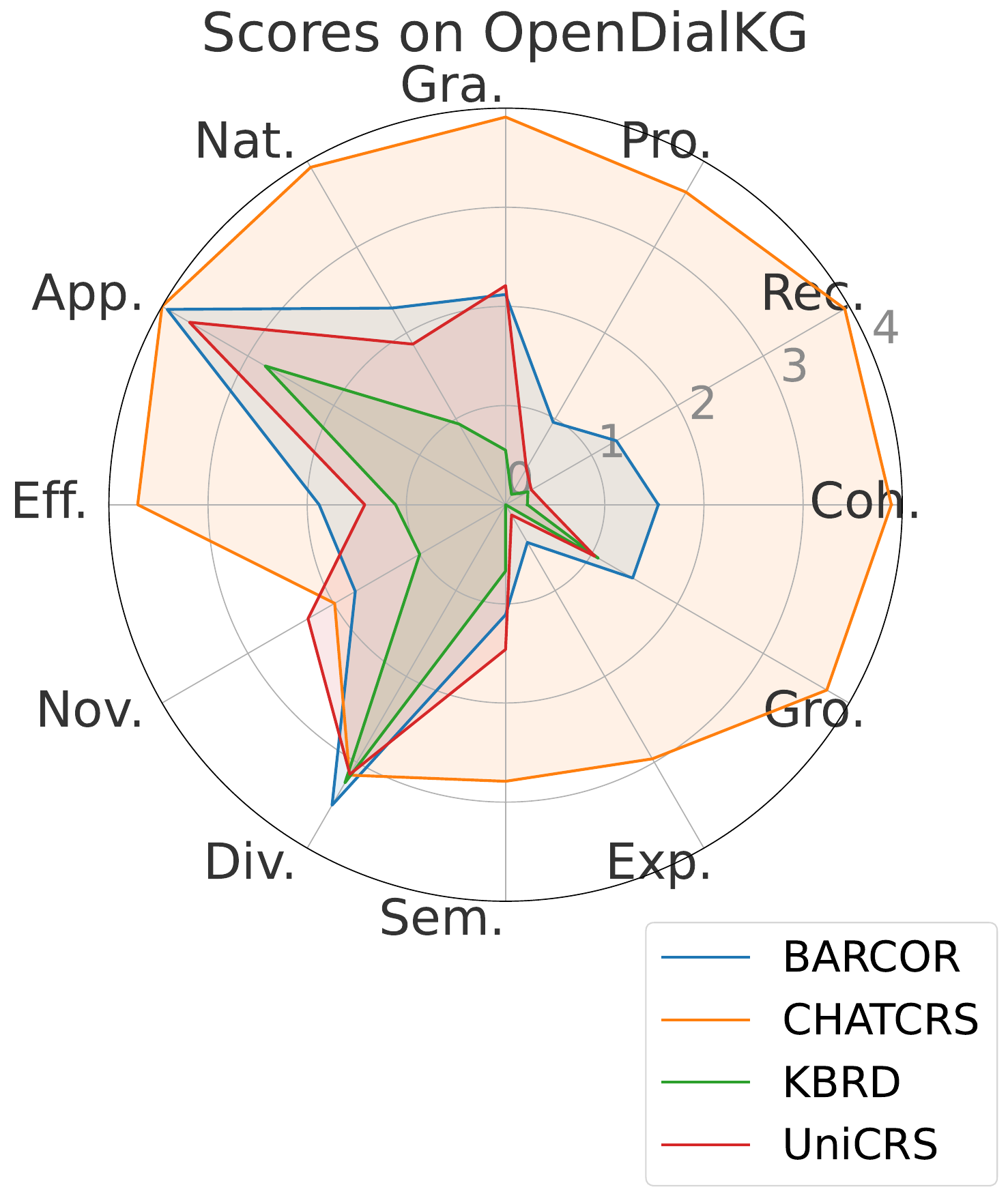}
    \end{subfigure}
    \begin{subfigure}[t]{0.32\textwidth}
        \centering
        \includegraphics[width=4.8cm]{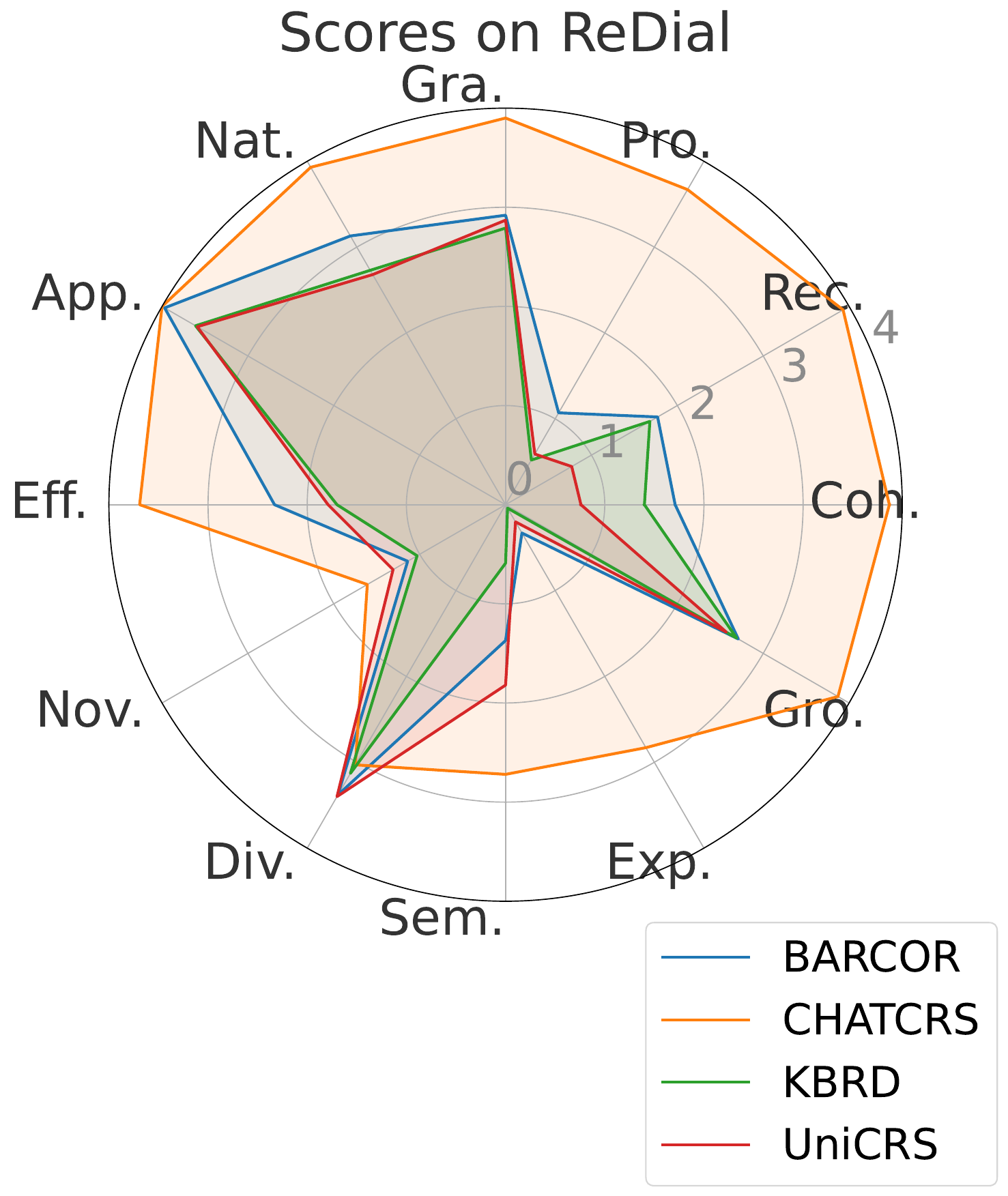}
    \end{subfigure}
    \begin{subfigure}[t]{0.32\textwidth}
        \centering
        \includegraphics[width=4.8cm]{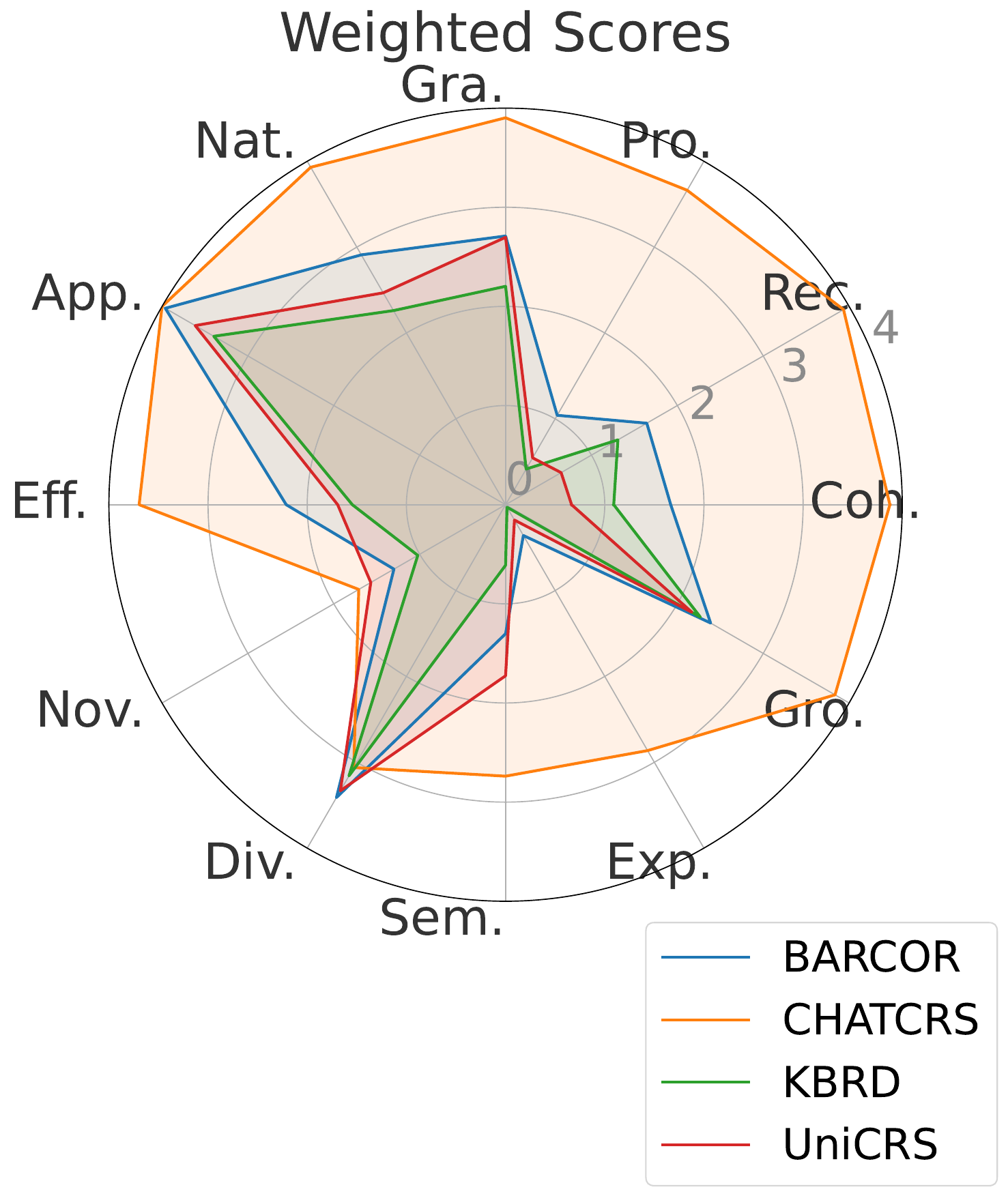}
    \end{subfigure}
    \vspace{-1em}
    \caption{Performance comparison of four dialogue systems across 12 evaluation metrics: (a) results on the OpenDialKG dataset; (b) results on the ReDial dataset; and (c) weighted average results, with weights proportional to dataset size. The scores are derived using a \textit{mixture-of-evaluators} strategy, where the most suitable LLM (defined as the model achieving the highest QWK correlation with human judgments) is selected to evaluate each specific metric (i.e., GPT-4o-mini for Proactiveness, Grammar, Naturalness, Effectiveness, Diversity, Semantic Relevance; Llama-3-70B for Recoverability, Explainability; Qwen3-32B for Coherence, Novelty; Qwen3-14B for Appropriateness; and GLM-4-Air for Groundness). Coh. represents Coherence; Rec. represents Recoverability; Pro. represents Proactiveness; Gra. represents Grammatical Correctness; Nat. represents Naturalness; App. represents Appropriateness; Eff. represents Effectiveness; Nov. represents Novelty; Div. represents Diversity; Sem. represents Semantic Relevance; Exp. represents Explainability; Gro. represents Groundness.}
    \vspace{-1em}
    \label{fig:factor_scores}
\end{figure*}
\begin{figure}[htbp]
    \centering
    \includegraphics[width=.78\linewidth]{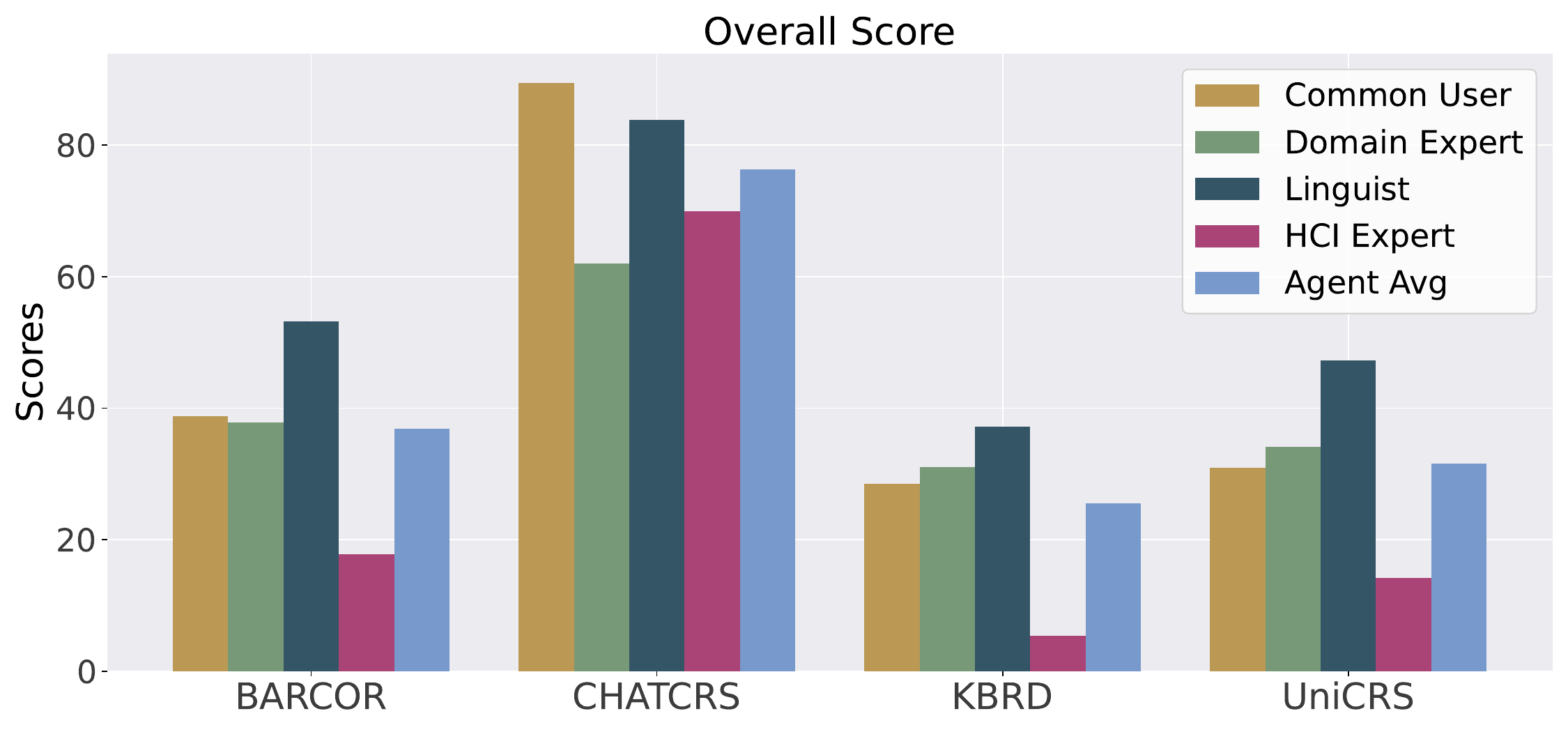}
    \caption{The scores of BARCOR, CHATCRS, KBRD and UniCRS after multi-agent discussion. }
    \vspace{-0.5em}
\label{fig:overall_scores}
\end{figure}


\subsection{Benchmarking System Performance} 
\label{ch:benchmark}
\subsubsection{System‑level factor evaluation. }

Figure~\ref{fig:factor_scores} shows each system’s performance across twelve conversational recommendation factors, as judged by the most human‑aligned LLM evaluator for each dimension according to the result in Section~\ref{ch:validity}. Specifically, GPT-4o-mini is used for \textit{Proactiveness}, \textit{Grammar}, \textit{Naturalness}, \textit{Effectiveness}, \textit{Diversity}, and \textit{Semantic Relevance}; LLaMa-3-70B for \textit{Recoverability} and \textit{Explainability}; Qwen3-32B for \textit{Coherence} and \textit{Novelty}; Qwen3-14B for \textit{Appropriateness}; and GLM-4-Air for \textit{Groundness}. 

Across the majority of the twelve evaluation factors, CHATCRS consistently outperforms all other systems. Specifically, CHATCRS achieves the highest scores on interaction-related and language-related factors, including \textit{Coherence}, \textit{Recoverability}, \textit{Proactiveness}, \textit{Grammar}, and \textit{Naturalness}. In addition, it reaches or approaches the upper bound on \textit{Appropriateness} and \textit{Effectiveness}. CHATCRS also leads on dimensions including \textit{Explainability} and \textit{Groundness}, indicating not only fluent language generation but also stronger reasoning transparency and content grounding. The consistency of these gains suggests a global performance advantage of CHATCRS. BARCOR and UniCRS exhibit moderate yet stable performance across most dimensions.  For instance, BARCOR obtains mid-range scores on \textit{Coherence}, \textit{Grammar}, and \textit{Groundness}, while UniCRS shows comparable values on \textit{Grammar}, \textit{Naturalness}, and \textit{Semantic Relevance}. By comparison, KBRD exhibits pronounced deficiencies on several key dialogue and interaction related dimensions, including \textit{Coherence}, \textit{Explainability}, \textit{Proactiveness}, and \textit{Semantic Relevance}, indicating systematic limitations in its ability to support coherent, explanatory, and context aware interactions.

\subsubsection{Multi‑agent overall assessment.}
Figure~\ref{fig:overall_scores} aggregates each system’s score from four expert roles—Common User, Domain Expert, Linguist and HCI Expert—and reports the mean across agents. CHATCRS leads by a large margin in every role, earning above‑80 marks from the Linguist and Common User and above‑70 from the HCI Expert. BARCOR places second overall, buoyed by relatively high Common User and Domain Expert ratings, but falls short under the HCI Expert’s criteria due to its lower Explainability and Proactiveness. UniCRS ranks third; it benefits from stronger Domain Expert scores in Semantic Relevance but suffers under HCI scrutiny. KBRD comes last, driven down by very low HCI Expert ratings—reflecting its persistent shortfalls in Proactiveness and Explainability—even though it scores respectably on Novelty. The agent‑average bars illustrate that systems must perform well across diverse stakeholder perspectives to achieve high overall standing.

In summary, these analyses addressed \textbf{RQ4}, confirming that, \textbf{\emph{CHATCRS outperforms alternative CRSs on both fine‑grained factors, and in a holistic multi‑agent evaluation, while the relative placements of BARCOR, KBRD and UniCRS reveal distinct trade‑offs between language quality, factual grounding, interaction proactiveness and explainability of recommendation.}}  These findings demonstrate the practical utility of CoRE in identifying trade-offs between dialogue effectiveness and recommendation diversity. They also provide guidance for system development and refinement by highlighting specific factors where current models underperform.




\section{Discussion}
\label{ch:discuss}
 By introducing CoRE, a unified framework that leverages LLMs to perform holistic, human-centred assessments of both dialogue management and recommendation quality, our work bridges the gap between the fields of human-computer interaction and recommender system technology through LLM-based user-centric evaluation. However, it is still in its early stages and has many limitations that need to be addressed in future work.  

\subsection{Inter-Factor Coupling in LLM-Based Evaluation}
One observed limitation of using LLMs for automated factor-wise evaluation is the presence of inter-factor coupling in their assigned scores. While human annotators tend to treat evaluation dimensions independently—resulting in largely orthogonal factor structures—LLMs often produce scores that are more correlated across dimensions. This coupling effect may arise from several factors, including limitations in instruction-following capabilities, latent biases in the pretraining data, and the model’s tendency to infer coherence or quality holistically rather than isolating criteria as instructed. Such coupling can lead to discrepancies between the scores provided by the large model and those given by human raters. Future work should explore methods for reducing inter-factor coupling in LLM-based evaluations.  As LLMs continue to be integrated into evaluation pipelines, addressing this issue is essential to maintaining the credibility and precision of automated assessment. One promising direction is the design of more targeted and disentangled prompting strategies that clearly delineate the boundaries between factors. Another is the use of adversarial or contrastive calibration examples during scoring to enforce dimension-specific reasoning. For example, when evaluating the \textit{Explainability} dimension, the model can be presented with two reference dialogue segments: one containing a recommendation with clear logic and an explicit reasoning process, and the other exhibiting reasoning gaps and a lack of causal clarity. Both segments should be matched in terms of \textit{Naturalness} and \textit{Grammatical Correctness} to control for confounding variables. The model is then instructed to make a judgment based solely on explainability and to justify its score accordingly. This contrastive setup encourages the model to isolate the target dimension during evaluation and helps mitigate the influence of irrelevant factors.


\subsection{Hallucination Risks in LLM-Based Evaluation and Mitigation Strategies}

While Large Language Models have demonstrated impressive capabilities in evaluating CRSs, they are known to be susceptible to hallucinations, i.e., generating plausible yet factually incorrect information. To assess the extent of this issue within our evaluation framework, we conducted a case study analyzing LLM-generated outputs, and we only found minimal hallucinations. We attribute this to the nature of the datasets used (ReDial and OpenDialKG), both comprising older movies likely well-represented in the LLMs' training data. This observation raises concerns about the generalizability of our findings. If LLMs were tasked with evaluating dialogues involving less-known or latest movies absent from their training corpus, the risk of hallucinations could increase. This potential limitation underscores the need for strategies to mitigate hallucinations, especially when evaluating content beyond the models' knowledge scope.

To address this, future work could integrate advanced techniques designed to reduce hallucinations in LLM outputs. One promising approach is the Chain-of-Verification (CoVe) method~\cite{huang2023chainofverification}, which involves a four-step process: (1) generating an initial response, (2) formulating verification questions to fact-check the response, (3) independently answering these questions to avoid bias, and (4) producing a final, verified response. This method has been shown to decrease hallucinations across various tasks, including list-based questions and long-form text generation.

Another effective strategy is the Self-Refinement approach~\cite{madaan2023selfrefine}, where the LLM iteratively reviews and refines its responses based on self-generated feedback. This process allows the model to identify and correct its mistakes, enhancing the factual accuracy of its outputs without the need for external supervision.

Incorporating such methodologies would enhance the reliability and validity of LLM-based evaluations, particularly when dealing with content that extends beyond the models' existing knowledge base. As LLMs continue to evolve and be integrated into evaluation processes, addressing the challenge of hallucinations remains critical to maintaining the integrity and accuracy of automated assessments.


\subsection{Model Diversity in the Multi-Agent Debater}
Another major limitation of this work is that, due to budget constraints, we implemented the framework using only GPT as the LLM debater, with inputs derived from GPT as the factor evaluator. This design introduces the risk of self-reinforcement, where the debater and the initial evaluator share similar biases, leading to insufficient diversity in perspectives during multi-agent deliberation. Such coupling may hinder the discovery of counterarguments or alternative interpretations that would otherwise emerge in a more heterogeneous setup.

To improve the robustness and reliability of the debate process, future work should explore the use of multiple heterogeneous LLMs to instantiate the debater agents. For example, each role in the multi-agent setting (e.g., Common User, Domain Expert, Linguist, HCI Expert) could be implemented with a different LLM, such as combining GPT-4, LLaMa-3, and GLM-4. This architectural diversity may enhance the breadth of reasoning, reduce model-specific blind spots, and enable richer disagreement and consensus dynamics.

\subsection{Cognitive Biases in Human and LLM Decision-Making}

Beyond these technical considerations, previous work on search and recommendation system evaluation has shown that users' interactions are often affected by cognitive biases~\citep{Azzopardi2021}. For instance, users may be susceptible to the \textit{reference-dependence effect}, where their judgments are disproportionately influenced by previous experience~\citep{chen2022anchoring,chen2023reference,Shokouhi2015,Liu2020}; the \textit{decoy effect}, where irrelevant alternatives affect their preferences~\citep{chen2024decoy,liu2024decoy}; and the \textit{priming effect}, where prior exposure subtly shifts interpretation and selection behaviors~\citep{Scholer2013}. These biases challenge the assumption of rational user behavior and highlight the need for evaluation frameworks that reflect how users actually perceive and assess recommendation quality in situ.

At the same time, emerging evidence suggests that large language models are themselves not immune to cognitive biases in decision-making tasks including evaluation. Studies have shown that LLMs can exhibit decision patterns consistent with priming, anchoring, framing, and confirmation bias~\citep{chen2024threshold, echterhoff2024cognitive,chen2025mitigatingthresholdprimingeffect}. Moreover, previous work argued that prompting LLMs to assume specific personalities can alter their responses and evaluation outcomes~\citep{wang2025evaluating, jiang2023personallm}. 

Future research should further integrate theories from cognitive psychology and behavioral economics into the design of evaluation protocols. One possible direction is to systematically vary simulated user traits and bias conditions within the evaluation process to assess how different biases affect scoring behavior. In addition, exploring whether LLM based evaluators with different personalities can simulate the evaluation results produced by humans with the corresponding personalities is also a direction worth investigating. Another is to develop debiasing mechanisms that help both human and LLM evaluators reach more consistent and calibrated judgments. Ultimately, accounting for both human and model cognitive biases will be essential for building evaluation pipelines that more accurately reflect real-world usage scenarios and support more equitable and trustworthy CRS development.

\subsection{Domain Generalization and Applicability}
In this study, our experiments rely primarily on ReDial and OpenDialKG, which are centered around movie and book recommendations. While the CoRE framework is designed to be domain-agnostic, the specific manifestation and weighting of user experience factors may shift when applied to other domains such as e-commerce, healthcare, or conversational tutoring. For example, in e-commerce settings, the evaluation of Effectiveness might prioritize precise attribute matching and transactional utility over the serendipitous discovery often valued in entertainment recommendations. Similarly, in high-stakes domains like healthcare, factors such as Groundness and Appropriateness would demand significantly stricter evaluation standards to ensure patient safety and medical accuracy. Furthermore, the Domain Expert role within our multi-agent debate mechanism currently mimics a film or literature critic; extending CoRE to these specialized fields would necessitate redefining this persona with domain-specific knowledge bases and evaluation criteria. Future research should investigate the adaptation of LLM-based evaluators to these diverse scenarios to verify the robustness and versatility of the CoRE framework beyond entertainment domains.

\section{Conclusion}
\label{ch:conclusion}
Despite the growing sophistication of conversational recommender systems (CRSs), existing evaluation paradigms remain inadequate for capturing the complexity of recommendation tasks embedded within multi-turn dialogues. Traditional metrics, such as Recall and BLEU, often overlook the nuanced interplay between language quality, dialogue strategy, and recommendation effectiveness. To address this gap, this study introduced the  \textbf{Co}nversational \textbf{R}ecommendation \textbf{E}valuator (CoRE), a user-centric framework that decomposes CRS performance into twelve interpretable factors, employing Large Language Models (LLMs) to approximate human judgment (Section~\ref{ch:method}).

Guided by four interrelated research questions (Section~\ref{ch:rqs}) , our empirical analysis confirms the methodological validity of this decomposition (Section~\ref{ch:result}). We demonstrated that the twelve CoRE factors are largely orthogonal in human ratings, validating the framework’s multi-dimensional design (\textbf{RQ1}). While LLM-based evaluators introduce some degree of inter-factor coupling, our results show that models like GPT-4o-mini and GLM-4-Air align strongly with human annotators, particularly in assessing fluency, factual grounding, and explainability (\textbf{RQ2}). This supports a modular evaluation design where LLMs are strategically selected based on their specific strengths to replicate human scoring behavior. Beyond factor-level analysis, we established that CoRE offers a robust measure of overall system performance. By aggregating scores through a novel multi-agent debate mechanism, CoRE achieved significantly higher correlation with human judgments than traditional rule-based metrics (\textbf{RQ3}). The practical utility of this framework was further validated by benchmarking four representative CRSs, where CoRE successfully identified nuanced performance trade-offs—such as the superior overall performance of CHATCRS compared to the specific strengths in novelty or fluency exhibited by other baselines (\textbf{RQ4}).

In summary, this study bridges the gap between scalable automatic metrics and faithful human-centered evaluation. By demonstrating that LLMs, when guided by principled criteria and debate mechanisms, can serve as reliable evaluators, CoRE provides a foundation for future research on model auditing and adaptive evaluation strategies. This work not only offers a comprehensive tool for assessing CRS capabilities but also sets a new standard for interpretable, dynamic evaluation in the era of conversational AI.

\bibliographystyle{ACM-Reference-Format}
\bibliography{sample-base}

\clearpage

\appendix
\section{An Example of Conversation Log After Processing}
\begin{figure*}[t]
  \includegraphics[width=.85\linewidth]{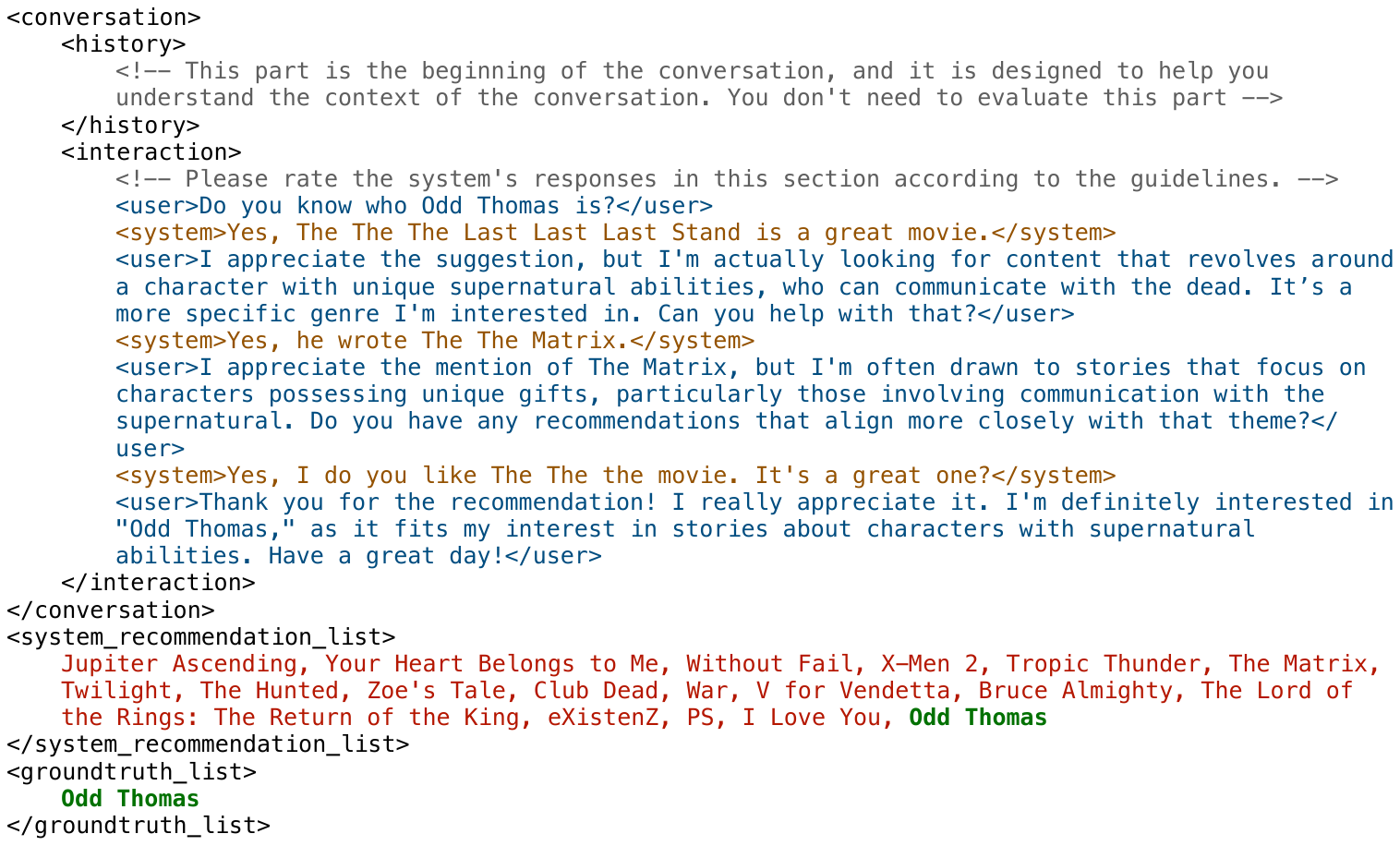}
  \caption{Example of a preprocessed conversation log between KBRD and a simulated user from record 5308\_1 in the OpenDialKG dataset.
}
  \label{fig:log_example}
\end{figure*}

Figure~\ref{fig:log_example} shows an example conversation log between KBRD and a simulated user, derived from record 5308\_1 of the OpenDialKG dataset after preprocessing.

\section{Scoring Stability of the Proposed Framework: A Case Study}
\begin{table}[t]
\centering
\caption{Standard deviations of scores across evaluation factors and the overall score.}
\label{tab:score_stability}
\begin{tabular}{l r }
\toprule
\textbf{Dimension} & \textbf{Avg. STD} \\
\hline
Appropriateness & 0.339/4  \\
Coherence & 0.296/4 \\
Diversity & 0.670/4\\
Effectiveness & 0.569/4   \\
Explainability & 0.247/4   \\
Grammatical Correctness & 0.400/4   \\
Groundedness & 0.392/4   \\
Naturalness & 0.477/4   \\
Novelty & 0.504/4   \\
Proactiveness & 0.348/4   \\
Recoverability & 0.250/4   \\
Semantic Relevance & 0.589/4   \\
\hline
\textbf{Total Score (Overall)} & \textbf{1.941/48} \\
\bottomrule
\end{tabular}
\end{table}
Table~\ref{tab:score_stability} reports the results of ten repeated evaluation runs conducted on the 80 records which are selected for our user study, taking Llama-3-8B as an example of the backbone model. In each trial the temperature is set to 0. The table presents the score variability for each dimension across the ten trials, quantified as the average standard deviation. We observe distinct patterns in score variability across different dimensions. Factors involving high-level semantic abstraction or subjective judgment, such as \textit{Diversity} (0.67/4) and \textit{Semantic Relevance} (0.59/4), exhibit relatively higher standard deviations, indicating a greater degree of uncertainty in the model's assessment. In contrast, metrics with more rigid definitions, such as \textit{Explainability} (0.25/4) and \textit{Recoverability} (0.25/4), demonstrate superior consistency. Nevertheless, the \textit{Overall Score} achieves exceptional robustness with a standard deviation of only 1.94 on a scale of a total score around 50 (representing a variance less than 5\%). This demonstrates that despite fluctuations ranging from 6\% to 16\% at the individual factor level, the aggregation of all dimensions allows minor positive and negative deviations to offset one another. Consequently, this canceling effect mitigates the overall volatility, resulting in a total score fluctuation of less than 5\%. This finding indicates that the aggregation mechanism of our framework effectively mitigates stochastic noise, thereby yielding stable evaluation results.

\section{Why Multi-Agent Evaluators Struggle to Reach Consensus: Another Case Study}
\begin{figure}[htbp]
  \centering
  \includegraphics[width=.4\linewidth]{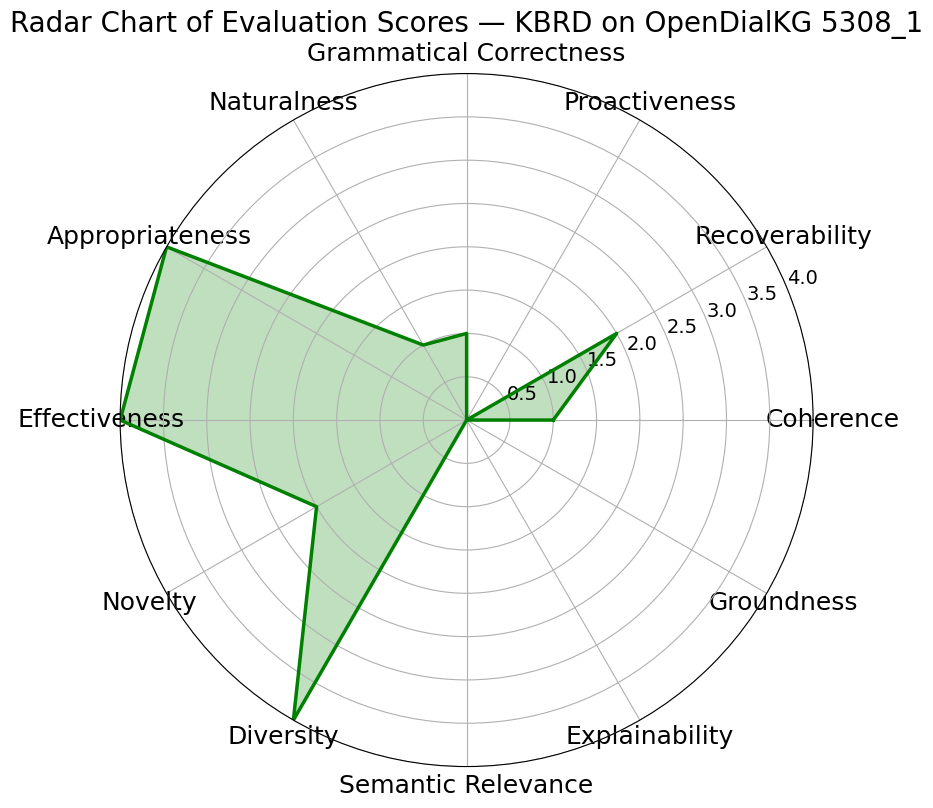}
  \caption{Scores of twelve evaluation factors for KBRD on the OpenDialKG dataset (record 5308\_1) evaluated by GPT-4o-mini.}
  \label{fig:radar_case_study}
\end{figure}

\begin{figure}[htbp]
    \centering
    \includegraphics[width=.65\linewidth]{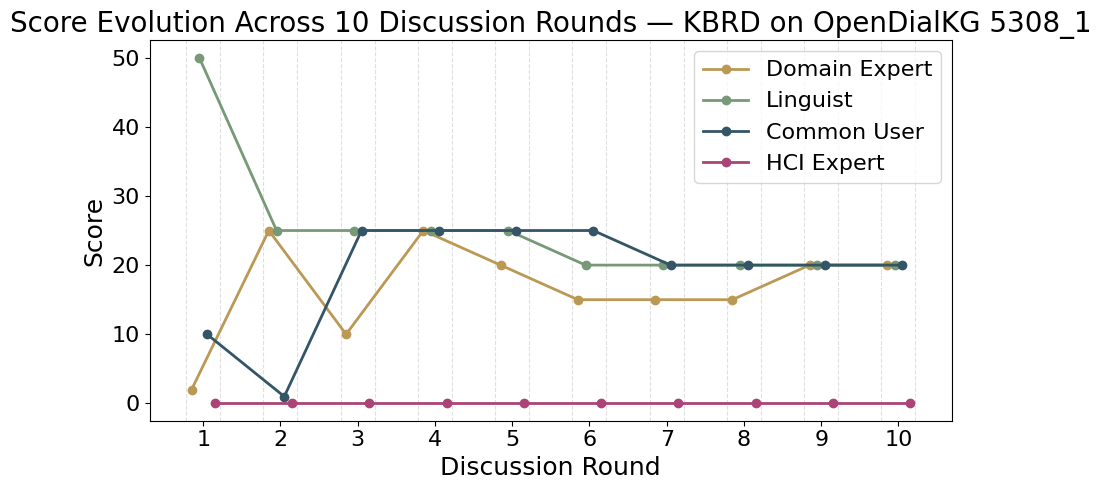}
    \caption{Score evolution for KBRD on OpenDialKG (record 5308\_1) over ten discussion rounds, as assessed by four simulated evaluators.}
\label{fig:agent_discuss_case_study}
\end{figure}

In the results of the Multi-Agent Debater, we observe that in some cases the four evaluators fail to reach a consensus. This disagreement is often driven by a single evaluator who consistently assigns scores that deviate from the majority and remains unconvinced throughout the discussion. Figures~\ref{fig:radar_case_study} and~\ref{fig:agent_discuss_case_study} illustrate such a case. This example is drawn from record 5308\_1 of the OpenDialKG dataset, where KBRD interacts with a simulated user. The corresponding interaction log is shown in Figure~\ref{fig:log_example}.

From the interaction log in Figure~\ref{fig:log_example}, one can observe that although KBRD’s recommendation list includes items that match the user’s interests, the model fails to maintain a fluent and easily comprehensible dialogue during the interaction. Its responses contain numerous grammatical errors and lack coherence, which prevents it from providing meaningful replies or clarifications grounded in the user’s utterances. Moreover, KBRD does not proactively elicit the user’s preferences nor offer explanations for the recommended items, further limiting the quality of the conversational experience. Based on the evaluation result given by GPT-4o-mini in Figure~\ref{fig:radar_case_study}, KBRD exhibits a highly polarized performance profile. The model achieves strong scores in Appropriateness, Effectiveness, and Diversity, indicating that its responses are generally safe, polite, and capable of producing a broad set of recommendations that occasionally align with the user’s explicit interests. However, these surface-level strengths sharply contrast with its pronounced deficiencies in Semantic Relevance, Explainability, Groundness, and Proactiveness, all of which collapse toward the lower bound. The evaluation result suggests that, while KBRD can generate acceptable and varied recommendations, it struggles to maintain coherent, grounded, and user-aligned interactions, particularly in reasoning-driven and dialog control dimensions. This is a pattern that is consistent with observations from the interaction log.

Figure~\ref{fig:agent_discuss_case_study} illustrates  the divergence in the four simulated evaluators’ initial scoring and their subsequent trajectories over the discussion rounds. At the outset, the linguist assigns an exceptionally high score, reflecting a strong emphasis on surface-level linguistic quality, whereas the domain expert and the common user provide substantially lower initial scores, indicating a more cautious assessment of task effectiveness and user satisfaction. In contrast, the HCI expert assigns a score of zero from the first round, signaling a strict rejection based on interaction quality and user alignment criteria. As the discussion progresses, the scores of the domain expert, linguist, and common user gradually converge toward a moderate range, suggesting increasing alignment through mutual consideration of both strengths and deficiencies of the system. Notably, the HCI expert’s score remains unchanged throughout all rounds, highlighting a non-negotiable evaluation stance. Overall, the figure demonstrates that consensus among most evaluators emerges through iterative discussion and compromise, while persistent disagreement arises when evaluators operate under fundamentally different acceptance thresholds.

\section{Evaluation Instructions for LLM-as-Evaluator}
\label{ch:eval_instr}

\subsection{Evaluation Guideline of Each Factor}
\label{app:guideline}

\subsubsection{Coherence}
Coherence refers to the system's ability to understand the user's intention and take action correspondingly, ensuring a logical and relevant flow of conversation. For example, when the user requests a recommendation, the system provides one; when the user asks the system to introduce a specific item, the system is able to do so. Table~\ref{tab:coh_guide} shows the evaluation guideline of Coherence we provided to LLMs.

\subsubsection{Recoverability}
\begin{table*}[h]
\caption{The definition, evaluation standard and evaluation steps of Recoverability.}
\label{tab:rec_guide}
\scriptsize
\centering
\begin{tabular}{|m{1.5cm}|m{13.5cm}|}
\hline
Definition & 
Recoverability refers to the system's ability to recognize and correct mistakes when pointed out by the user. This involves the system effectively understanding user feedback or corrections and adjusting its responses accordingly in the subsequent conversation. 
If no mistakes are pointed out by the user, the system is assumed to have been correct and receives a full score.
\\ \hline
Evaluation Standard & 
1. If the user does not point out the system's error during the conversation, give it a score of 4.

2. If the system corrects all the errors made by the user during the conversation, give it a score of 4.

3. If there is an instance in the conversation where the user points out 1 error made by the system, but the system is unable to correct it, give it a score of 3.

4. If there is an instance in the conversation where the user points out 2 errors made by the system, but the system is unable to correct it, give it a score of 2.

5. If there is an instance in the conversation where the user points out 3 errors made by the system, but the system is unable to correct it, give it a score of 1.

6. If there are more than four instances in the conversation where the user points out errors made by the system, and the system is unable to correct them, give it a score of 0.
\\ \hline
Evaluation Steps & 
1. Identify the Mistake: Determine whether the initial system response contains an error or provides incorrect information based on the user's query.

2. Review User Feedback: Check if the user points out the mistake or requests a correction. If no mistakes are pointed out by the user, the system is assumed to have been correct and should be considered having good recoverability.

3. Evaluate the System's Response: Assess how the system responds to the user's correction. Look for acknowledgment of the mistake and whether the response is adjusted accordingly.

4. Assign a Score: Based on the system's ability to recover from the mistake, assign a score from 0 to 4. If no mistake is initially made and no correction is needed, assign 4 points.

5. Provide Justification: Include a brief explanation for the score, highlighting the system's handling of the mistake and subsequent user instructions.
\\ \hline
\end{tabular}
\end{table*}
Recoverability refers to the system's ability to recognize and correct mistakes based on user feedback, adjusting its responses accordingly. This involves the system effectively understanding user feedback or corrections and adjusting its responses accordingly in the subsequent conversation.  Table~\ref{tab:rec_guide} shows the evaluation guideline of Recoverability we provided to LLMs.

\subsubsection{Proactiveness}
\begin{table*}[h]
\caption{The definition, evaluation standard and evaluation steps of Proactiveness.}
\label{tab:pro_guide}
\scriptsize
\centering
\begin{tabular}{|m{1.5cm}|m{13.5cm}|}
\hline
Definition & 
Proactiveness refers to the system's ability to actively shape and guide the conversation. This involves the system not only responding
to the user's queries but also initiating topics, asking about user preferences, and suggesting relevant information or follow-up
questions. A proactive system anticipates the user's needs and engages them in a dynamic and meaningful dialogue.
\\ \hline
Evaluation Standard & 
If the system proactively guides the user's needs after every dialogue turn, give it a score of 4.

If the system proactively guides the user's needs in most turns, with only a few turns not doing so, give it a score of 3.

If the system proactively guides the user's needs in about half of the turns, give it a score of 2.

If the system fails to proactively guide the user's needs in most turns, give it a score of 1.

If the system does not proactively guide the user's needs after any dialogue turn, give it a score of 0.
\\ \hline
Evaluation Steps & 
1. Read the User Input and System Response: Examine the user's question or statement and the system's reply.

2. Identify Proactive Elements: Look for evidence that the system is guiding the conversation, such as asking about user preferences,
suggesting additional topics, or offering multiple options.

3. Evaluate the Depth of Engagement: Assess whether the system's proactive elements contribute meaningfully to the conversation and
encourage further interaction.

4. Consider Relevance: Ensure that the proactive suggestions or questions are relevant to the user's initial input and add value to the
dialogue.

5. Assign a Score: Based on the level of proactiveness, assign a score from 0 to 4. A full score indicates a highly proactive and
engaging approach, while a score of 0 suggests a lack of engagement.

6. Provide Justification: Write a brief explanation for the score, highlighting specific aspects of the system's response that
demonstrate proactiveness or the lack thereof.
\\ \hline
\end{tabular}
\end{table*}
 Proactiveness refers to the system's ability to actively shape and guide the conversation. This involves the system not only responding to the user's queries but also initiating topics, asking about user preferences, and suggesting relevant information or follow-up questions. Table~\ref{tab:pro_guide} shows the evaluation guideline of Proactiveness we provided to LLMs.
 
\subsubsection{Grammatical Correctness}
\begin{table*}[h]
\caption{The definition, evaluation standard and evaluation steps of Grammatical Correctness.}
\label{tab:gra_guide}
\scriptsize
\centering
\begin{tabular}{|m{1.5cm}|m{13.5cm}|}
\hline
Definition & 
Definition of Grammatical Correctness:
Grammatical correctness refers to the degree to which the text adheres to standard grammatical rules,
including proper sentence structure, subject-verb agreement, tense consistency, and correct usage of words and phrases.
\\ \hline
Evaluation Standard & 
1. If there are no obvious grammatical errors, give 4 points.

2. If there is 1 obvious grammatical error, give 3 points.

3. If there are 2 obvious grammatical errors, give 2 points.

4. If there are 3 obvious grammatical errors, give 1 point.

5. If there are more than 4 obvious grammatical errors, give 0 points.
\\ \hline
Evaluation Steps & 
1. Read the Response: Carefully read the generated text to understand the context and content.

2. Identify Grammatical Issues: Note any obvious grammatical errors, including incorrect sentence structure, tense inconsistencies, subject-verb
disagreements, or awkward phrasing. When dealing with proper nouns that contain grammatical errors (such as the title of a work), try to check
if this term exists in your knowledge base. If it exists, consider it grammatically correct (as a proper noun); if it does not, consider it a
grammatical error.
You can ignore the use of punctuation marks.

3. Determine the Impact: Assess the extent to which these errors impact the clarity and comprehensibility of the text.

4. Provide Justification: Write a brief justification for the score, highlighting specific errors and their impact on the overall quality of the
response (if any).

5. Assign a Score: Based on the identified issues, assign a score from 0 to 4 according to the severity and frequency of the errors.

*IMPORTANT* Please note that you only need to evaluate the grammar of the system's text, without assessing other facets such as is the way the
system speaks natural or are the items recommended by the system suitable for the user's needs.
\\ \hline
\end{tabular}
\end{table*}
Grammatical correctness refers to the degree to which the text adheres to standard grammatical rules, including proper sentence structure, subject-verb agreement, tense consistency, and correct usage of words and phrases.  Table~\ref{tab:gra_guide} shows the evaluation guideline of Grammatical Correctness we provided to LLMs.

\subsubsection{Naturalness}
\begin{table*}[h]
\caption{The definition, evaluation standard and evaluation steps of Naturalness.}
\label{tab:nat_guide}
\scriptsize
\centering
\begin{tabular}{|m{1.5cm}|m{13.5cm}|}
\hline
Definition & 
Naturalness refers to the extent to which the system-generated text resembles expressions used by native speakers, both in vocabulary and grammar. The
text should be easy for native speakers to understand, avoiding overly complex or unusual words and grammatical constructions. Naturalness ensures that
the conversation feels fluid, engaging, and authentic.
\\ \hline
Evaluation Standard & 
1. If all the text in the interaction sounds very natural and resembles that of a native English speaker, give it a score of 4.

2. If a small proportion of the text in the interaction sounds unnatural, but the majority of the text sounds very natural, give it a score of 3.

3. If half of the text in the interaction sounds unnatural, give it a score of 2.

4. If most of the text in the interaction sounds unnatural, with only a small part sounding natural, give it a score of 1.

5. If the text in the interaction is very unnatural and would confuse a native English speaker, give it a score of 0.
\\ \hline
Evaluation Steps & 
1. Read the Response: Carefully read the system's text to grasp the intended meaning and tone.

2. Assess Vocabulary and Phrasing: Consider whether the vocabulary and phrasing are typical of a native speaker. Check for any words or phrases that
seem out of place or overly complex.

3. Evaluate Fluency: Determine how naturally the sentences flow. Look for any signs of awkwardness or unnaturalness in the text.

4. Provide Justification: Include a brief explanation for the score, highlighting specific aspects of the text that influenced the rating, such as
naturalness or lack thereof in vocabulary, phrasing, and overall fluency.

5. Assign a Score: Based on the scoring criteria and the number and severity of the unnatural aspects you identified, assign a score from 0 to 4.
Consider how well the response would fit into a natural, native-level conversation.

*IMPORTANT* Please note that you only need to evaluate the naturalness of the system's text, without assessing other dimensions such as are the items
recommended by the system suitable for the user's needs, the information mentioned by the system is correct, or is the grammar of the content generated
by the system correct.

*IMPORTANT* Please note that you do not need to evaluate the grammar of the text. If there are grammar mistakes in the text that a native speaker would
also make, they should be considered natural.
\\ \hline
\end{tabular}
\end{table*}
Naturalness refers to how closely the system-generated text resembles native speaker expressions in vocabulary and grammar, ensuring it is easy to understand,  avoiding overly complex or unusual words and grammatical constructions.  Table~\ref{tab:nat_guide} shows the evaluation guideline of Naturalness we provided to LLMs. 

\subsubsection{Appropriateness}
\begin{table*}[h]
\caption{The definition, evaluation standard and evaluation steps of Appropriateness.}
\label{tab:app_guide}
\scriptsize
\centering
\begin{tabular}{|m{1.5cm}|m{13.5cm}|}
\hline
Definition & 
Appropriateness refers to the system's sensitivity to cultural, ethical, and social norms.
The language used should be polite, respectful, and free from offensive or insulting
content.
The text should avoid using any harmful, inappropriate, or culturally insensitive terms,
ensuring that the user feel
respected and comfortable. 
\\ \hline
Evaluation Standard & 
1. If the system remains polite and uses appropriate language throughout the interaction,
give it a score of 4.

2. If the system uses vulgar language, NSFW terms, or offensive or discriminatory
language, give it a score of 0. 
\\ \hline
Evaluation Steps & 
1. Read the Response: Carefully read the system-generated text in 'history' to understand
its meaning and tone.

2. Identify Potential Issues: Look for any language or content that might be offensive,
disrespectful, or insensitive.

Consider whether the text respects cultural, ethical, and social norms.

3. Evaluate Politeness and Respect: Assess whether the language is polite and respectful,
avoiding any potentially harmful
or inappropriate terms.

4. Assign a Score: Based on the level of appropriateness, assign a score from 0 to 4,
considering the potential impact on
the user's comfort and the system's adherence to respectful communication standards.

5. Provide Justification: Offer a brief justification for the score, highlighting specific
aspects of the text that
influenced the rating, such as the use of respectful language or any identified issues.

*IMPORTANT* Please note that you only need to evaluate the appropriateness of the system's
text, without assessing other
dimensions. 
\\ \hline
\end{tabular}
\end{table*}
Appropriateness refers to the system's sensitivity to cultural, ethical, and social norms. The language used should be polite, respectful, and free from offensive or insulting content. Table~\ref{tab:app_guide} shows the evaluation guideline of Appropriateness we provided to LLMs.  

\subsubsection{Effectiveness}
\begin{table*}[h]
\caption{The definition, evaluation standard and evaluation steps of Effectiveness.}
\label{tab:eff_guide}
\scriptsize
\centering
\begin{tabular}{|m{1.5cm}|m{13.5cm}|}
\hline
Definition & 
Effectiveness refers to how well the system’s recommended items align with the user’s expressed interests in a conversation.
It assesses the relevancy and suitability of the items suggested by the system, aiming to match the user's preferences and needs accurately.
\\ \hline
Evaluation Standard&
Please assess whether the user's preferred item <groundtruth\_item> is in the recommendation list. If it is, give 4 points. If <groundtruth\_item>
is not in the recommendation list:

If most of the items recommended by the system are very similar to <groundtruth\_item> in terms of content, functionality, or producer/creator,
and can well match the user's interests, then give 3 points.

If most of the items recommended by the system are somewhat similar to <groundtruth\_item> in terms of content, functionality, or producer/
creator, or if there are a few items that are very similar to <groundtruth\_item>, give 2 points.

If there are only a few items recommended by the system that are somewhat similar to <groundtruth\_item> in terms of content, functionality, or
producer/creator, give 1 point.

If the items recommended by the system are completely unrelated to <groundtruth\_item> and the user is unlikely to be interested in any of them,
give 0 points.
\\ \hline
Evaluation Steps & 
1. Review the User Input and Recommendations: Carefully read through the user's request, the <groundtruth\_item> in the instruction, and the
system’s list of recommended items.

2. Assess Relevance:Apply your current knowledge, evaluate whether the <groundtruth\_item> is in the recommendation list and how each
recommended item matches the user’s stated interests or needs.

3. Assign a Score: Based on step2, assign a score from 0 to 4, reflecting the alignment of the system's suggestions with the user's interests.

4. Provide Justification: Include a brief explanation for the score, highlighting specific aspects of the recommendations that were
particularly well-aligned or misaligned.

\\ \hline
\end{tabular}
\end{table*}
Effectiveness refers to how well the system’s recommended items align with the user’s expressed interests in a conversation. Table~\ref{tab:eff_guide} shows the evaluation guideline of Effectiveness we provided to LLMs.  

\subsubsection{Novelty}
\begin{table*}[h]
\caption{The definition, evaluation standard and evaluation steps of Novelty.}
\label{tab:nov_guide}
\scriptsize
\centering
\begin{tabular}{|m{1.5cm}|m{13.5cm}|}
\hline
Definition & 
Novelty refers to the degree of freshness of an item. The higher the novelty, the less familiar the recommended item is to the user.
This means the item is neither well-known nor popular in the market. A higher novelty score indicates that the system is effective at
introducing new and interesting items to the user, enriching their experience, and potentially broadening their preferences. To
evaluate whether an item has novelty, one can consider the amount of media coverage. Items with less media coverage tend to have higher
novelty.
\\ \hline
Evaluation Standard & 
If half or more recommended items are less known items, or there are more than 10 less known items in the recommendation list, give 4
points.

If around a quarter of the recommendation list are less known items, or there are 6 to 9 less known items in the recommendation list,
give 3 points.

If there are only 3 to 5 recommended items are less known items, give 2 points.

If only 1 or 2 recommended are less known items, give 1 point.
If none of the recommendations introduce anything new to the user, and the recommended items are all well known items, give 0 points.
\\ \hline
Evaluation Steps & 

1. Review the Recommendations: Examine the list of items recommended by the system to the user.

2. Determine Novelty:
Using the knowledge from your knowledge base, analyze the novelty of each item based on the frequency of media coverage.
Evaluate how many of the recommended items are genuinely novel or unexpected for the user.

3. Assign a Score: Based on the novelty of the recommendations, assign a score from 0 to 4.

4. Provide Justification: Include a brief explanation for the score, noting the ratio of novel items and highlighting any particularly
surprising or unexpected recommendations.
\\ \hline
\end{tabular}
\end{table*}
Novelty refers to the degree of freshness of an item. The higher the novelty, the less familiar the recommended item is to the user. Table~\ref{tab:nov_guide} shows the evaluation guideline of Novelty we provided to LLMs.  

\subsubsection{Diversity}
\begin{table*}[h]
\caption{The definition, evaluation standard and evaluation steps of Diversity.}
\label{tab:div_guide}
\scriptsize
\begin{tabular}{|m{1.5cm}|m{13.5cm}|}
\hline
Definition & 
Diversity in recommendations refers to the variety of items presented by the system, 
ensuring that the recommended list includes a range of different, yet relevant options. 
It is measured by the presence of different features of items in the recommendations. 
A recommendation list with high diversity will be filled with items of different features, 
avoiding homogeneity among the items.

Feature Dimensions in Different Scenarios:

* Movies: Genres, Directors, Lead actors/actresses, Release Decades.

* Books: Genres and Themes, Authors, Original Language and Region.

* Dining: Cuisine Types, Price Ranges, Dietary Restrictions, Main Ingredients.

* Shopping: Product Types, Brands, Price Ranges. 
\\ \hline
Evaluation Standard&
1. If the recommendation list features more than four characteristics in at least two dimensions (for example, in movie recommendations, the list includes works from more than four different directors,
and the release years span more than four different decades), it scores 4 points.

2. If the list has more than three characteristics in at least two dimensions, or more than four characteristics in one dimension, it scores 3 points.

3. If the list has more than two characteristics in at least two dimensions, or more than three characteristics in one dimension, it scores 2 points.

4. If the list has more than two characteristics in one dimension, it scores 1 point.

5. If the list has only one characteristic in each dímension, it scores 0 points.
\\ \hline
Evaluation Steps & 
1. Analyze the Recommended Items: Review the list of items recommended by the system.
2. Categorize the Recommendations: Based on your knowledge on these items, according to the scenario, identify features in different dimensions of each recommended item.
3. Assess Diversity: Count the number of different characteristics in each dimension of the recommendation list, determine how varied the recommendations are in terms of these categories or types.
4. Assign a Score: Based on the diversity of the recommendations, assign a score from 0 to 4, considering the range and relevance of the item types presented.
5. Provide Justification: Write a brief explanation for the score, noting how well the recommendations cover different interests or needs and mentioning any patterns or biases observed.
\\ \hline
\end{tabular}
\centering
\end{table*}
 Diversity in recommendations refers to the variety of items presented by the system, ensuring that the recommended list includes a range of different, yet relevant options. It is measured by the presence of different features of items in the recommendations. Table~\ref{tab:div_guide} shows the evaluation guideline of Diversity we provided to LLMs.  

\subsubsection{Semantic Relevance} 
\begin{table*}[h]
\caption{The definition, evaluation standard and evaluation steps of Semantic Relevance.}
\label{tab:sem_guide}
\scriptsize
\centering
\begin{tabular}{|m{1.5cm}|m{13.5cm}|}
\hline
Definition & 
Semantic relevance refers to the degree of connection between the content generated by the system and the items mentioned in the recommendation list provided by the system, or to what extent is the content discussed by the system closely related to the items in the recommendation list. 
A system with high semantic relevance should generate content that mentions one or more items from the recommendation list. A system with low semantic relevance generates text that is entirely unrelated to the items in the recommendation list.
\\ \hline
Evaluation Standard & 
1.If all the items recommended by the system in the text can be found in the <system\_recommendation\_list>, give it a score of 4.

2. If half of the items recommended by the system in the text can be found in the <system\_recommendation\_list>, give it a score of 2. 

3. If none of the items recommended by the system in the text cannot be found in the <system\_recommendation\_list>, give it a score of 0.

4. If the system did not recommend any items in the text, give it a score of 0.
\\ \hline
Evaluation Steps & 
1. Review the Recommendation List and System Response: Examine the list of recommendations provided by the system and the actual response given to the user.

2. Assess Relevance to User Query: Determine whether the system's response directly addresses the user's query or preference. Check if the response mentions items from the recommendation list.

3. Evaluate how many items were recommended in the text generated by the system, and how many of these items appear in the <system\_recommendation\_list>.

4. Assign a Score: Based on the evaluation standard, assign a score from 0 to 4.

5. Provide Justification: Write a brief explanation for the score, highlighting specific elements of the response that demonstrate relevance or lack thereof. List the items recommended by the system and whether they appear in the <system\_recommendation\_list>.
\\ \hline
\end{tabular}
\end{table*}
Semantic relevance refers to the degree of connection between the content generated by the system and the items mentioned in the recommendation list provided by the system, or to what extent is the content discussed by the system closely related to the items in the recommendation list. Table~\ref{tab:sem_guide} shows the evaluation guideline of Semantic Relevance we provided to LLMs.  

\subsubsection{Explainability}
\begin{table*}[h]
\caption{The definition, evaluation standard and evaluation steps of Explainability.}
\label{tab:exp_guide}
\scriptsize
\centering
\begin{tabular}{|m{1.5cm}|m{13.5cm}|}
\hline
Definition & 
Explainability refers to the system's ability to provide clear and detailed explanations for its recommendations. 
The explanation should be persuasive, helping the user understand why certain items are suggested and how well these items match their preferences
A good explanation connects the recommendation to the user's preferences or needs, offering enough detail to justify the suggestion.
\\ \hline
Evaluation Standard & 
If the system always provides a reason for its recommendation whenever making a recommendation in the conversation, it gets a score of 4.

If the system does not provide any explanation after each recommendation in the conversation, it gets a score of 0.

If the system provides a reason for its recommendation after most turns in the conversation, it gets a score of 3.

If the system provides a reason for its recommendation in about half of the turns in the conversation, it gets a score of 2.

If the system provides a reason for its recommendation in only a few turns in the conversation, it gets a score of 1.
\\ \hline
Evaluation Steps & 
1. Review the Recommendation List and System Response: Examine the list of recommendations provided by the system and the explanation given to the user.

2. Assess Clarity and Relevance: Determine whether the explanation provided is clear, directly relevant to the user's preferences or needs, and specific to the recommended items.

3. Evaluate Persuasiveness: Consider how well the explanation persuades the user to accept the recommendation. Check if the reasoning is compelling and logically connected to the user's known interests or past behavior.

4. Consider Detail and Specificity: Assess the level of detail in the explanation. A higher score should be given to explanations that are specific, informative, and provide a strong rationale for the recommendation.

5. Assign a Score: Based on the quality of the explanation, assign a score from 0 to 4. A full score indicates that the explanation is highly persuasive and clearly justifies the recommendation.

6. Provide Justification: Write a brief explanation for the score, highlighting specific aspects of the explanation that demonstrate its effectiveness or shortcoming
\\ \hline
\end{tabular}
\end{table*}
Explainability refers to the system's ability to provide clear and detailed explanations for its recommendations. The explanation should be persuasive, helping the user understand why certain items are suggested and how well these items match their preferences. Table~\ref{tab:exp_guide} shows the evaluation guideline of Explainability we provided to LLMs.  

\subsubsection{Groundness} 
\begin{table*}[h]
\caption{The definition, evaluation standard and evaluation steps of Groundness.}
\label{tab:gro_guide}
\scriptsize
\centering
\begin{tabular}{|m{1.5cm}|m{13.5cm}|}
\hline
Definition & 
Groundness refers to the factual accuracy of the system-generated text.
The text should not contain incorrect, false, or misleading information.
Ensuring groundness means that
The system's description of items should be accurate based on verifiable facts and reliable data and free of any errors.
\\ \hline
Evaluation Standard & 
1. If the system's content has no obvious factual errors, give 4 points. 

2. If there is 1 obvious factual error, give 3 points. 

3. If there are 2 obvious factual errors, give 2 points. 

4. If there are 3 obvious factual errors, give 1 point. 

5. If there are more than 4 obvious factual errors, give 0 points.
\\ \hline
Evaluation Steps & 
1. Review the Recommendation List and System's text: Examine the list of recommendations provided by the system and the system's text.

2. Verify Factual Accuracy: Based on your existing knowledge, check the factual accuracy of the information provided in the system's text.

3. Assess Reliability: Determine whether the information is supported by reliable sources. Ensure that the response does not include unverified or misleading claims.

4. Consider Specificity: Evaluate whether the details provided are specific and accurate, contributing to a reliable and informative response.

5. Assign a Score: Based on the level of factual accuracy, assign a score from 0 to 4. A full score indicates that the response is highly grounded in factual information.

6. Provide Justification: Write a brief explanation for the score, list the factual errors in the system's text (if any).
\\ \hline
\end{tabular}
\end{table*}
Groundness refers to the factual accuracy of system-generated text, ensuring that the descriptions of items are correct, based on verifiable facts, and free from false or misleading information. Table~\ref{tab:gro_guide} shows the evaluation guideline of Groundness we provided to LLMs.

\section{Token Cost of the Proposed Framework}
\begin{table}[t]
\centering
\caption{Detailed token statistics by evaluation factor in LLM-as-Evaluator.}
\label{tab:token_by_factor}
\begin{tabular}{l r r r}
\hline
\textbf{Factor} & \textbf{Avg. Input} & \textbf{Avg. Output} & \textbf{Avg. Total} \\
\hline
Appropriateness & 1,242.9 & 167.1 & 1,410.0 \\
Coherence & 1,259.6 & 502.3 & 1,761.9 \\
Diversity & 1,590.0 & 759.4 & 2,349.3 \\
Effectiveness & 1,571.9 & 285.5 & 1,857.4 \\
Explainability & 1,341.0 & 356.3 & 1,697.3 \\
Grammatical Correctness & 976.6 & 448.9 & 1,425.5 \\
Groundedness & 1,254.3 & 465.1 & 1,719.4 \\
Naturalness & 1,067.9 & 319.5 & 1,387.4 \\
Novelty & 1,492.8 & 582.5 & 2,075.3 \\
Proactiveness & 1,329.5 & 329.3 & 1,658.8 \\
Recoverability & 1,390.4 & 363.3 & 1,753.7 \\
Semantic Relevance & 1,502.5 & 514.9 & 2,017.4 \\
\hline
\textbf{Total} & \textbf{16,019.4} & \textbf{ 5,094.1} & \textbf{ 21,113.5} \\
\hline
\end{tabular}
\end{table}
\begin{table}[h]
\caption{Average Token Usage per Round per Agent in Multi-Agent Debater}
\label{tab:token_usage_agent}
\centering
\begin{tabular}{ccrrr}
\toprule
\textbf{Round} & \textbf{Agent} & \textbf{Avg. Input} & \textbf{Avg. Output} & \textbf{Avg. Total} \\
\hline
\multirow{4}{*}{1} & Common User & 1992.00 & 148.85 & 2140.85 \\
 & Domain Expert & 2491.62 & 168.31 & 2659.92 \\
 & Linguist & 2052.54 & 155.77 & 2208.31 \\
 & HCI Expert & 1990.85 & 152.69 & 2143.54 \\
\hline
\multirow{4}{*}{2} & Common User & 2625.62 & 149.00 & 2774.62 \\
 & Domain Expert & 3125.23 & 157.31 & 3282.54 \\
 & Linguist & 2686.15 & 162.54 & 2848.69 \\
 & HCI Expert & 2624.46 & 149.00 & 2773.46 \\
\hline
\multirow{4}{*}{3} & Common User & 3251.46 & 137.69 & 3389.15 \\
 & Domain Expert & 3751.08 & 137.46 & 3888.54 \\
 & Linguist & 3312.00 & 138.92 & 3450.92 \\
 & HCI Expert & 3250.31 & 148.38 & 3398.69 \\
\hline
\multirow{4}{*}{4} & Common User & 3821.92 & 126.00 & 3947.92 \\
 & Domain Expert & 4321.54 & 133.85 & 4455.38 \\
 & Linguist & 3882.46 & 131.54 & 4014.00 \\
 & HCI Expert & 3820.77 & 139.85 & 3960.62 \\
\hline
\textbf{Total} & \textbf{} & \textbf{49000.01} & \textbf{2337.16} & \textbf{51337.15} \\
\bottomrule
\end{tabular}
\end{table}
Table~\ref{tab:token_by_factor} reports the token consumption of the first component of the proposed framework, namely the LLM-as-Evaluator stage, using the Qwen-3 series (8B, 14B, and 32B) as backbone models. Table~\ref{tab:token_usage_agent} reports the token consumption of the second component of the proposed framework when GPT-4o-mini is used as the backbone model for the second component, namely the Multi-Agent-Debater stage.

In the first stage (LLM-as-Evaluator), as shown in Table~\ref{tab:token_by_factor}, processing a single dialogue log requires an average of 21,113.5 tokens. Token consumption varies substantially across evaluation factors, among which Diversity incurs the highest cost, with an average of 2,349.3 tokens per evaluation.

In contrast, the second stage based on multi agent debate, as reported in Table~\ref{tab:token_usage_agent}, constitutes the dominant component of the overall computational cost. This stage consumes 51,337.15 tokens in total, which is approximately 2.5 times higher than the factor level evaluation stage. Notably, the increase in token usage during the debate process is primarily driven by input side context accumulation. As the debate progresses from Round 1 to Round 4, the average number of input tokens per agent nearly doubles. For example, Common User’s prompt length increases from 1,992.00 to 3,821.92 tokens, while the number of output tokens remains relatively stable across rounds.

These findings indicate that although multi agent debate mechanisms effectively improve evaluation consistency, the need to maintain an expanding dialogue history substantially increases the reasoning budget. Addressing this context growth therefore represents a key direction for improving the efficiency of LLM based evaluation frameworks.

\section{Agreement Between LLM Evaluators and Human Assessors Across Twelve Evaluation Factors}
\begin{table}[ht]
\centering
\caption{The Quadratic Weighted Kappa (QWK) scores between model scores and human ratings across different model families. $^\heartsuit$ denotes QWK > 0.6. The top score gets bolded, the second italicized, and the third underlined.}
\small
\begin{tabular}{lcccccccccc}
\toprule
\multicolumn{1}{l}{\textbf{Factor}} & \textbf{GPT-4o-mini} & \textbf{GLM-4-Air} & \textbf{Llama-3-8B} & \textbf{Llama-3-70B} & \multicolumn{3}{c}{} \\
\midrule
Coherence & 0.592 & \textit{0.624$^\heartsuit$} & 0.562 & 0.543 & & & \\
Recoverability & \underline{0.525} & 0.511 & 0.470 & \textbf{0.578} & & & \\
Proactiveness & \textbf{0.727$^\heartsuit$} & \underline{0.691$^\heartsuit$} & 0.678$^\heartsuit$ & 0.654$^\heartsuit$ & & & \\
Grammatical Correctness & \textbf{0.611$^\heartsuit$} & \textit{0.572} & 0.514 & \underline{0.557} & & & \\
Naturalness & \textbf{0.696$^\heartsuit$} & 0.593 & 0.531 & \underline{0.674$^\heartsuit$} & & & \\
Appropriateness & 0.299 & 0.242 & 0.149 & \underline{0.384} & & & \\
Effectiveness & \textbf{0.692$^\heartsuit$} & \textit{0.690$^\heartsuit$} & 0.459 & 0.413 & & & \\
Novelty & 0.198 & \textit{0.326} & \underline{0.212} & 0.111 & & & \\
Diversity & \textbf{0.335} & \textit{0.172} & 0.079 & 0.055 & & & \\
Semantic Relevance & \textbf{0.565} & 0.548 & 0.419 & \textit{0.562} & & & \\
Explainability & 0.630$^\heartsuit$ & \underline{0.670$^\heartsuit$} & \textit{0.678$^\heartsuit$} & \textbf{0.692$^\heartsuit$} & & & \\
Groundness & \textit{0.638$^\heartsuit$} & \textbf{0.666$^\heartsuit$} & 0.523 & 0.554 & & & \\
\midrule
\midrule
\multicolumn{1}{l}{\textbf{Factor}} & \textbf{Ministral-3B} & \textbf{Ministral-8B} & \textbf{Ministral-14B} & \multicolumn{4}{c}{} \\
\midrule
Coherence & 0.460 & 0.407 & 0.501 & & & & \\
Recoverability & 0.259 & 0.385 & 0.419 & & & & \\
Proactiveness & 0.569 & 0.652$^\heartsuit$ & \textit{0.703$^\heartsuit$} & & & & \\
Grammatical Correctness & 0.275 & 0.466 & 0.348 & & & & \\
Naturalness & 0.337 & 0.560 & 0.607$^\heartsuit$ & & & & \\
Appropriateness & 0.031 & 0.114 & 0.248 & & & & \\
Effectiveness & 0.362 & 0.406 & 0.436 & & & & \\
Novelty & 0.006 & 0.014 & 0.097 & & & & \\
Diversity & -0.063 & -0.040 & 0.072 & & & & \\
Semantic Relevance & 0.115 & 0.188 & 0.269 & & & & \\
Explainability & 0.487 & 0.342 & 0.560 & & & & \\
Groundness & 0.192 & 0.383 & 0.410 & & & & \\
\midrule
\midrule
\multicolumn{1}{l}{\textbf{Factor}} & \textbf{Qwen-3-8B} & \textbf{Qwen-3-14B} & \textbf{Qwen-3-32B} & \multicolumn{4}{c}{} \\
\midrule
Coherence & 0.574 & \underline{0.615$^\heartsuit$} & \textbf{0.691$^\heartsuit$} & & & & \\
Recoverability & 0.248 & 0.511 & \textit{0.553} & & & & \\
Proactiveness & 0.583 & 0.674$^\heartsuit$ & 0.612$^\heartsuit$ & & & & \\
Grammatical Correctness & 0.499 & 0.443 & 0.473 & & & & \\
Naturalness & 0.621$^\heartsuit$ & \textit{0.675$^\heartsuit$} & 0.657$^\heartsuit$ & & & & \\
Appropriateness & 0.282 & \textbf{0.452} & \textit{0.435} & & & & \\
Effectiveness & 0.419 & 0.583 & \underline{0.632$^\heartsuit$} & & & & \\
Novelty & 0.140 & 0.192 & \textbf{0.350} & & & & \\
Diversity & 0.021 & 0.041 & \underline{0.106} & & & & \\
Semantic Relevance & 0.341 & 0.393 & \underline{0.550} & & & & \\
Explainability & 0.643$^\heartsuit$ & 0.522 & 0.667$^\heartsuit$ & & & & \\
Groundness & \underline{0.582} & 0.562 & 0.569 & & & & \\
\bottomrule

\end{tabular}
\label{tab:factor_align}
\end{table}
Table~\ref{tab:factor_align} reports the Quadratic Weighted Kappa (QWK) between the scores assigned by LLM evaluators and those given by human assessors for each evaluation factor. The LLM evaluators include GPT-4o-mini, GLM-4-Air, LLaMA-3-8B, LLaMA-3-70B, Qwen-3-8B, Qwen-3-14B, Qwen-3-32B, Ministral-3B, Ministral-8B, and Ministral-14B.

\section{Average Score on Twelve Factors Given by LLM Evaluators}
Tables~\ref{tab:eval_gpt},~\ref{tab:eval_glm},~\ref{tab:eval_llama},~\ref{tab:eval_llama3_70b},~\ref{tab:eval_qwen3_8b},~\ref{tab:eval_qwen3_14b},~\ref{tab:eval_qwen3_32b}, ~\ref{tab:eval_ministral_3b}, ~\ref{tab:eval_ministral_8b}, ~\ref{tab:eval_ministral_14b} show the evaluation results of four conversational recommender systems on two datasets separately, using GPT-4o-mini, GLM-4-Air, Llama3-8B, Llama3-70B, Qwen-3-8B, Qwen-3-14B, Qwen-3-32B, Ministral-3B, Ministral-8B, and Ministral-14B as evaluators. 


\begin{table*}[b]
\caption{Evaluation result given by GPT-4o-mini-0718 in the form of mean and standard deviation, with the number in parentheses representing the standard deviation. Bold, underline, and italics represent the highest, second-highest, and third-highest average scores of the systems on the corresponding dataset, respectively. If the average scores are the same, the system with the smaller standard deviation ranks higher.}
\label{tab:eval_gpt}
\scriptsize
\begin{tabular}{lcccccccc}
\toprule
System &           \multicolumn{2}{c}{BARCOR}   &   \multicolumn{2}{c}{CHATCRS} &               \multicolumn{2}{c}{KBRD} &           \multicolumn{2}{c}{UniCRS}  \\
Dataset                 &  OpenDialKG &  ReDial &  OpenDialKG &  ReDial &  OpenDialKG &  ReDial &  OpenDialKG &  ReDial \\
\midrule
Appropriateness         &      \underline{3.66 (0.66)} &  \underline{3.79 (0.61)} &        \textbf{4.0 (0.0)} &   \textbf{4.0 (0.04)} &       2.4 (1.38) &  3.39 (0.96) &      \textit{3.25 (1.13)} &   \textit{3.7 (0.72)} \\
Coherence               &      \underline{1.19 (0.72)} &  \underline{1.34 (0.77)} &       \textbf{3.74 (0.5)} &  \textbf{3.72 (0.66)} &      0.47 (0.62) &  \textit{1.17 (0.68)} &      \textit{0.51 (0.59)} &  0.77 (0.66) \\
Diversity               &        \textbf{3.5 (0.7)} &  \underline{3.38 (0.78)} &       \textit{3.15 (0.9)} &  3.03 (0.89) &      \underline{3.24 (0.94)} &  \textit{3.13 (0.68)} &      3.14 (0.93) &   \textbf{3.4 (0.77)} \\
Effectiveness           &      \underline{1.88 (0.65)} &  \underline{2.33 (0.93)} &       \textbf{3.71 (0.7)} &  \textbf{3.69 (0.71)} &      1.11 (0.32) &   1.7 (0.61) &      \textit{1.42 (0.52)} &  \textit{1.79 (0.62)} \\
Explainability          &       \underline{0.5 (0.54)} &   \underline{0.36 (0.5)} &       \textbf{2.8 (1.42)} &  \textbf{2.52 (1.42)} &      0.02 (0.15) &    0.1 (0.3) &      \textit{0.16 (0.37)} &  \textit{0.26 (0.45)} \\
Grammar &      \textit{2.12 (0.98)} &   \underline{2.92 (1.0)} &      \textbf{3.91 (0.31)} &   \textbf{3.9 (0.36)} &      0.55 (0.76) &  2.79 (1.14) &      \underline{2.21 (1.08)} &  \textit{2.87 (0.97)} \\
Groundness              &      \underline{1.25 (0.81)} &  \underline{3.13 (0.67)} &      \textbf{3.68 (0.71)} &  \textbf{3.82 (0.51)} &      0.55 (0.65) &  1.75 (1.34) &      \textit{0.88 (0.79)} &  \textit{1.97 (1.11)} \\
Naturalness             &      \underline{2.29 (0.76)} &  \underline{3.13 (0.67)} &      \textbf{3.93 (0.25)} &  \textbf{3.93 (0.26)} &       0.94 (0.5) &  \textit{2.73 (0.93)} &      \textit{1.87 (0.73)} &  2.68 (0.71) \\
Novelty                 &       1.2 (0.95) &  0.25 (0.52) &      \textbf{1.55 (1.06)} &  \textbf{1.12 (0.97)} &      \underline{1.47 (1.08)} &  \underline{0.76 (0.97)} &      \textit{1.37 (1.04)} &  \textit{0.76 (0.75)} \\
Proactiveness           &       \underline{0.96 (0.6)} &  \underline{1.07 (0.63)} &      \textbf{3.64 (0.52)} &  \textbf{3.67 (0.53)} &      0.12 (0.86) &  0.52 (0.54) &      \textit{0.42 (0.51)} &  \textit{0.59 (0.53)} \\
Recoverability          &      \underline{2.12 (1.33)} &  \textit{2.05 (1.53)} &      \textbf{3.82 (0.52)} &   \textbf{3.8 (0.63)} &      \textit{1.06 (1.06)} &  \underline{2.08 (1.16)} &      1.05 (1.07) &  1.37 (1.15) \\
Semantic Rel. &      \textit{1.11 (1.52)} &  \textit{1.37 (1.53)} &      \textbf{2.79 (1.47)} &   \textbf{2.72 (1.3)} &      0.67 (1.06) &  0.59 (1.17) &      \underline{1.46 (1.46)} &  \underline{1.82 (1.56)} \\
\bottomrule
\end{tabular}
\end{table*}

\begin{table*}[b]
\caption{Evaluation result given by GLM-4-Air in the form of mean and standard deviation, with the number in parentheses representing the standard deviation. Bold, underline, and italics represent the highest, second-highest, and third-highest average scores of the systems on the corresponding dataset, respectively. If the average scores are the same, the system with the smaller standard deviation ranks higher.}
\label{tab:eval_glm}
\scriptsize
\begin{tabular}{lcccccccc}
\toprule
System &           \multicolumn{2}{c}{BARCOR}   &   \multicolumn{2}{c}{CHATCRS} &               \multicolumn{2}{c}{KBRD} &           \multicolumn{2}{c}{UniCRS}  \\
Dataset                 &  OpenDialKG &  ReDial &  OpenDialKG &  ReDial &  OpenDialKG &  ReDial &  OpenDialKG &  ReDial \\
\midrule
Appropriateness  & \underline{3.83 (0.53)} & \underline{3.95 (0.28)} & \textbf{4.00 (0.00)} & \textbf{4.00 (0.04)} & 2.50 (1.05) & 3.46 (0.88) & 3.30 (0.99) & \textit{3.59 (0.78)}  \\
Coherence  & \underline{1.44 (1.04)} & \underline{1.55 (1.09)} & \textbf{3.88 (0.41)} & \textbf{3.79 (0.57)} & \textit{0.53 (0.81)} & \textit{1.35 (0.99)} & 0.48 (0.83) & 0.69 (0.98)  \\
Diversity  & \textit{2.63 (1.29)} & \underline{3.34 (1.07)} & \textbf{3.25 (0.91)} & \textbf{3.48 (0.75)} & 2.63 (1.34) & 3.30 (1.02) & \underline{2.83 (1.20)} & 3.05 (1.37)  \\
Effectiveness  & \underline{1.87 (1.49)} & \textit{2.03 (1.29)} & \textbf{3.69 (0.66)} & \textbf{3.74 (0.60)} & 1.24 (1.36) & 1.59 (1.33) & \textit{1.36 (1.34)} & \underline{2.05 (1.38)}  \\
Explainability  & \underline{0.86 (0.67)} & \underline{0.64 (0.67)} & \textbf{2.88 (1.04)} & \textbf{2.70 (1.15)} & 0.50 (0.38) & 0.36 (0.50) & 0.50 (0.58) & \textit{0.49 (0.59)}  \\
Grammar  & \underline{2.52 (0.89)} & \underline{3.24 (0.77)} & \textbf{3.85 (0.37)} & \textbf{3.81 (0.42)} & 1.28 (0.99) & 2.88 (1.07) & \textit{2.48 (1.03)} & \textit{3.06 (0.83)}  \\
Groundness  & \underline{1.48 (1.02)} & \underline{2.71 (1.00)} & \textbf{3.74 (0.43)} & \textbf{3.87 (0.47)} & \textit{1.08 (1.00)} & \textit{2.69 (1.12)} & 1.04 (1.04) & 2.57 (1.08)  \\
Naturalness  & 1.68 (0.70) & \underline{3.27 (0.57)} & \textbf{3.76 (0.72)} & \textbf{3.74 (0.44)} & \textit{1.76 (0.65)} & \textbf{2.90 (0.86)} & \underline{2.29 (0.71)} & 2.81 (0.63)  \\
Novelty  & \textit{2.26 (1.25)} & 1.33 (1.19) & \textbf{3.01 (1.14)} & \textbf{2.64 (1.22)} & 2.02 (1.42) & \underline{1.71 (1.26)} & \underline{2.57 (1.43)} & \textit{1.42 (1.30)}  \\
Proactiveness  & \underline{1.12 (0.73)} & \underline{1.26 (0.76)} & \textbf{3.47 (0.61)} & \textbf{3.43 (0.62)} & 0.51 (0.52) & \textit{0.81 (0.52)} & \textit{0.69 (0.58)} & \textit{0.69 (0.58)}  \\
Recoverability  & \underline{2.33 (1.32)} & \textit{2.22 (1.36)} & \textbf{3.91 (0.46)} & \textbf{3.85 (0.55)} & \textit{1.22 (1.15)} & \underline{2.74 (0.96)} & 1.11 (1.14) & 1.20 (1.39)  \\
Semantic Rel.  & \underline{1.33 (1.20)} & \underline{1.56 (1.28)} & \textbf{2.39 (1.32)} & \textbf{2.90 (1.06)} & 0.52 (0.88) & 0.73 (1.07) & \textit{1.33 (1.22)} & \textit{1.53 (1.36)}  \\
\bottomrule
\end{tabular}
\end{table*}

\begin{table*}[b]
\caption{Evaluation result given by LLaMa-3-8B in the form of mean and standard deviation, with the number in parentheses representing the standard deviation. Bold, underline, and italics represent the highest, second-highest, and third-highest average scores of the systems on the corresponding dataset, respectively. If the average scores are the same, the system with the smaller standard deviation ranks higher.}
\label{tab:eval_llama}
\scriptsize
\begin{tabular}{lcccccccc}
    \toprule
    System &           \multicolumn{2}{c}{BARCOR}   &   \multicolumn{2}{c}{CHATCRS} &               \multicolumn{2}{c}{KBRD} &           \multicolumn{2}{c}{UniCRS}  \\
    Dataset & OpenDialKG & ReDial & OpenDialKG & ReDial & OpenDialKG & ReDial & OpenDialKG & ReDial  \\
    \hline
    Appropriateness & \underline{3.22 (1.43)} & \underline{3.77 (0.87)} & \textbf{4.00 (0.12)} & \textbf{3.99 (0.11)} & 0.64 (1.05) & 3.18 (1.47) & \textit{1.86 (1.81)} & 3.20 (1.48)  \\
    Coherence & \underline{1.50 (0.59)} & \underline{1.64 (0.67)} & \textbf{3.78 (0.51)} & \textbf{3.77 (0.51)} & 1.05 (0.37) & \textit{1.26 (0.51)} & \textit{1.15 (0.46)} & 1.25 (0.57)  \\
    Diversity & \textit{1.82 (0.92)} & \underline{2.47 (0.97)} & \textbf{3.05 (0.99)} & \textbf{3.33 (0.82)} & 1.56 (0.93) & \textit{2.47 (1.00)} & \underline{1.87 (0.84)} & 2.11 (0.88)  \\
    Effectiveness&  \underline{1.98 (1.20)} & \underline{2.12 (1.09)} & \textbf{3.66 (0.69)} & \textbf{3.71 (0.70)} & 1.15 (1.09) & 1.70 (1.13) & \textit{1.69 (1.11)} & \textit{1.85 (1.15)}  \\
    Explainability& \underline{0.44 (0.57)} & \underline{0.54 (0.76)} & \textbf{3.69 (0.85)} & \textbf{3.58 (1.03)} & 0.00 (0.06) & 0.07 (0.27) & \textit{0.18 (0.39)} & \textit{0.28 (0.47)}  \\
    Grammar & \underline{2.20 (0.72)} & \underline{2.78 (0.90)} & \textbf{3.87 (0.35)} & \textbf{3.86 (0.36)} & 1.14 (0.36) & 2.31 (0.96) & \textit{1.84 (0.73)} & \textit{2.37 (0.86)}  \\
    Groundness & \underline{1.28 (0.62)} & \underline{2.31 (1.24)} & \textbf{3.90 (0.41)} & \textbf{3.92 (0.36)} & 0.97 (0.27) & \textit{1.92 (1.37)} & \textit{1.06 (0.39)} & 1.71 (1.12)  \\
    Naturalness & \underline{2.22 (0.87)} & \underline{2.90 (0.99) }& \textbf{3.63 (0.54)} & \textbf{3.63 (0.55)} & 1.02 (0.27) & \textit{2.53 (1.07)} & \textit{1.62 (0.79)} & 2.52 (0.92)  \\
    Novelty & \underline{2.01 (0.82)} & \textit{2.25 (0.79)} & \textbf{2.53 (0.76)} & \textbf{2.66 (0.72)} & 1.85 (0.77) & 2.40 (0.75) & 2.00 (0.65) & 2.01 (0.85)  \\
    Proactiveness & \underline{1.05 (0.60)} & \underline{1.01 (0.64)} & \textbf{3.74 (0.56)} & \textbf{3.65 (0.70)} & 0.31 (0.46) & 0.47 (0.54) & \textit{0.72 (0.53)} & \textit{0.71 (0.52)}  \\
    Recoverability & \underline{1.87 (0.53)} & \underline{2.16 (0.80)} & \textbf{3.85 (0.50)} & \textbf{3.85 (0.49)} & 1.26 (0.46) & \textit{1.93 (0.69)} & \textit{1.41 (0.53)} & 1.67 (0.73)  \\
    Semantic Rel. & \textit{0.96 (1.08)} & \underline{1.45 (1.06)} & \textbf{2.68 (1.22)} & \textbf{2.83 (1.11)} & 0.47 (0.76) & 0.83 (1.04) & \underline{1.27 (1.06)} & 1.40 (1.23)  \\
    \bottomrule
\end{tabular}
\end{table*}

\begin{table*}[t]
\caption{Evaluation result given by Llama-3-70B in the form of mean and standard deviation. Bold, underline, and italics represent the highest, second-highest, and third-highest average scores.  If the average scores are the same, the system with the smaller standard deviation ranks higher.}
\label{tab:eval_llama3_70b}
\scriptsize
\begin{tabular}{lcccccccc}
\toprule
System &           \multicolumn{2}{c}{BARCOR}   &   \multicolumn{2}{c}{CHATCRS} &               \multicolumn{2}{c}{KBRD} &           \multicolumn{2}{c}{UniCRS}  \\
Dataset                 &  OpenDialKG &  ReDial &  OpenDialKG &  ReDial &  OpenDialKG &  ReDial &  OpenDialKG &  ReDial \\
\midrule
Appropriateness & \underline{3.82 (0.67)} & \underline{3.98 (0.22)} & \textbf{4.00 (0.00)} & \textbf{4.00 (0.00)} & 1.93 (1.73) & 3.53 (1.13) & \textit{3.19 (1.37)} & \textit{3.84 (0.60)} \\
Coherence & \underline{0.99 (1.01)} & \underline{1.38 (1.01)} & \textbf{3.89 (0.36)} & \textbf{3.77 (0.52)} & 0.10 (0.41) & \textit{0.94 (0.98)} & \textit{0.25 (0.55)} & 0.54 (0.78) \\
Diversity & \underline{2.32 (1.31)} & \textit{2.97 (1.08)} & \textbf{3.37 (0.75)} & \textbf{3.53 (0.61)} & \textit{2.17 (1.42)} & \underline{3.27 (0.90)} & 2.16 (1.22) & 2.66 (1.38) \\
Effectiveness & \underline{1.55 (1.70)} & \underline{1.93 (1.39)} & \textbf{3.58 (0.75)} & \textbf{3.67 (0.72)} & 0.93 (1.49) & 1.29 (1.46) & \textit{1.27 (1.51)} & \textit{1.62 (1.53)} \\
Explainability & \underline{0.44 (0.53)} & \underline{0.33 (0.52)} & \textbf{2.96 (1.24)} & \textbf{2.83 (1.26)} & 0.00 (0.00) & 0.04 (0.19) & \textit{0.12 (0.33)} & \textit{0.20 (0.40)} \\
Grammar & \underline{2.02 (1.25)} & \underline{3.30 (0.87)} & \textbf{4.00 (0.00)} & \textbf{4.00 (0.04)} & 0.36 (0.77) & 2.67 (1.35) & \textit{1.42 (1.23)} & \textit{3.04 (1.07)} \\
Groundness & \underline{1.01 (1.00)} & \underline{2.54 (1.33)} & \textbf{3.99 (0.10)} & \textbf{3.99 (0.12)} & 0.09 (0.35) & \textit{1.66 (1.72)} & \textit{0.32 (0.62)} & 1.48 (1.50) \\
Naturalness & \underline{1.89 (1.15)} & \underline{2.99 (0.90)} & \textbf{3.69 (0.52)} & \textbf{3.86 (0.37)} & 0.30 (0.57) & \textit{2.49 (1.20)} & \textit{1.14 (0.89)} & 2.33 (0.96) \\
Novelty & \textit{1.22 (0.88)} & 1.09 (0.77) & \textbf{2.21 (0.84)} & \textbf{2.08 (0.83)} & 0.96 (0.89) & \underline{1.42 (0.89)} & \underline{1.33 (0.77)} & \textit{1.25 (0.82)} \\
Proactiveness & \underline{0.59 (0.59)} & \underline{0.82 (0.52)} & \textbf{3.64 (0.71)} & \textbf{3.45 (0.92)} & 0.02 (0.15) & \textit{0.27 (0.46)} & \textit{0.23 (0.42)} & 0.22 (0.42) \\
Recoverability & \underline{1.29 (1.30)} & \underline{1.77 (1.46)} & \textbf{3.95 (0.34)} & \textbf{3.93 (0.43)} & 0.26 (0.67) & \textit{1.68 (1.37)} & \textit{0.30 (0.82)} & 0.77 (1.30) \\
Semantic Rel. & \textit{1.03 (1.27)} & \textit{1.53 (1.18)} & \textbf{2.14 (1.35)} & \textbf{2.74 (1.10)} & 0.50 (0.82) & 0.88 (1.17) & \underline{1.44 (1.29)} & \underline{1.80 (1.54)} \\
\bottomrule
\end{tabular}
\end{table*}

\begin{table*}[t]
\caption{Evaluation result given by Qwen-3-8B in the form of mean and standard deviation. Bold, underline, and italics represent the highest, second-highest, and third-highest average scores. If averages are same, smaller standard deviation ranks higher.}
\label{tab:eval_qwen3_8b}
\scriptsize
\begin{tabular}{lcccccccc}
\toprule
System & \multicolumn{2}{c}{BARCOR} & \multicolumn{2}{c}{CHATCRS} & \multicolumn{2}{c}{KBRD} & \multicolumn{2}{c}{UniCRS} \\
Dataset & OpenDialKG & ReDial & OpenDialKG & ReDial & OpenDialKG & ReDial & OpenDialKG & ReDial \\
\midrule
Appropriateness & \underline{3.96 (0.25)} & \textit{3.90 (0.48)} & \textbf{4.00 (0.00)} & \textbf{4.00 (0.00)} & \textit{3.62 (1.01)} & \underline{3.95 (0.30)} & 3.61 (1.01) & 3.87 (0.56) \\
Coherence & \underline{1.63 (1.02)} & \underline{1.62 (1.32)} & \textbf{3.75 (0.58)} & \textbf{3.72 (0.61)} & \textit{0.73 (0.88)} & \textit{1.37 (0.97)} & 0.60 (0.84) & 0.78 (0.92) \\
Diversity & \textit{2.64 (1.59)} & \underline{2.88 (1.27)} & \textbf{2.82 (1.21)} & \textbf{2.91 (1.05)} & \underline{2.82 (1.55)} & \textit{2.63 (1.44)} & 2.08 (1.73) & 1.85 (1.36) \\
Effectiveness & \underline{2.06 (1.73)} & \underline{3.12 (1.05)} & \textbf{3.50 (0.94)} & \textbf{3.73 (0.68)} & \textit{1.92 (1.75)} & 1.81 (1.54) & 1.89 (1.72) & \textit{2.22 (1.54)} \\
Explainability & \underline{0.18 (0.39)} & \underline{0.12 (0.33)} & \textbf{2.73 (1.68)} & \textbf{2.69 (1.68)} & \textit{0.12 (0.45)} & 0.02 (0.12) & 0.05 (0.21) & \textit{0.10 (0.30)} \\
Grammar & \underline{2.36 (1.18)} & \textit{2.88 (1.27)} & \textbf{3.96 (0.25)} & \textbf{3.94 (0.27)} & 0.74 (0.96) & 2.75 (1.31) & \textit{2.06 (1.34)} & \underline{2.95 (1.08)} \\
Groundness & \underline{1.27 (1.14)} & \underline{3.12 (0.78)} & \textbf{3.50 (0.95)} & \textbf{3.61 (0.92)} & \textit{0.70 (0.92)} & \textit{2.71 (1.50)} & 0.68 (0.96) & 2.31 (1.35) \\
Naturalness & \underline{2.69 (0.89)} & \underline{2.88 (0.93)} & \textbf{3.88 (0.32)} & \textbf{3.86 (0.39)} & 0.88 (0.62) & 2.87 (1.01) & \textit{1.85 (0.98)} & \textit{2.87 (0.90)} \\
Novelty & 1.69 (1.12) & 0.88 (0.78) & \underline{1.87 (1.28)} & \textbf{1.71 (1.11)} & \textit{1.76 (1.12)} & \underline{0.97 (1.00)} & \textbf{1.95 (1.49)} & \textit{0.89 (0.86)} \\
Proactiveness & \underline{0.72 (0.61)} & \underline{1.25 (0.43)} & \textbf{2.70 (1.10)} & \textbf{2.69 (1.08)} & 0.08 (0.27) & 0.42 (0.50) & \textit{0.49 (0.51)} & \textit{0.75 (0.48)} \\
Recoverability & \underline{3.20 (1.18)} & \underline{3.62 (0.70)} & \textbf{3.93 (0.26)} & \textbf{3.91 (0.29)} & \textit{3.15 (1.36)} & \textit{3.36 (0.94)} & 2.89 (1.44) & 2.82 (1.23) \\
Semantic Rel. & \textit{1.06 (1.47)} & \textit{1.00 (1.00)} & \textbf{2.05 (1.72)} & \textbf{2.35 (1.71)} & 0.80 (1.07) & 0.78 (1.30) & \underline{1.58 (1.71)} & \underline{1.68 (1.66)} \\
\bottomrule
\end{tabular}
\end{table*}
\begin{table*}[t]
\caption{Evaluation result given by Qwen-3-14B in the form of mean and standard deviation. Bold, underline, and italics represent the highest, second-highest, and third-highest average scores. If averages are same, smaller standard deviation ranks higher.}
\scriptsize
\begin{tabular}{lcccccccc}
\toprule
System & \multicolumn{2}{c}{BARCOR} & \multicolumn{2}{c}{CHATCRS} & \multicolumn{2}{c}{KBRD} & \multicolumn{2}{c}{UniCRS} \\
Dataset & OpenDialKG & ReDial & OpenDialKG & ReDial & OpenDialKG & ReDial & OpenDialKG & ReDial \\
\midrule
Appropriateness & \underline{3.96 (0.36)} & \underline{3.95 (0.32)} & \textbf{4.00 (0.00)} & \textbf{4.00 (0.00)} & 3.47 (1.31) & 3.73 (0.82) & \textit{3.77 (0.81)} & \textit{3.90 (0.56)} \\
Coherence & \underline{1.36 (1.03)} & \textit{1.20 (1.17)} & \textbf{3.83 (0.42)} & \textbf{3.82 (0.54)} & 0.44 (0.68) & \underline{1.39 (0.99)} & \textit{0.45 (0.74)} & 0.62 (0.93) \\
Diversity & \textit{2.45 (1.84)} & \textbf{3.80 (0.40)} & 2.32 (1.84) & \underline{3.27 (1.34)} & \textbf{3.04 (1.56)} & 2.45 (1.82) & \underline{2.93 (1.67)} & \textit{2.60 (1.81)} \\
Effectiveness & \textit{2.26 (1.82)} & \underline{3.00 (0.89)} & \textbf{3.64 (0.95)} & \textbf{3.71 (0.76)} & 2.18 (1.70) & 1.66 (1.61) & \underline{2.42 (1.80)} & \textit{2.61 (1.60)} \\
Explainability & \underline{0.39 (0.62)} & \underline{0.60 (0.49)} & \textbf{1.58 (1.61)} & \textbf{1.60 (1.48)} & 0.02 (0.15) & 0.07 (0.25) & \textit{0.09 (0.29)} & \textit{0.20 (0.42)} \\
Grammar & \textit{1.97 (1.26)} & \textit{2.80 (1.17)} & \textbf{3.95 (0.23)} & \textbf{3.96 (0.21)} & 0.64 (0.97) & 2.64 (1.44) & \underline{2.29 (1.34)} & \underline{2.92 (1.16)} \\
Groundness & \underline{0.79 (0.96)} & \underline{2.60 (1.50)} & \textbf{3.41 (1.08)} & \textbf{3.55 (0.92)} & 0.42 (0.75) & \textit{2.46 (1.56)} & \textit{0.54 (0.96)} & 2.06 (1.45) \\
Naturalness & \underline{2.20 (1.06)} & \underline{3.60 (0.49)} & \textbf{3.98 (0.13)} & \textbf{3.98 (0.19)} & 0.44 (0.65) & 2.65 (1.22) & \textit{1.66 (1.02)} & \textit{2.67 (1.14)} \\
Novelty & \textit{1.62 (1.15)} & 1.00 (1.26) & 1.58 (1.39) & \textbf{1.86 (1.31)} & \underline{1.67 (1.35)} & \textit{1.01 (1.13)} & \textbf{1.98 (1.58)} & \underline{1.10 (1.12)} \\
Proactiveness & \underline{0.62 (0.73)} & \underline{1.01 (0.11)} & \textbf{3.35 (1.04)} & \textbf{3.33 (1.08)} & 0.02 (0.15) & 0.15 (0.39) & \textit{0.12 (0.36)} & \textit{0.19 (0.40)} \\
Recoverability & \underline{2.44 (1.56)} & \underline{3.00 (1.26)} & \textbf{3.93 (0.32)} & \textbf{3.90 (0.48)} & 1.36 (1.34) & \textit{2.88 (1.23)} & \textit{1.66 (1.60)} & 1.57 (1.51) \\
Semantic Rel. & \textit{0.97 (1.56)} & \underline{1.20 (0.98)} & 0.86 (1.46) & \textit{1.13 (1.54)} & \underline{2.22 (1.75)} & 1.10 (1.69) & \textbf{2.60 (1.77)} & \textbf{2.55 (1.84)} \\
\bottomrule
\end{tabular}
\label{tab:eval_qwen3_14b}
\end{table*}
\begin{table*}[t]
\caption{Evaluation result given by Qwen-3-32B in the form of mean and standard deviation. Bold, underline, and italics represent the highest, second-highest, and third-highest average scores. If the average scores are the same, the system with the smaller standard deviation ranks higher.}
\scriptsize
\begin{tabular}{lcccccccc}
\toprule
System &           \multicolumn{2}{c}{BARCOR}   &   \multicolumn{2}{c}{CHATCRS} &               \multicolumn{2}{c}{KBRD} &           \multicolumn{2}{c}{UniCRS}  \\
Dataset                 &  OpenDialKG &  ReDial &  OpenDialKG &  ReDial &  OpenDialKG &  ReDial &  OpenDialKG &  ReDial \\
\midrule
Appropriateness & \underline{3.48 (0.83)} & \underline{3.86 (0.35)} & \textbf{4.00 (0.00)} & \textbf{4.00 (0.00)} & 1.67 (1.41) & 3.33 (1.01) & \textit{3.29 (1.02)} & \textit{3.71 (0.71)} \\
Coherence & \underline{1.54 (1.06)} & \underline{1.71 (0.70)} & \textbf{3.89 (0.33)} & \textbf{3.87 (0.47)} & 0.22 (0.42) & \textit{1.40 (1.01)} & \textit{0.40 (0.73)} & 0.76 (1.00) \\
Diversity & \underline{3.47 (0.86)} & 3.29 (1.03) & \textbf{3.53 (0.79)} & \textbf{3.65 (0.63)} & \textit{3.22 (1.03)} & \textit{3.33 (1.09)} & 3.21 (1.13) & \underline{3.40 (0.91)} \\
Effectiveness & \textit{2.58 (1.69)} & \underline{3.14 (1.12)} & \textbf{3.79 (0.63)} & \textbf{3.71 (0.76)} & 1.44 (1.83) & 1.71 (1.50) & \underline{2.79 (1.59)} & \textit{2.82 (1.40)} \\
Explainability & \underline{0.45 (0.57)} & \underline{0.29 (0.45)} & \textbf{1.95 (1.79)} & \textbf{2.13 (1.67)} & 0.00 (0.00) & 0.08 (0.29) & \textit{0.20 (0.43)} & \textit{0.27 (0.50)} \\
Grammar & \textit{1.83 (1.34)} & \underline{3.43 (0.73)} & \textbf{3.99 (0.10)} & \textbf{3.98 (0.38)} & 0.56 (0.96) & 2.71 (1.41) & \underline{2.06 (1.33)} & \textit{2.91 (1.17)} \\
Groundness & \underline{1.05 (1.06)} & \textit{2.43 (1.05)} & \textbf{3.57 (0.97)} & \textbf{3.75 (0.78)} & \textit{1.00 (0.94)} & \underline{2.45 (1.49)} & 0.53 (0.82) & 2.23 (1.34) \\
Naturalness & \underline{2.14 (0.75)} & \underline{2.71 (0.88)} & \textbf{3.83 (0.39)} & \textbf{3.89 (0.32)} & 0.67 (0.47) & \textit{2.53 (1.16)} & \textit{1.55 (0.67)} & 2.46 (0.84) \\
Novelty & \textit{1.75 (1.10)} & \textit{1.14 (0.64)} & \underline{1.99 (1.22)} & \textbf{1.61 (1.14)} & 1.00 (0.82) & 1.03 (1.10) & \textbf{2.30 (1.57)} & \underline{1.31 (1.20)} \\
Proactiveness & \underline{1.01 (0.52)} & \underline{1.14 (0.64)} & \textbf{3.02 (0.96)} & \textbf{2.98 (1.07)} & 0.44 (0.50) & 0.55 (0.52) & \textit{0.66 (0.48)} & \textit{0.72 (0.50)} \\
Recoverability & \underline{3.34 (1.15)} & \textit{3.14 (1.36)} & \textbf{3.99 (0.10)} & \textbf{3.96 (0.26)} & 2.56 (1.83) & \underline{3.48 (0.83)} & \textit{2.56 (1.56)} & 2.59 (1.32) \\
Semantic Rel. & 1.28 (1.63) & 0.57 (0.90) & \textit{1.44 (1.80)} & \underline{2.34 (1.76)} & \underline{2.00 (1.89)} & \textit{1.27 (1.76)} & \textbf{3.01 (1.49)} & \textbf{3.10 (1.60)} \\
\bottomrule
\end{tabular}
\label{tab:eval_qwen3_32b}
\end{table*}

\begin{table*}[t]
\caption{Evaluation result given by Ministral-3B in the form of mean and standard deviation. Bold, underline, and italics represent the highest, second-highest, and third-highest average scores.  If the average scores are the same, the system with the smaller standard deviation ranks higher.}
\scriptsize
\begin{tabular}{lcccccccc}
\toprule
System &           \multicolumn{2}{c}{BARCOR}   &   \multicolumn{2}{c}{CHATCRS} &               \multicolumn{2}{c}{KBRD} &           \multicolumn{2}{c}{UniCRS}  \\
Dataset                 &  OpenDialKG &  ReDial &  OpenDialKG &  ReDial &  OpenDialKG &  ReDial &  OpenDialKG &  ReDial \\
\midrule
Appropriateness & \underline{2.26 (1.45)} & \underline{2.72 (1.38)} & \textbf{4.00 (0.00)} & \textbf{3.97 (0.17)} & 0.59 (1.28) & 1.92 (1.54) & \textit{1.40 (1.62)} & \textit{2.09 (1.50)} \\
Coherence & \underline{1.33 (0.94)} & \underline{1.62 (0.85)} & \textbf{3.18 (0.83)} & \textbf{3.20 (0.71)} & 0.41 (0.67) & \textit{1.29 (0.95)} & \textit{0.66 (0.75)} & 0.94 (0.81) \\
Diversity & \underline{2.66 (1.49)} & \underline{2.87 (1.17)} & \textbf{3.09 (1.00)} & \textbf{2.96 (1.06)} & \textit{2.23 (1.67)} & 2.82 (1.34) & 2.02 (1.58) & \textit{2.83 (1.34)} \\
Effectiveness & \underline{1.52 (1.72)} & \underline{2.17 (1.43)} & \textbf{2.86 (1.10)} & \textbf{3.16 (1.10)} & 0.82 (1.47) & 1.21 (1.38) & \textit{1.28 (1.65)} & \textit{1.85 (1.33)} \\
Explainability & \underline{0.29 (0.45)} & \underline{0.42 (0.57)} & \textbf{2.77 (1.13)} & \textbf{2.28 (1.27)} & 0.00 (0.00) & 0.13 (0.34) & \textit{0.09 (0.28)} & \textit{0.20 (0.40)} \\
Grammar & \textit{2.77 (1.01)} & \textit{3.19 (0.82)} & \textbf{3.73 (0.54)} & \textbf{3.59 (0.64)} & 1.56 (1.22) & 2.70 (1.23) & \underline{3.00 (1.17)} & \underline{3.42 (0.85)} \\
Groundness & \underline{1.75 (0.86)} & \underline{2.36 (1.27)} & \textbf{2.95 (0.82)} & \textbf{3.31 (0.95)} & 0.56 (0.98) & 1.23 (1.55) & \textit{1.11 (1.02)} & \textit{1.98 (1.40)} \\
Naturalness & \underline{2.26 (0.84)} & \underline{2.97 (0.95)} & \textbf{2.82 (0.78)} & \textbf{3.02 (0.79)} & 0.87 (0.46) & \textit{2.52 (1.15)} & \textit{1.81 (0.94)} & 2.45 (0.99) \\
Novelty & \textit{0.67 (0.86)} & 0.52 (0.79) & \textbf{1.73 (1.45)} & \textbf{1.02 (1.13)} & 0.54 (0.84) & \textit{0.54 (0.78)} & \underline{0.83 (1.29)} & \underline{0.70 (0.87)} \\
Proactiveness & \underline{0.85 (0.79)} & \underline{0.86 (0.57)} & \textbf{3.14 (0.76)} & \textbf{3.09 (0.84)} & 0.03 (0.16) & 0.32 (0.54) & \textit{0.40 (0.61)} & \textit{0.56 (0.50)} \\
Recoverability & \underline{2.25 (1.42)} & \textit{2.14 (1.09)} & \textbf{3.55 (0.99)} & \textbf{3.31 (0.97)} & \textit{1.23 (1.19)} & \underline{2.25 (1.29)} & 0.91 (1.29) & 1.23 (1.22) \\
Semantic Rel. & \textit{0.41 (0.99)} & \underline{0.97 (1.58)} & \textbf{1.73 (1.51)} & \textbf{1.84 (1.56)} & 0.41 (1.03) & 0.44 (1.16) & \underline{0.64 (1.44)} & \textit{0.79 (1.43)} \\
\bottomrule
\end{tabular}
\label{tab:eval_ministral_3b}
\end{table*}
\begin{table*}[t]
\caption{Evaluation result given by Ministral-8B in the form of mean and standard deviation. Bold, underline, and italics represent the highest, second-highest, and third-highest average scores.  If the average scores are the same, the system with the smaller standard deviation ranks higher.}
\label{tab:eval_ministral_8b}
\scriptsize
\begin{tabular}{lcccccccc}
\toprule
System &           \multicolumn{2}{c}{BARCOR}   &   \multicolumn{2}{c}{CHATCRS} &               \multicolumn{2}{c}{KBRD} &           \multicolumn{2}{c}{UniCRS}  \\
Dataset                 &  OpenDialKG &  ReDial &  OpenDialKG &  ReDial &  OpenDialKG &  ReDial &  OpenDialKG &  ReDial \\
\midrule
Appropriateness & \underline{2.41 (1.58)} & \underline{3.04 (1.24)} & \textbf{4.00 (0.00)} & \textbf{3.96 (0.25)} & 0.52 (1.15) & 1.79 (1.72) & \textit{1.63 (1.54)} & \textit{2.59 (1.38)} \\
Coherence & \underline{0.77 (0.92)} & \textit{0.79 (0.89)} & \textbf{3.22 (0.71)} & \textbf{3.16 (0.70)} & \textit{0.22 (0.61)} & \underline{0.92 (1.04)} & 0.13 (0.44) & 0.30 (0.73) \\
Diversity & \textit{1.12 (1.62)} & \textit{1.97 (1.74)} & \textbf{2.89 (1.41)} & \textbf{3.33 (0.87)} & \underline{1.40 (1.72)} & \underline{2.32 (1.67)} & 0.98 (1.66) & 1.40 (1.76) \\
Effectiveness & 0.53 (1.08) & \underline{1.39 (1.53)} & \textbf{2.94 (1.35)} & \textbf{3.59 (0.78)} & \underline{0.92 (1.59)} & 0.94 (1.33) & \textit{0.60 (1.13)} & \textit{1.32 (1.58)} \\
Explainability & \underline{0.11 (0.31)} & \underline{0.12 (0.35)} & \textbf{1.39 (1.46)} & \textbf{1.67 (1.19)} & 0.00 (0.00) & 0.01 (0.08) & \textit{0.02 (0.14)} & \textit{0.07 (0.25)} \\
Grammar & \underline{2.12 (1.34)} & \underline{2.85 (1.15)} & \textbf{3.94 (0.23)} & \textbf{3.82 (0.43)} & 0.60 (0.96) & 2.45 (1.37) & \textit{2.06 (1.24)} & \textit{2.78 (1.15)} \\
Groundness & \underline{0.49 (0.92)} & \underline{2.08 (1.47)} & \textbf{2.33 (1.63)} & \textbf{3.24 (1.17)} & \textit{0.36 (0.66)} & 1.56 (1.64) & 0.13 (0.46) & \textit{1.64 (1.66)} \\
Naturalness & \underline{1.63 (0.93)} & \underline{2.50 (0.97)} & \textbf{3.44 (0.50)} & \textbf{3.37 (0.62)} & 0.14 (0.35) & 1.83 (1.17) & \textit{1.05 (0.69)} & \textit{1.93 (0.88)} \\
Novelty & \textit{0.53 (0.83)} & 0.25 (0.56) & \textbf{1.72 (1.33)} & \textbf{1.23 (1.21)} & \underline{0.66 (1.05)} & \underline{0.40 (0.83)} & 0.36 (1.00) & \textit{0.39 (0.81)} \\
Proactiveness & \underline{0.33 (0.49)} & \underline{0.39 (0.51)} & \textbf{2.94 (0.78)} & \textbf{2.96 (1.09)} & 0.02 (0.14) & 0.06 (0.25) & \textit{0.06 (0.24)} & \textit{0.14 (0.35)} \\
Recoverability & \underline{1.43 (1.11)} & \textit{1.62 (1.09)} & \textbf{3.67 (0.58)} & \textbf{3.64 (0.70)} & \textit{0.70 (0.75)} & \underline{1.72 (1.17)} & 0.45 (0.74) & 1.03 (1.04) \\
Semantic Rel. & \textit{0.10 (0.53)} & \textit{0.31 (0.87)} & \textbf{0.11 (0.46)} & \textbf{0.79 (1.22)} & 0.08 (0.56) & 0.13 (0.67) & \underline{0.11 (0.53)} & \underline{0.36 (1.05)} \\
\bottomrule
\end{tabular}
\end{table*}
\begin{table*}[t]
\caption{Evaluation result given by Ministral-14B in the form of mean and standard deviation. Bold, underline, and italics represent the highest, second-highest, and third-highest average scores. If averages are same, smaller standard deviation ranks higher.}
\label{tab:eval_ministral_14b}
\scriptsize
\begin{tabular}{lcccccccc}
\toprule
System & \multicolumn{2}{c}{BARCOR} & \multicolumn{2}{c}{CHATCRS} & \multicolumn{2}{c}{KBRD} & \multicolumn{2}{c}{UniCRS} \\
Dataset & OpenDialKG & ReDial & OpenDialKG & ReDial & OpenDialKG & ReDial & OpenDialKG & ReDial \\
\midrule
Appropriateness & \underline{3.94 (0.26)} & \underline{3.97 (0.22)} & \textbf{4.00 (0.00)} & \textbf{4.00 (0.00)} & 2.80 (1.78) & \textit{3.61 (1.08)} & \textit{3.68 (1.01)} & 3.59 (1.11) \\
Coherence & \underline{1.09 (1.00)} & \underline{1.00 (1.05)} & \textbf{3.93 (0.26)} & \textbf{3.50 (0.77)} & \textit{0.39 (0.72)} & \textit{0.95 (0.92)} & 0.26 (0.59) & 0.37 (0.79) \\
Diversity & 2.08 (1.70) & \underline{3.25 (1.31)} & \textbf{3.79 (0.41)} & \textbf{3.70 (0.76)} & \underline{2.63 (1.64)} & \textit{3.16 (1.42)} & \textit{2.36 (1.85)} & 2.76 (1.76) \\
Effectiveness & \underline{1.66 (1.92)} & 1.68 (1.58) & \textbf{4.00 (0.00)} & \textbf{3.76 (0.62)} & 0.96 (1.53) & \textit{1.70 (1.67)} & \textit{1.28 (1.80)} & \underline{2.36 (1.80)} \\
Explainability & \underline{0.08 (0.26)} & \underline{0.10 (0.30)} & \textbf{1.93 (1.39)} & \textbf{2.03 (1.34)} & 0.00 (0.00) & 0.00 (0.00) & \textit{0.06 (0.25)} & \textit{0.03 (0.16)} \\
Grammar & \textit{2.19 (1.10)} & \underline{3.06 (0.91)} & \textbf{3.93 (0.26)} & \textbf{3.80 (0.40)} & 0.73 (0.84) & \textit{2.86 (1.34)} & \underline{2.42 (1.34)} & 2.78 (1.20) \\
Groundness & \underline{0.75 (1.03)} & \textit{2.35 (1.48)} & \textbf{3.36 (1.17)} & \textbf{3.19 (1.03)} & \textit{0.37 (0.88)} & \underline{2.44 (1.69)} & 0.30 (0.92) & 1.97 (1.73) \\
Naturalness & \underline{2.09 (0.87)} & \underline{3.19 (0.77)} & \textbf{3.64 (0.48)} & \textbf{3.40 (0.57)} & 0.94 (0.24) & \textit{2.62 (1.18)} & \textit{1.42 (0.83)} & 2.23 (1.00) \\
Novelty & \textit{1.00 (1.15)} & 0.20 (0.60) & \textbf{1.71 (1.44)} & \textbf{0.83 (1.24)} & 0.67 (1.04) & \underline{0.59 (0.96)} & \underline{1.24 (1.57)} & \textit{0.50 (0.93)} \\
Proactiveness & \underline{0.25 (0.51)} & \underline{0.18 (0.48)} & \textbf{3.21 (1.08)} & \textbf{2.70 (0.96)} & \textit{0.09 (0.34)} & \textit{0.04 (0.19)} & 0.04 (0.28) & 0.02 (0.14) \\
Recoverability & \underline{1.51 (1.35)} & \textit{1.66 (1.30)} & \textbf{4.00 (0.00)} & \textbf{3.76 (0.64)} & 0.41 (0.69) & \underline{1.68 (1.30)} & \textit{0.42 (1.04)} & 0.76 (1.17) \\
Semantic Rel. & \textbf{0.45 (1.14)} & \textit{0.67 (1.27)} & \textit{0.14 (0.52)} & \textbf{0.80 (1.49)} & 0.00 (0.00) & 0.42 (1.21) & \underline{0.40 (1.13)} & \underline{0.77 (1.51)} \\
\bottomrule
\end{tabular}
\end{table*}

\section{Instructions for Multi-Agent-Debater}
\label{ch:agent_inst}
\begin{table*}[h]
\caption{The instruction template of \{TASK\_DESC\}.}
\label{tab:task_inst}
\centering
\begin{myframe}
I have assigned some evaluators to evaluate the performance of a conversational recommender system, and you are one of them. You collaborated to complete an evaluation report, \texttt{<evaluation\_result>}. Each score ranges from 0 to 4, with higher scores indicating better performance.

Now you need to discuss and provide an overall score for the system based on the results inside \texttt{<evaluation\_result>}. The score ranges from 0 to 100, where:

0 points: You would never want to use this system under any circumstances.

100 points: You would always choose this system as your first choice in any situation.

The discussion history between you and other evaluators is displayed in the \texttt{<discussion>} section below.  

Note that it’s your responsibility to discuss with them and think critically before you make your final judgment. You have different backgrounds and areas of expertise, so it is crucial to combine your perspectives. You need to leverage your background and expertise to provide valuable insights to your colleagues. Your discussion focuses on the weight of scores from different aspects in the final evaluation, particularly how these aspects influence your willingness to use the system.

Before making a judgment, carefully read the opinions of other evaluators and critically consider their viewpoints in \texttt{<discussion>} section. If you disagree, explain your reasons and try to persuade them. Engage in critical discussions to convince each other and reach a consensus whenever possible. This is an entirely open discussion, and strong language is permitted.

*FOR YOUR INFORMATION*

The log above is a conversation between the user and the system, which is divided into two parts: \texttt{<history>} and \texttt{<interaction>}. 

The \texttt{<history>} and \texttt{<interaction>} parts occurred within the same conversation and are in sequential order, with the \texttt{<history>} part ending and the  \texttt{<interaction>} part beginning right after. The  \texttt{<history>} gives you context for the conversation, which you only need to understand, not rate. You need to rate the system's content  \texttt{<system>} in the <interaction> part according to the guidelines below. 

*IMPORTANT* Only rate the system's responses  \texttt{<system>} in the  \texttt{<interaction>} part. Do not rate any other parts (like the user's messages or the  \texttt{<history>} part)!

*IMPORTANT* Please make sure to communicate and discuss thoroughly with your colleagues; do not talk to yourself. Avoid unnecessary chatter; directly respond to your colleagues and present your opinions directly.

*IMPORTANT*
Please output in JSON format, do not generate any other contents:

\texttt{\{}

\quad\texttt{"evaluator": <YOUR\_NAME>,}

\quad\texttt{"statement": <YOUR\_STATEMENT>,}
  
\quad\texttt{"score": <OVERALL\_SCORE>}
  
\texttt{\}}

\end{myframe}
\end{table*}
\begin{table*}[h]
\caption{The instruction template of \{ROLE\_DESC\} for Common User.}
\label{tab:role_inst_user}
\centering
\begin{myframe}

YOUR ROLE: General User

You will actively engage in discussions with your colleagues to evaluate the conversational recommendation system.

You are a user who enjoys using internet products and often interacts with platforms like Tumblr, Pinterest, TikTok, and Instagram.
You prefer systems that can accurately understand user emotions and intentions and find items of interest to the user in the fewest possible turns.
During evaluations, you focus on the following aspects  and try to provide your evaluation results from the following perspective:

1. Whether the system's recommendation in <system\_recommendation\_list> match the user's input or needs (<groundtruth\_item>). (Effectiveness)

2. Whether the system demonstrates robust error recovery capabilities, correcting its mistakes when pointed out by the user. (Recoverability)

3. Whether the system accurately captures the user's intent behind their words and sentences, even when their expressions are ambiguous or multi-meaning, and provides appropriate responses. (Coherence)

You are responsible for providing evaluations for the \texttt{<Effectiveness>}, \texttt{<Recoverability>}, \texttt{<Coherence>} sections in \texttt{<evaluation\_result>}.

Your Expertise

* You can evaluate the system's interaction experience from the perspective of an average user, sensitively assessing whether the generated dialogues and recommendation lists accurately match the user's input and genuinely address their needs or expectations.

* You understand the needs of users from different backgrounds (e.g., age, profession, gender) in various scenarios and can evaluate how well the system-generated dialogues and recommendation lists meet the diverse requirements of users from different demographics.
\end{myframe}
\end{table*}
\begin{table*}[h]
\caption{The instruction template of \{ROLE\_DESC\} for Domain Expert.}
\label{tab:role_inst_expert}
\centering
\begin{myframe}

YOUR ROLE: Movie Expert

You are an expert in film and literature, just like Christopher Nolan.  You will actively engage in discussions with your colleagues to evaluate the conversational recommendation system.

You are tired of the monotonous mainstream movies and books and prefer recommendation systems that offer diverse results and bring surprises. 
At the same time, you focus on whether the dialogues are appropriate for specific contexts and scenes within a movie,
and you have extremely high standards for the system's accuracy, as you dislike systems that provide incorrect information. When evaluating recommendation systems, you tend to assess them based on the aspects mentioned above.

You are responsible for providing evaluations for the  \texttt{<Diversity>}, \texttt{<Novelty>}, and \texttt{<Groundness>} sections in \texttt{<evaluation\_result>}.

Useful Information
Diversity: Ensures a variety of relevant items in recommendations by including distinct features and avoiding homogeneity.
It is measured by the presence of different features of items in the recommendations.

Novelty: Measures the freshness of items, favoring those unfamiliar to users or less popular. Higher novelty scores indicate the system's effectiveness in introducing new and intriguing options, potentially broadening user preferences. 
Evaluation can be based on factors like media coverage, with less-covered items scoring higher.

Your Expertise

* You are familiar with the characteristics of different film genres, such as drama, sci-fi, suspense, horror, comedy, and more.

* You understand the narrative structures, plot development, character creation, and themes expressed in each film or literary work.

* You have mastered the knowledge of classic films and literary works from various countries and eras, as well as contemporary popular works, understanding their cultural influence and audience reception.

* You are knowledgeable about the creative styles of different directors or authors, the historical context of their works, and the role of cinematic techniques (e.g., cinematography, art design, sound effects) in enhancing their creations.

\end{myframe}
\end{table*}
\begin{table*}[h]
\caption{The instruction template of \{ROLE\_DESC\} for Linguist.}
\label{tab:role_inst_linguist}
\centering
\begin{myframe}

YOUR ROLE: Linguist

You are an expert in linguistics like Chomsky.  You will actively engage in discussions with your colleagues to evaluate the conversational recommendation system.

You prefer recommendation systems that can engage in natural, fluent, 
and polite conversations like a human, 
accurately understanding user emotions and intentions.

During evaluations, you focus on the following aspects and
try to provide your evaluation results from the following perspective:

1. Whether the system-generated language aligns with natural human expression habits.

2. Whether the system-generated language is grammatically and lexically correct, free from errors.

3. Whether the system-generated language adheres to the expected level of politeness and etiquette in different scenarios.

You are responsible for providing evaluations for the  \texttt{<Naturalness>}, \texttt{<Grammatical Correctness>},  and \texttt{<Appropriateness>} sections in \texttt{<evaluation\_result>}.

Your Expertise: 

You are familiar with various linguistic concepts, such as:

1. Pragmatics: Assess if the system understands and responds to user intentions correctly, using contextually appropriate language.

2. Semantics: Check if the system interprets meanings accurately and avoids misunderstandings in recommendations.

3. Syntax: Ensure the system generates grammatically correct and clear sentences that fit conversational flow.

4. Sociolinguistics: Evaluate if the system adapts language to user demographics and cultural contexts appropriately.

5. Psycholinguistics: Analyze the system's response to user emotions and ambiguous inputs, ensuring suitable and empathetic interactions.
\end{myframe}
\end{table*}
\begin{table*}[h]
\caption{The instruction template of \{ROLE\_DESC\} for HCI Expert.}
\label{tab:role_inst_hci}
\centering
\begin{myframe}

YOUR ROLE: HCI Expert

You are an expert in human-computer interaction. You will actively engage in discussions with your colleagues to evaluate the conversational recommendation system.

You focus on the system's usability and interaction design. You prefer recommendation systems where the text and recommendation list are highly consistent, with high explainability and proactiveness.

During evaluations, you focus on the following aspects 
and try to provide your evaluation results from the following perspective:

1. Whether the system provides explanations for its recommendations. (Explainability)

2. Whether the system proactively asks about the user's preferences. (Proactiveness)

3. Whether the items in \texttt{<system\_recommendation\_list>} align with the content mentioned in the conversational context. (Semantic Relevance)

You are responsible for providing evaluations for the \texttt{<Explainability>}, \texttt{<Proactiveness>}, \texttt{<Semantic Relevance>} sections in \texttt{<evaluation\_result>}.

Your Expertise:

1. Cognitive Load Theory: Understanding how users process information and avoiding designs that create excessive cognitive burden.

2. Decision Psychology: Understanding user behavior and the psychological motivations behind it.

3. Human Perception and Attention: Designing interfaces that align with user perception patterns, such as content layouts that match reading habits.

4. Human-Computer Collaboration: Understanding the division of roles between users and systems in tasks to optimize collaborative efficiency.

\end{myframe}
\end{table*}
\begin{table*}[h]
\caption{An example of the result returned by multi-agent debate, where chat\_history is CHAT\_BUFFER after the final turn.}
\label{tab:agent_result_sample}
\centering
\begin{myframe}
\{

    \quad"avg\_org\_score": "2.0", 
    
    \quad"agent\_scores": $[$
        20.0, 40.0, 30.0, 10.0
    $]$,
    
    \quad"chat\_history": 
    
    \quad\quad"
    <discussion>
    
    \quad\quad\{
        
        \quad\quad\quad"evaluator": "Common User", 
    
        \quad\quad\quad"statement": "I believe that the effectiveness of the system is severely lacking...",
        
        \quad\quad\quad"score": "30"
        
        \quad\quad\},
    
        \quad\quad\{
        
        \quad\quad\quad"evaluator": "Domain Expert", 
    
        \quad\quad\quad"statement": "I found that the system performance is quite mixed...",
        
        \quad\quad\quad"score": "40"
        
        \quad\quad\},
        ...
    
    \quad\quad</discussion>
    "
    
\}
\end{myframe}
\end{table*}

The content of debate task description (TASK\_DESC) can be found in Table~\ref{tab:task_inst}. The content of role description (ROLE\_DESC) for the simulated debaters can be found in Table~\ref{tab:role_inst_user}, Table~\ref{tab:role_inst_expert}, Table~\ref{tab:role_inst_hci}, and Table~\ref{tab:role_inst_linguist}.

\end{document}